\documentclass[secnumarabic,amssymb, nobibnotes, aps, prd]{revtex4-2}
\usepackage{graphicx}
\usepackage{amsmath}
\usepackage{subcaption}
\usepackage{xcolor}
\usepackage{multirow}

\begin{document}
	\title{Thermodynamics of Rotating AdS black holes in Kaniadakis statistics}
	\author{Bidyut Hazarika$^1$}
	
	\email{$rs_bidyuthazarika@dibru.ac.in$}
	\author{Amijit Bhattacharjee$^1$}
	\email{$rs_amijitbhattacharjee@dibru.ac.in$}
	\author{Prabwal Phukon$^{1,2}$}
	\email{prabwal@dibru.ac.in}	
	\affiliation{$1.$Department of Physics, Dibrugarh University, Dibrugarh, Assam,786004.\\$2.$Theoretical Physics Division, Centre for Atmospheric Studies, Dibrugarh University, Dibrugarh, Assam,786004.}
	\begin{abstract}
In this study, we investigate the thermodynamic properties and phase transitions in rotating anti-de Sitter (AdS) black holes by applying the Kaniadakis(KD) entropy framework.   To achieve this, we analyze three prominent rotating AdS black hole systems: the Kerr AdS black hole, the Kerr-Sen AdS black hole, and the Kerr-Newman AdS black hole. We assess their thermodynamic quantities, phase transitions,  thermodynamic topology and  thermodynamic geometry within the Kaniadakis statistical framework.  We observed that Kaniadakis entropy introduces an entropy bound beyond which the black hole solutions become thermally unstable, unlike the Gibbs-Boltzmann framework, where stability persists across infinite Bekenstein-Hawking  entropy range.This bound is controlled by the Kaniadakis parameter \(\kappa\), with smaller \(\kappa\) values allowing stability over a broader entropy range. To illustrate these changes, we examine the free energy landscape, which highlights alterations in the phase structure and the stability of black holes.. Thermodynamic topology further indicates that the topological class of these black holes changes from 1 to 0 when transitioning from GB to KD statistics. Along with changes in the topological charge, the number of creation and annihilation points also changes. Notably, the topological charge remains independent of all thermodynamic parameters in both GB and KD statistics.We discuss the thermodynamic geometry of rotating AdS black holes using two different formalisms: Ruppeiner and Geometrothermodynamic (GTD). Our analysis uncovers unequal number of singularities in the scalar curvature within both frameworks. In the Ruppeiner formalism, these singularities do not coincide with the discontinuities observed in the heat capacity curves. In contrast, the GTD formalism shows that the singularities in the scalar curvature align with the discontinuities in the heat capacity curves.
	\end{abstract}
	\maketitle
	\tableofcontents
	\section{Introduction}
	Black holes, among the most intriguing objects predicted by Einstein's general theory of relativity.   The pioneering works on black hole thermodynamics  revealed that black holes are thermodynamic quantities characterized by quantities such as entropy, temperature,  governed by the principles of thermodynamics \cite{Bekenstein:1973ur,Hawking:1974rv,Hawking:1975vcx,Bardeen:1973gs}. This foundational work paved the way for further discoveries in the field \cite{Wald:1979zz,bekenstein1980black,Wald:1999vt,Carlip:2014pma,Wall:2018ydq,Candelas:1977zz,Mahapatra:2011si}.A particularly intriguing aspect of black hole thermodynamics is the study of phase transitions \cite{Davies:1989ey,Hawking:1982dh,curir_rotating_1981,Curir1981,Pavon:1988in,Pavon:1991kh,OKaburaki,Cai:1996df,Cai:1998ep,Wei:2009zzf,Bhattacharya:2019awq,Kastor:2009wy,Dolan:2010ha,Dolan:2011xt,Dolan:2011jm,Dolan:2012jh,Kubiznak:2012wp,Kubiznak:2016qmn,Bhattacharya:2017nru}. Black holes undergo various types of phase transitions, including Davies-type transitions \cite{Davies:1989ey}, Hawking–Page transitions \cite{Hawking:1982dh}, extremal transitions (involving shifts between non-extremal and extremal states) \cite{curir_rotating_1981,Curir1981,Pavon:1988in,Pavon:1991kh,OKaburaki,Cai:1996df,Cai:1998ep,Wei:2009zzf,Bhattacharya:2019awq}, and transitions analogous to van der Waals behavior \cite{Kastor:2009wy,Dolan:2010ha,Dolan:2011xt,Dolan:2011jm,Dolan:2012jh,Kubiznak:2012wp,Kubiznak:2016qmn,Bhattacharya:2017nru}.These transitions have been extensively examined through various theoretical frameworks, such as extended black hole thermodynamics \cite{14,15,16,17,18,19,20,21,22,23,24,25,new}, restricted phase space thermodynamics, and holographic thermodynamics \cite{rp1,rp2,rp3,rp4,rp5,rp6,rp7,rp8,rp9,rp10,rp11}.  However, a critical question arises: how do these thermodynamic descriptions evolve under alternative statistical frameworks? Moreover, can these frameworks offer new insights into the stability of black holes?\\

An important avenue of exploration involves generalizations of the Bekenstein-Hawking entropy, which scales with the horizon area. The Boltzmann-Gibbs entropy is defined as an extensive entropy that scales with the size of the system. But the extensivity of Boltzmann-Gibbs entropy is due to ignoring the long range forces that arise between the thermodynamic sub-systems. These forces are often ignored due to the fact that the size of the system exceeds the interaction range. However, these long-range forces are extremely important in strong gravitational systems such as: Black holes. Moreover, the Bekenstein-Hawking entropy is  non-extensive in nature and follows a non-additive composition rule. Therefore, the Boltzmann-Gibbs statistical mechanics may not be an amicable choice for studying black hole thermodynamics. Therefore, in order to  better  understand the underlying nature of black hole entropy, several extensions of Boltzmann-Gibbs statistics have been discussed in literature\cite{Cirto,Quevedo,Tsallis,Barrow,Reny,Sharma}.  \\

One of the many such proposals is the Kaniadakis entropy~\cite{Kania0,Kania1,Kania2,Scarf1,Scarf2} which is a relativistic generalization of the Boltzmann-Gibbs entropy. We know that the laws of special relativity emerge as a one-parameter deformation of the corresponding non-relativistic laws. These generalizations are imposed by the lorentz-transformations, which affect various physical observables such as momentum and energy. For instance, the relativistic momentum is given as: $P =\frac{p}{\sqrt{1- x^2}}$ where, $P$ is the relativistic momentum and $x=\frac{v}{c}$ is the deformation parameter. Likewise, Kaniadakis statistics is also based upon the understanding that the lorentz-transformations in special relativity can actually bring forth a continuous one parameter deformation of the classical Boltzmann-Gibbs-Shannon entropy. This deformed relativistic entropy is both coherent and self-consistent  in nature as it obeys the standard features of ordinary statistical theory. In order to achieve this relativistic deformation of the Boltzmann-Gibbs entropy, Kaniadakis introduced a deformed logarithmic function in the Boltzmann-Gibbs entropy formula along with a deformed exponential function in the Maxwell-Boltzmann distribution. These deformed functions are given as:
 $$\ln_K (x) = \frac{x^K - x^{-K}}{2 K} $$
 
 $$ exp_K (x)= \left(\sqrt{1 + K^2 x^2} + K x\right)^\frac{1}{K}$$

 where x is a chosen variable and K is the deformation parameter known as the Kaniadakis parameter. Both these above mentioned deformed functions reduce to their original form as the deformed parameter, K approaches appropriate limit i.e. $K \rightarrow 0$. The Kaniadakis entropy is then defined as:
 $$S_{K}=-\sum_{i=1} ^ {n} P_{i} \ln_K (P_{i})$$
 where,
  $$\ln_K (P_{i}) = \frac{P_{i}^K - P_{i}^{-K}}{2 K} $$
  In the limit $K \rightarrow 0$, the Kaniadakis entropy reduces to the Boltzmann-Gibbs entropy. On further simplification one can easily show that for the case of black holes Kaniadakis entropy takes the form which is of the form\cite{ka0,Morad,LympKan,Drepanou:2021jiv,She} 
\begin{equation}
S=\,\frac{1}{\kappa}\sinh\left(\kappa\,S_{BH}\right)\,.
\label{Kstat}
\end{equation}
where $S$ is the KD entropy,$S_{BH}$ is the conventional GB black hole entropy and $\kappa$ is the KD parameter.
Deviations from 
Boltzmann-Gibbs statistics are quantified by the
dimensionless parameter $0<\kappa<1$. Recently, a significant amount of research across various domains of physics has been conducted within the framework of Kaniadakis (KD) entropy \cite{ka0,ka1,ka11,ka2,ka3,ka3,ka4,ka5,ka6,ka7,ka8}. \\ 

 Our choice of Kaniadakis entropy stems from its foundation in relativistic statistical mechanics, distinguishing it from other generalized entropies like Tsallis, Rényi, and Barrow, which arise from phenomenological modifications of Boltzmann-Gibbs entropy to capture non-extensive effects, fractal structures, or long-range interactions. Unlike these approaches, Kaniadakis entropy naturally emerges from a relativistic deformation of logarithmic and exponential functions, preserving thermodynamic consistency while incorporating relativistic effects. This intrinsic link to relativistic kinetic theory provides a strong physical basis for its application to black holes, particularly in AdS spacetimes where holography is crucial. In this work, we systematically explore the impact of Kaniadakis entropy on phase transitions, stability, and critical phenomena in Kerr-AdS and Kerr-Newman-AdS spacetimes, revealing distinct thermodynamic behavior unexplored in previous studies. Given the universal role of entropy modifications in gravitational systems and their relevance to quantum gravity corrections, our findings extend beyond theoretical interest, offering new insights into black hole thermodynamics, holographic duals, and potential observational signatures.\\
 
It is important to acknowledge that although we employ Kaniadakis entropy as our entropy of choice here, it is worth noting that there are various other entropies such as Tsallis \cite{Tsallis},  Barrow \cite{Barrow}, Renyi \cite{Reny} which are being used to study different physical systems and hold important insights in their respective domains. There has also been attempts to provide a generalized construct which under appropriate conditions could lead to any of the above mentioned entropies. One such generalized entropy construct with different parameters  have been introduced in \cite{gen1,gen3} that generalizes all the above mentioned entropies. These generalized entropy constructs are formulated such that they reduce to all the required entropy paradigms under appropriate limits. For example, the six-parameter entropy formulation proposed in \cite{gen3} is given as:

\begin{equation}
S_G(\alpha_{\pm}, \beta_{\pm}, \gamma_{\pm}) = \frac{1}{\alpha_{+} + \alpha_{-}} \left[ \left( 1 + \frac{\alpha_{+}}{\beta_{+}} S^{\gamma_{+}} \right)^{\beta_{+}} - \left( 1 + \frac{\alpha_{-}}{\beta_{-}} S^{\gamma_{-}} \right)^{-\beta_{-}} \right],
\end{equation}
where all parameters $(\alpha_{\pm}, \beta_{\pm}, \gamma_{\pm})$ are positive. This expression reduces to the above mentioned entropy formulations under specific parameter limits. For example,by taking $\beta_{\pm} \to 0$, $\gamma_{\pm} = 1$, and $\alpha_{\pm} = K$, the generalized entropy simplifies to
    \begin{equation}
    S_G \to \frac{1}{K} \sinh(KS),
    \end{equation}
   which corresponds to the Kaniadakis entropy.\\

Recent studies have utilized Kaniadakis statistics to investigate the criticality, phase transitions, and geometrothermodynamics of charged AdS black holes \cite{ka0}. Building on these advancements, our work focuses on exploring the implications of Kaniadakis entropy in the context of rotating AdS black holes. Specifically, we identify a critical entropy bound in case of rotating AdS black hole  governed by the KD entropy parameter $\kappa$, beyond which black hole solutions become unstable. Unlike the GB framework, where these AdS black holes remain globally stable across an infinite range of entropy, the Kaniadakis framework imposes a finite stability range.  Our analysis reveals that smaller values of $\kappa$ correspond to higher entropy bounds, suggesting an expanded stability range as $\kappa$ decreases.  We aim to investigate the impact of the Kaniadakis (KD) entropy bound on the thermodynamic properties and phase transitions of rotating anti-de Sitter (AdS) black holes. Specifically, we focus on three well-known black hole systems: the Kerr-AdS black hole, the Kerr-Sen-AdS black hole, and the Kerr-Newman-AdS black hole. In our analysis, we apply two widely used approaches to study the influence of KD entropy on the thermodynamic phase space. First, we examine the thermodynamic topology of these black holes using KD entropy and compare the results with those obtained under traditional Gibbs-Boltzmann (GB) statistics. Thermodynamic topology provides an effective means of analyzing temperature-dependent phase transitions within the system while also offering insights into the local and global stability of black holes. Additionally, we explore the thermodynamic geometry within the KD framework to gain a understanding of the underlying microstructures of these black holes. Our findings reveal that the introduction of an entropy bound leads to significant deviations in both thermodynamic topology and thermodynamic geometry compared to the conventional thermodynamic behavior. The motivation behind the study is to highlight these changes and provides a comprehensive comparison with the traditional thermodynamics of these black holes.\\

Several modified entropy formalisms, including Rényi, Tsallis, and Kaniadakis entropy, have been proposed to incorporate non-extensive statistical effects into gravitational systems. However, these extensions introduce fundamental inconsistencies in black hole thermodynamics.  As discussed in \cite{gen2,gen4}, a major issue with non-standard entropy is its effect on the first law of black hole thermodynamics. If we assume an alternative entropy function but keep the standard definitions of black hole mass and Hawking temperature, the basic thermodynamic equations may no longer hold. This can result in incorrect predictions for the Hawking temperature or black hole energy, which contradicts the standard quantum field theory derivation of black hole radiation. Specifically, if the black hole mass is identified as the thermodynamic energy, the modified entropy may require an unphysical temperature to maintain the first law, or vice versa. The core problem with generalized entropies is that black hole thermodynamics is built on two fundamental quantities: the Hawking temperature, derived from quantum field theory in curved spacetime, and the ADM mass, which represents the total energy of the black hole as determined by energy conservation and Birkhoff’s theorem. Any change to the entropy must still satisfy the first law of thermodynamics,$dE = T dS$,while remaining consistent with these definitions. However, many generalized entropy models fail to meet this requirement. This leads to inconsistencies where either the black hole energy is incorrectly modified, or the temperature deviates from its expected quantum field theory value.  These issues suggest that using non-standard entropy in black hole thermodynamics requires additional constraints or modifications to ensure internal consistency. Without such corrections, these entropy models may not provide a physically valid description of black hole thermodynamics.\\

A possible resolution to this problem, as discussed in \cite{gen4} is to consider hairy black holes, where additional field contributions modify the ADM mass in a non-trivial way. For example, in the case of a Reissner-Nordström black hole with an electric field, or black holes coupled with scalar fields, the energy inside the black hole horizon is no longer determined solely by the standard ADM mass. This allows entropy modifications to emerge naturally without violating thermodynamic consistency. Additionally, in modified gravity theories, extra fields or curvature corrections introduce "hair" contributions that alter the first law of black hole thermodynamics, providing a framework where generalized entropy may be applicable.\\

While we have adopted the Kaniadakis entropy formalism in this work, our aim is not to replace the well-established Bekenstein-Hawking entropy but rather to explore its thermodynamic implications in black hole systems. To ensure that our results remain physically relevant and do not introduce significant deviations from standard black hole thermodynamics, we have considered very small values of the Kaniadakis entropy parameter $\kappa$. This allows for examining modifications while staying close to the conventional entropy framework. Notably, in the case of the considered rotating black holes, we have observed that the behaviour of Kaniadakis entropy closely resembles that of Bekenstein-Hawking entropy upto a certain limit, which we have identified as the "entropy bound." We carefully choose the parameter $\kappa$ such that it remains within this entropy bound, ensuring consistency with expected thermodynamic behaviour. However, our analysis reveals that for rotating black holes, exceeding this bound leads to the emergence of unstable black hole solutions. This suggests that while Kaniadakis entropy provides an interesting modification, its applicability in black hole thermodynamics must be constrained within physically meaningful limits.\\

While generalized entropy models are theoretically intriguing, there is currently no direct observational support indicating a departure from the standard area law for black hole entropy. Future observations—such as black hole shadow studies, gravitational wave signals from primordial black holes, and cosmological constraints—could offer valuable insights into potential deviations from Einsteinian gravity and the significance of alternative entropy formulations. However, until such evidence emerges, the application of generalized entropy to black hole thermodynamics should be approached with caution, ensuring consistency with fundamental thermodynamic principles.\\

The paper is organized as follows: Section II focuses on calculating the thermodynamic quantities of Kerr-AdS black holes and analyzing their thermodynamic phase space using the Kaniadakis (KD) entropy framework. In Sections III and IV, we explore the thermodynamic topology and thermodynamic geometry of Kerr-AdS black holes within the KD entropy framework. Sections V and VI extend this analysis to Kerr-Sen-AdS black holes, examining their thermodynamic topology and geometry, respectively. Similarly, Sections VII and VIII repeat this study for Kerr-Newman-AdS black holes. Finally, Section IX presents our concluding remarks.

	\section{Thermodynamics of Kerr-Ads Black holes}
\begin{figure}[h]	
		\centering
		\begin{subfigure}{0.40\textwidth}
			\includegraphics[width=\linewidth]{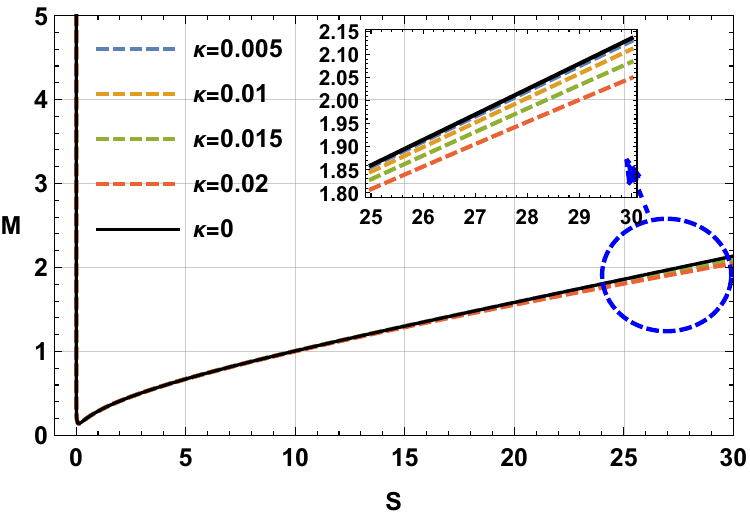}
			\caption{J=0.02,l=5}
			\label{1a}
		\end{subfigure}
		\hspace{0.5cm}
		\begin{subfigure}{0.40\textwidth}
			\includegraphics[width=\linewidth]{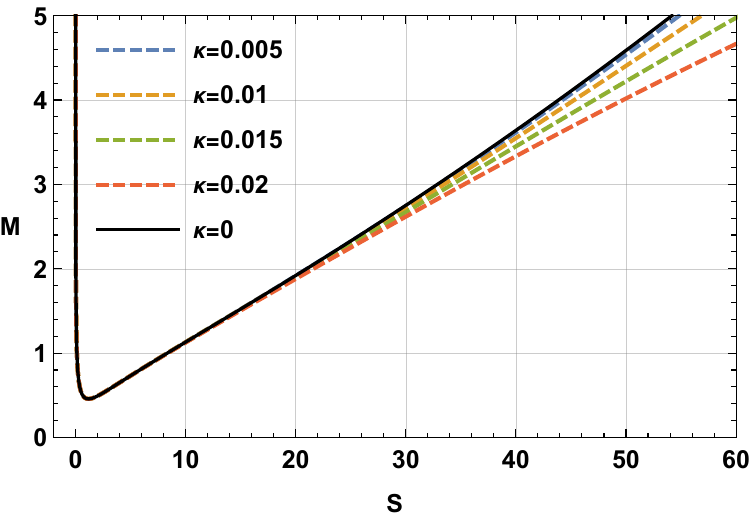}
			\caption{J=0.2,l=3.5}
			\label{1b}
		\end{subfigure}
		\begin{subfigure}{0.40\textwidth}
			\includegraphics[width=\linewidth]{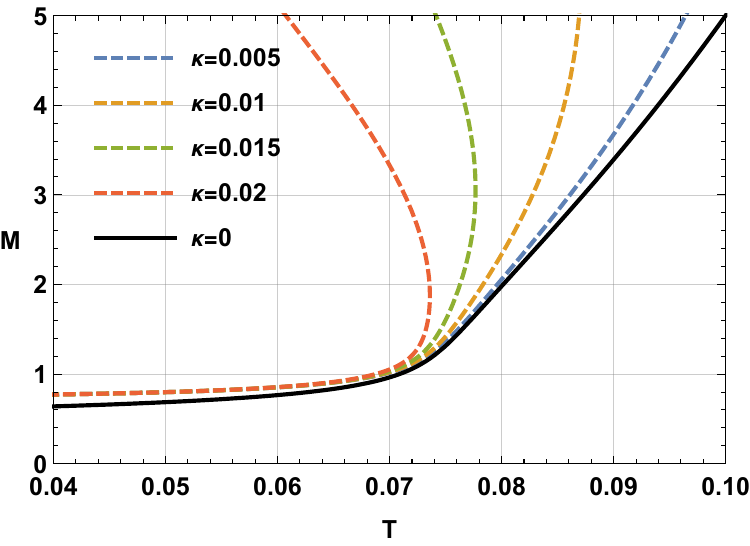}
			\caption{J=0.02,l=5}
			\label{1c}
		\end{subfigure}
		\hspace{0.5cm}
		\begin{subfigure}{0.40\textwidth}
			\includegraphics[width=\linewidth]{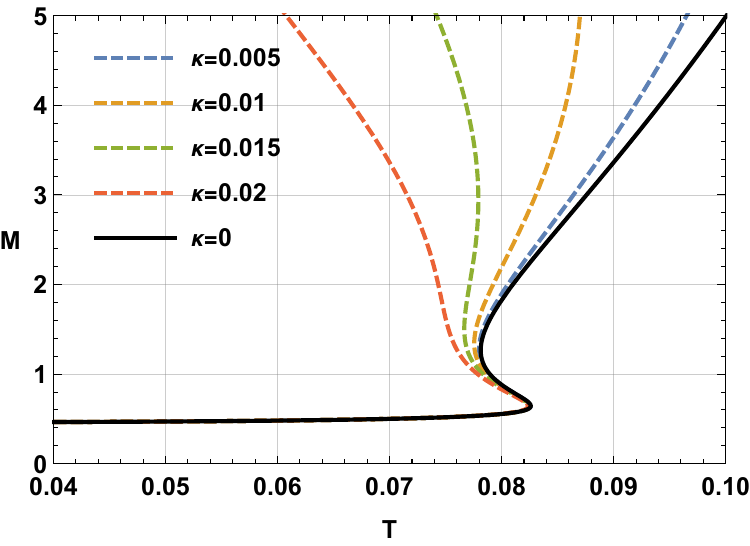}
			\caption{J=0.2,l=3.5}
			\label{1d}
		\end{subfigure}
		
		\caption{Kerr AdS black hole: Figure (a) and (b) represent $M$ vs $S$ plot for different values of $\kappa,J$ and $l.$ The impact of $\kappa$ parameter is more visible with higher values of $S$. Figure (c) and (d) represents $M$ vs $T$ plot for different value of $\kappa,J$ and $l.$
		}
		\label{1}
	\end{figure}

	We start with the ADM mass of the Kerr-Ads black hole ,which is 
	\begin{equation}
		M=\frac{\left(a^2+r_+^2\right) \left(\frac{r_+^2}{l^2}+1\right)}{2 \Xi^2 r_+}
	\end{equation}
	Where $$ \Xi=1-\frac{a^2}{l^2}, \hspace{0.5cm} a=\frac{J}{M}$$ 
	Here $a$ is the angular momentum per unit mass and $l$ is the AdS length
	From the expression of entropy, the radius $r_+$ can be written as :
	\begin{equation}
		r_+=\frac{\sqrt{1-\frac{a^2}{l^2}} \sqrt{\left(a^2-l^2\right) S_{\text{BH}}+\pi  a^2 l^2}}{\sqrt{\pi } \sqrt{a^2-l^2}}
		\label{kerrradius}
	\end{equation}
	Using eq.\ref{kerrradius}, the ADM mass in terms of entropy is finally obtained as :
	\begin{equation}
		M_{BH}=\frac{\sqrt{S_{\text{BH}}+\pi  l^2} \sqrt{\pi  l^2 S_{\text{BH}}^2+S_{\text{BH}}^3+4 \pi ^3 J^2 l^2}}{2 \pi ^{3/2} l^2 \sqrt{S_{\text{BH}}}}
		\label{kerrmass}
	\end{equation}
	Using equation.\ref{Kstat}, $S_{BH}$ can be written as 
	$$S_{BH}=\frac{1}{\kappa} arcsinh(\kappa S)$$
	Replacing the entropy $S_{BH}$ by Kaniadakis(KD) entropy $S$ in the expression for mass and the new mass in KD statistics is obtained as :
	\begin{equation}
		M_{KD}=\frac{\sqrt{\left(\pi  \kappa  l^2+\sinh ^{-1}(\kappa  S)\right) \left(4 \pi ^3 \kappa ^3 J^2 l^2+\pi  \kappa  l^2 \sinh ^{-1}(\kappa  S)^2+\sinh ^{-1}(\kappa  S)^3\right)}}{2 \pi ^{3/2} \kappa ^{3/2} l^2 \sqrt{\sinh ^{-1}(\kappa  S)}}
		\label{kerrmasskd}
	\end{equation}
	At $\kappa=0$, $M_{KD}$ becomes equal to the ADM mass in GB statistics written in eq.\ref{kerrmass}.
	The temperature can be evaluated from eq.\ref{kerrmass} as :
	\begin{multline}
		T=\frac{d M_{KD}}{d S}\\
		=-\frac{4 \pi ^4 \kappa ^6 J^2 l^4-\pi ^2 \kappa ^4 l^4 \sinh ^{-1}(\kappa  S)^2-4 \pi  \kappa ^3 l^2 \sinh ^{-1}(\kappa  S)^3-3 \kappa ^2 \sinh ^{-1}(\kappa  S)^4}{4 \pi ^{3/2} \kappa ^4 l^2 \sqrt{\kappa ^2 S^2+1} \left(\frac{\sinh ^{-1}(\kappa  S)}{\kappa }\right)^{3/2} \sqrt{\left(\pi  \kappa  l^2+\sinh ^{-1}(\kappa  S)\right) \left(4 \pi ^3 \kappa ^3 J^2 l^2+\pi  \kappa  l^2 \sinh ^{-1}(\kappa  S)^2+\sinh ^{-1}(\kappa  S)^3\right)}}
		\label{tempkerkd}
	\end{multline}
	The effect of the KD parameter $\kappa$  on the black hole mass is more prominent for a higher range of values of KD entropy $S$. The range of KD entropy within which the black hole mass in KD statistics resembles the black hole mass in GB statistics varies depending on the value of $\kappa$. As it can be seen from FIG.\ref{1a} and FIG.\ref{1b}, the smaller the value of $\kappa$, the larger the range of KD entropy $S$ within which the $M_{KD}$ will behave like $M_{BH}$. In FIG.\ref{1a} and FIG.\ref{1b}, the new mass of the black hole in KD statistics is plotted for different values of $\kappa$ while keeping $J=0.02,l=5$ and $J=0.02,l=5$  fixed respectively. In the figures FIG.\ref{1a} and FIG.\ref{1b}, there is a black solid line that shows the $M$ vs $S$ plot, when the parameter $\kappa$ is set to zero. This line serves as a reference line for comparison. When the parameter $\kappa$ increases from zero, the plots of $M$ vs $S$ start to deviate from this black solid line. These deviations are more observable for larger values of $\kappa$. This means that as $\kappa$ gets larger, the impact on the black hole mass becomes noticeable at lower entropy levels compared to smaller values of $\kappa$. In FIG.\ref{1c} and FIG.\ref{1d}, mass is plotted against temperature. Depending on the values of $J$ and $l$, the number of branches in $M$ vs $T$ graph changes for a fixed non-zero value of $\kappa$. For example in FIG.\ref{1c}, when $J=0.02,l=5$, there are two branches and FIG.\ref{1d} for $J=0.2,l=3.5$, there are four branches. \\
	
\begin{figure}[h!]	
		\centering
		\begin{subfigure}{0.40\textwidth}
			\includegraphics[width=\linewidth]{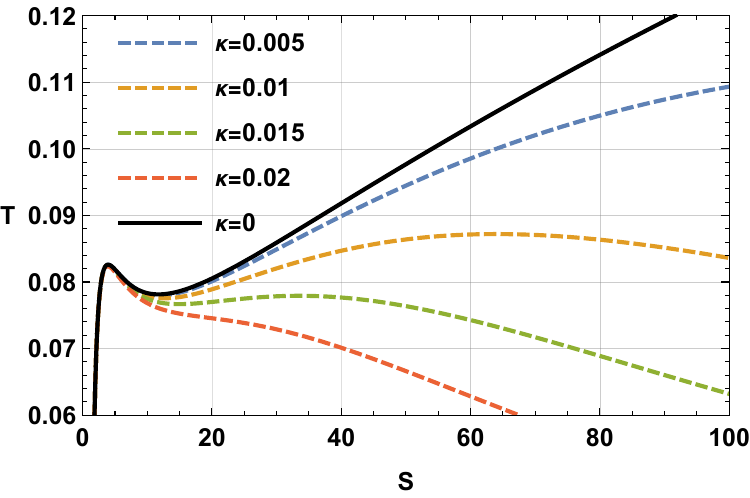}
			\caption{J=0.2, l=3.5}
			\label{2a}
		\end{subfigure}
		\hspace{0.5cm}
		\begin{subfigure}{0.40\textwidth}
			\includegraphics[width=\linewidth]{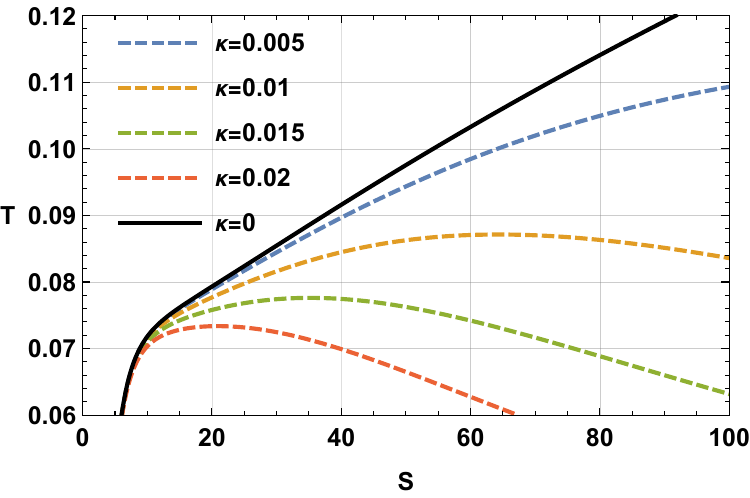}
			\caption{J=0.55, l=5}
			\label{2b}
		\end{subfigure}
		\begin{subfigure}{0.40\textwidth}
			\includegraphics[width=\linewidth]{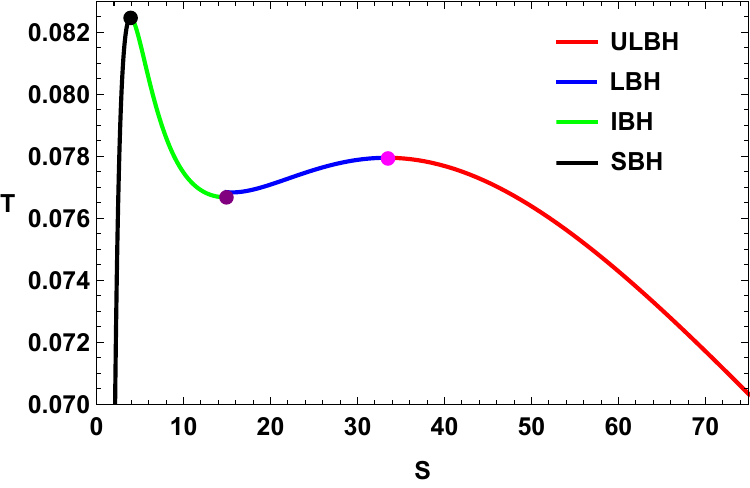}
			\caption{J=0.2, l=3.5, $\kappa$=0.015}
			\label{2c}
		\end{subfigure}
		\hspace{0.5cm}
		\begin{subfigure}{0.40\textwidth}
			\includegraphics[width=\linewidth]{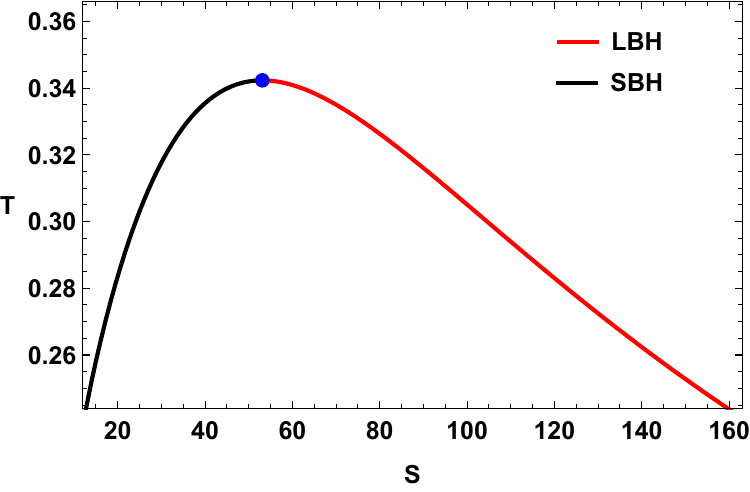}
			\caption{J=0.55, l=1.5, $\kappa$=0.015}
			\label{2d}
		\end{subfigure}
		
		\caption{Kerr AdS black hole: T vs S plots for different values of $J$ and $l.$ Here SBH, IBH,LBH and ULBH represent small black holes, intermediate black holes, large black holes and ultra large black holes.
		}
		\label{2}
	\end{figure}

\begin{figure}[h!]	
		\centering
		\begin{subfigure}{0.40\textwidth}
			\includegraphics[width=\linewidth]{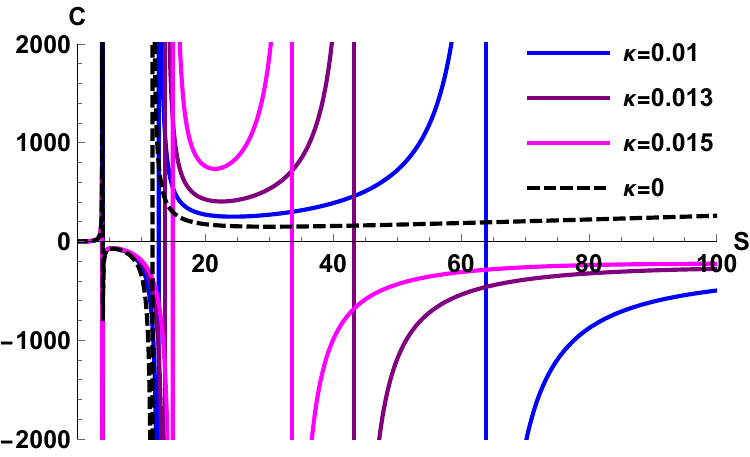}
			\caption{C vs S plot($J=0.2$ and $l=3.5$) for the range $0<S<100$}
			\label{3a}
		\end{subfigure}
		\hspace{0.5cm}
		\begin{subfigure}{0.40\textwidth}
			\includegraphics[width=\linewidth]{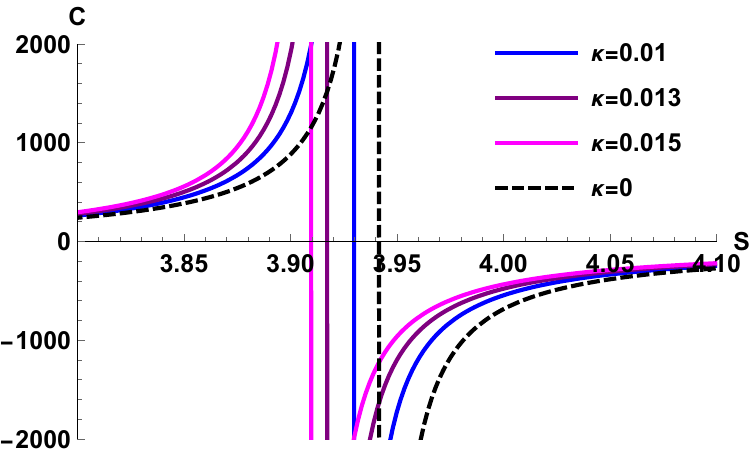}
			\caption{Closer look at $3.8<S<4.1$ range($J=0.2$ and $l=3.5$)}
			\label{3b}
		\end{subfigure}
		\begin{subfigure}{0.40\textwidth}
			\includegraphics[width=\linewidth]{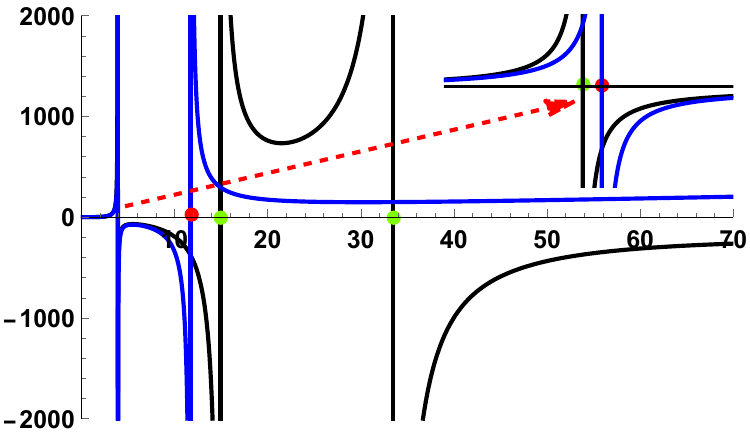}
			\caption{For the blue solid line, $\kappa=0$ and for black solid line $\kappa=0.015$}
			\label{3c}
		\end{subfigure}
		\hspace{0.5cm}
		\begin{subfigure}{0.40\textwidth}
			\includegraphics[width=\linewidth]{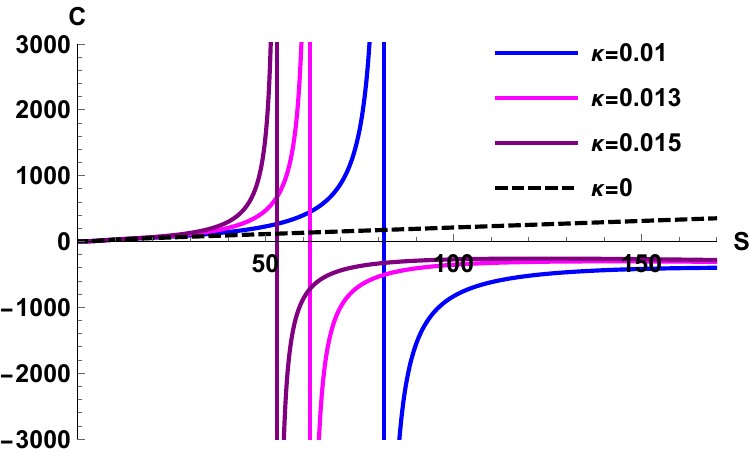}
			\caption{Here $J=0.55,l=1.5$}
			\label{3d}
		\end{subfigure}
		\caption{Kerr AdS black hole: C vs S plots. 
		}
		\label{3}
	\end{figure}

\begin{figure}[h!]	
		\centering
		\begin{subfigure}{0.40\textwidth}
			\includegraphics[width=\linewidth]{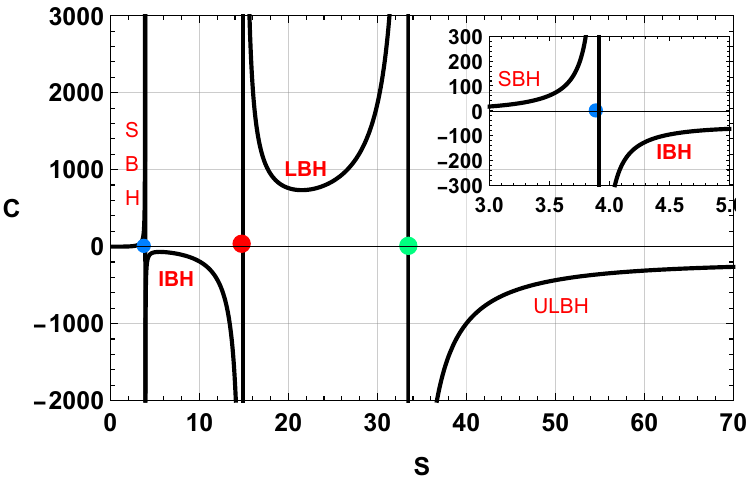}
			\caption{C vs S plot at $J=0.2, \kappa=0.015$ and $l=3.5$}
			\label{4a}
		\end{subfigure}
		\hspace{0.5cm}
		\begin{subfigure}{0.40\textwidth}
			\includegraphics[width=\linewidth]{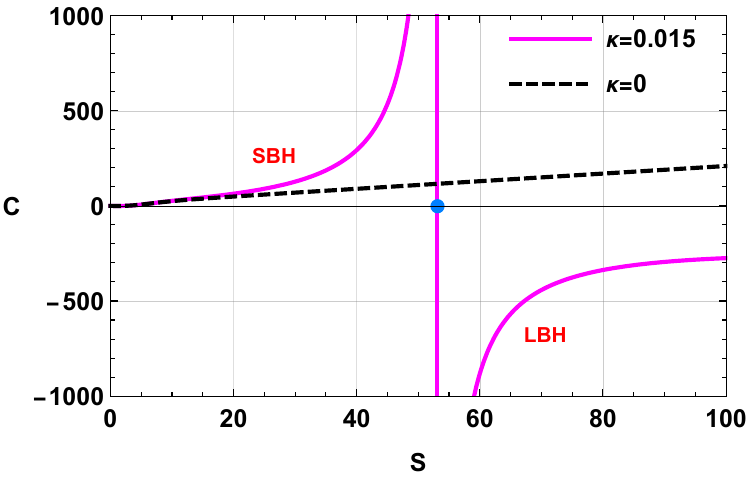}
			\caption{C vs S at $J=0.55,\kappa=0.015$ and $l=1.5$}
			\label{4b}
		\end{subfigure}
		\caption{Kerr AdS black hole: C vs S plots. 
		}
		\label{4}
	\end{figure}

	Similarly, Temperature $T$ is plotted against $S$ in FIG.\ref{2}. Here we have presented two cases. For the first case, in FIG.\ref{2a}, we have considered $J=0.2$ and $l=5$. Here we see four black hole phases. 
	In GB statistics, for critical values of $J$ and $l$, Kerr-Ads black holes exhibit Van der Waals phase transitions as represented by the black solid line in FIG.\ref{2a}. In KD statistics, an additional black hole branch which is the ultra-large black hole branch appears as illustrated by the red solid line in FIG.\ref{2c}. The appearance of the fourth phase is dependent on the value of the $\kappa$ along with critical values of $J$ and $l.$When the value of $\kappa$ is large, the ultra-large black hole starts to show up at smaller values of KD entropy $S.$ In FIG.\ref{2c}, we have considered $J=0.2,l=3.5$ and $\kappa=0.015$ for which we observe four black hole branches. In FIG.\ref{2c} when $S<3.91152$(black dot), a small black hole branch(SBH) is observed represented by a black solid line. Again, within the range  $3.91152 < S < 14.9327$(purple dot), an intermediate black hole branch(IBH) is observed which is represented by the green solid line. A large black hole branch(LBH) is observed within the range $14.9327< S< 33.4984$(magenta dot) represented by a blue solid line and finally an ultra large black hole branch(ULBH) is found for the range $S>33.4984.$
	On the opposite hand, for $J=0.2$ and $l=3.5$,  we observe two black hole branches as shown in FIG.\ref{2b}. In FIG.\ref{2d} we consider $J=0.55,l=5$ and $\kappa=0.015$, for which only two black hole branches are observed:  a large black hole branch(LBH) for the range $S<53.1069$(blue dot)and a small black hole(SBH) branch for the range $S>53.1069$. The SBH and LBH is represented by the black solid line and the red solid line respectively in FIG.\ref{2d}.\\
	To know the thermal stability of these black holes, we calculate the specific heat $(C)$ of the black hole using the formula :
	\begin{equation}
		C=\frac{d M}{d T}=\frac{d M}{d S} \left(\frac{d T}{d S}\right)^{-1}
	\end{equation}
	The expression for $C$ comes out to be :
	\begin{equation}
		C=\frac{\mathcal{A}}{\mathcal{B}}
	\end{equation}
	Where,
	\begin{multline}
		\mathcal{A}=-2 \left(\kappa ^2 S^2+1\right) \sinh ^{-1}(\kappa  S) (\pi  \kappa  l^2+\sinh ^{-1}(\kappa  S)) (4 \pi ^3 \kappa ^3 J^2 l^2+\pi  \kappa  l^2 \sinh ^{-1}(\kappa  S)^2+\\
		\sinh ^{-1}(\kappa  S)^3) (4 \pi ^4 \kappa ^4 J^2 l^4-\pi ^2 \kappa ^2 l^4 \sinh ^{-1}(\kappa  S)^2-4 \pi  \kappa  l^2 \sinh ^{-1}(\kappa  S)^3-3 \sinh ^{-1}(\kappa  S)^4)
	\end{multline}
	and 
	\begin{multline}
		\mathcal{B}=\kappa (48 \pi ^8 \kappa ^8 J^4 l^8 \sqrt{\kappa ^2 S^2+1}+6 \pi ^2 \kappa ^2 l^2 \sinh ^{-1}(\kappa  S)^6 (-4 \pi  \kappa ^2 J^2 S+l^2 \sqrt{\kappa ^2 S^2+1}-2 \pi  \kappa ^2 l^4 S) 
		\\-2 \pi ^3 \kappa ^3 l^2 \sinh ^{-1}(\kappa  S)^5 (\pi  \kappa ^2 l^6 S-24 J^2 (\sqrt{\kappa ^2 S^2+1}-\pi  \kappa ^2 l^2 S))
		\\-\pi ^4 \kappa ^4 l^4 \sinh ^{-1}(\kappa  S)^4 (l^4 \sqrt{\kappa ^2 S^2+1}-24 J^2 (5 \sqrt{\kappa ^2 S^2+1}-\pi  \kappa ^2 l^2 S))+96 \pi ^5 \kappa ^5 J^2 l^6 \sqrt{\kappa ^2 S^2+1} \sinh ^{-1}(\kappa  S)^3
		\\+8 \pi ^6 \kappa ^6 J^2 l^6 \sinh ^{-1}(\kappa  S)^2 \left(4 \pi  \kappa ^2 J^2 S+3 l^2 \sqrt{\kappa ^2 S^2+1}\right)+32 \pi ^7 \kappa ^7 J^4 l^6 (\pi  \kappa ^2 l^2 S+2 \sqrt{\kappa ^2 S^2+1}) \sinh ^{-1}(\kappa  S)
		\\+(3 \sqrt{\kappa ^2 S^2+1}-20 \pi  \kappa ^2 l^2 S) \sinh ^{-1}(\kappa  S)^8+8 \pi  \kappa  l^2 (\sqrt{\kappa ^2 S^2+1}-3 \pi  \kappa ^2 l^2 S) \sinh ^{-1}(\kappa  S)^7-6 \kappa  S \sinh ^{-1}(\kappa  S)^9)
	\end{multline}
	In FIG.\ref{3}, specific heat is plotted against KD entropy S for fix value of $J=0.2$ and $l=3.5.$ Black solid line in FIG.\ref{3a} represents the specific heat at GB statistics($\kappa=0$). As it is illustrated in FIG.\ref{3a}, the black line shows only three phases. On the opposite hand, the introduction of KD statistics produces an additional branch for a higher range of entropy. How the phase formation and the singularities are dependent on the $\kappa$ value and entropy range is clearly demonstrated in FIG.\ref{3a}. FIG.\ref{3b} is an enlarged version of FIG.\ref{3a} for the entropy range $3.8<S<4.1$. This figure shows how the point at which specific heat diverges,  varies with $\kappa$ value. In FIG.\ref{3c} comparison between $C$ VS $S$ plot for $\kappa-0$ and $\kappa=0.015$ is shown for $J=0.2,l=3.5.$ The black solid line is the $C$ vs $S$ plot for $\kappa=0.015$, where specific heat diverges at three points(represented by green dots.) On the other hand, the blue solid line represents $C$ vs $S$ plot in GB statistics ($\kappa=0$). Here only three phases are visible for the same set of values of $J$ and $l.$ Hence the specific heat diverges only at two points. In FIG.\ref{3d}, the $C$ vs $S$ graph is shown for $J=0.55, l=1.5$. For this set of values, in the $\kappa=0$ case, the specific heat does not diverge at any point hence there is no phase transition but for non-zero values of $\kappa$, we observe at least one point at which the specific heat diverges.
	It is well known that each point at which specific heat diverges represents a phase transition point. This is explained in FIG.\ref{4a}, where we see four black hole phases: a small black hole branch(SBH), an intermediate black hole branch(IBH), a large black hole branch(LBH), and an ultra-large black hole branch(ULBH). 
	There are three phase transition points at which the specific heat diverges represented by blue dot($S=3.91152$) red dot($S=14.9327$)  and green dot($S=33.4984$). Within the range $S<3.91152$, a small black hole is observed with positive specific heat, hence it is a thermally stable branch. Within the range $3.91152<S<14.9327$, an intermediate black hole branch is observed with negative specific heat, hence the branch is thermally not stable. Again within the range $14.9327<S<33.4984$,  a thermally stable large black hole branch is observed as the specific heat in that range is found to be positive. Finally, a thermally unstable ultra-large black hole is found within the $S>33.4984$ range where the specific heat is negative. In FIG.\ref{4b}, $C$ is plotted agaist $S$ for $J=0.55$ and $l=1.5$. Here we observe a Davies-type phase transition point at $S=53.1069$(represented by the blue dot) where the specific heat diverges. For the range $S<53.1069$, a stable small black hole branch(SBH) is seen and for $S>53.1069$, an unstable large black hole branch(LBH) is seen. On the contrary, in GB statistics we do not get any Davies-type phase transition as shown by the black dashed line in FIG.\ref{4b}.
	\begin{figure}[h!]	
		\centering
		\begin{subfigure}{0.40\textwidth}
			\includegraphics[width=\linewidth]{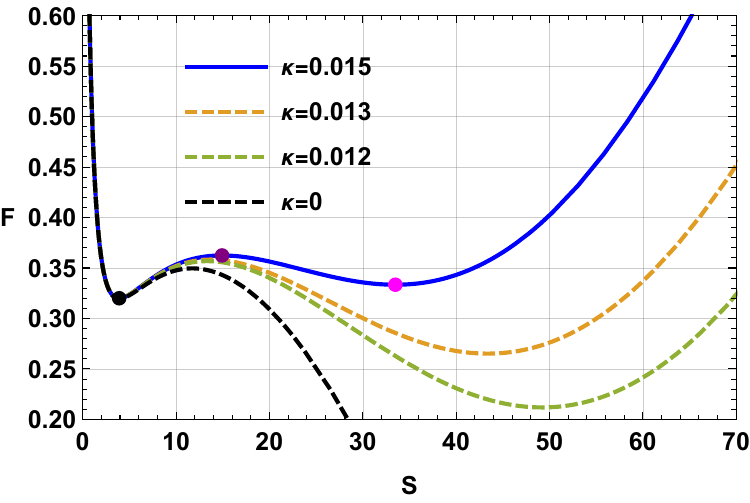}
			\caption{F vs S plot at $J=0.2$ and $l=3.5$}
			\label{5a}
		\end{subfigure}
		\hspace{0.5cm}
		\begin{subfigure}{0.40\textwidth}
			\includegraphics[width=\linewidth]{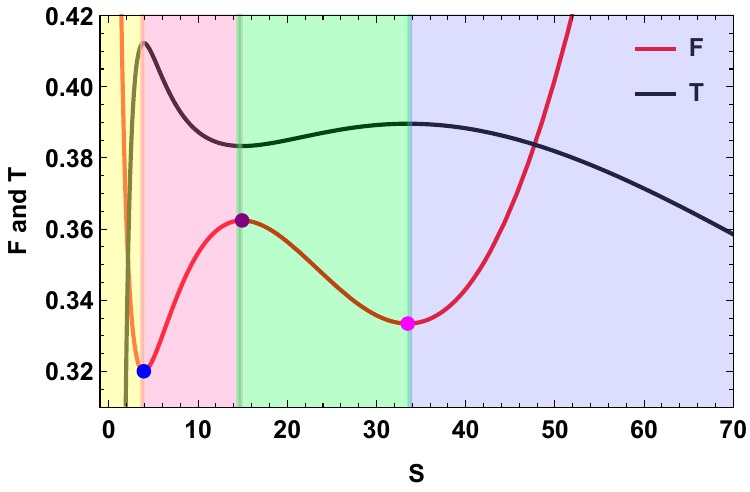}
			\caption{BH Phases in F/T vs S plot($J=0.2,l=3.5$)}
			\label{5b}
		\end{subfigure}
		\begin{subfigure}{0.40\textwidth}
			\includegraphics[width=\linewidth]{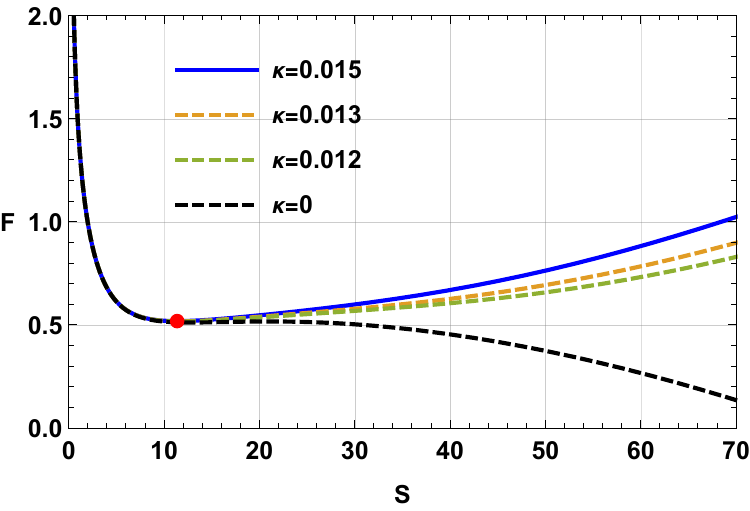}
			\caption{F vs S plot at $J=0.55$ and $l=5$}
			\label{5c}
		\end{subfigure}
		\hspace{0.5cm}
		\begin{subfigure}{0.40\textwidth}
			\includegraphics[width=\linewidth]{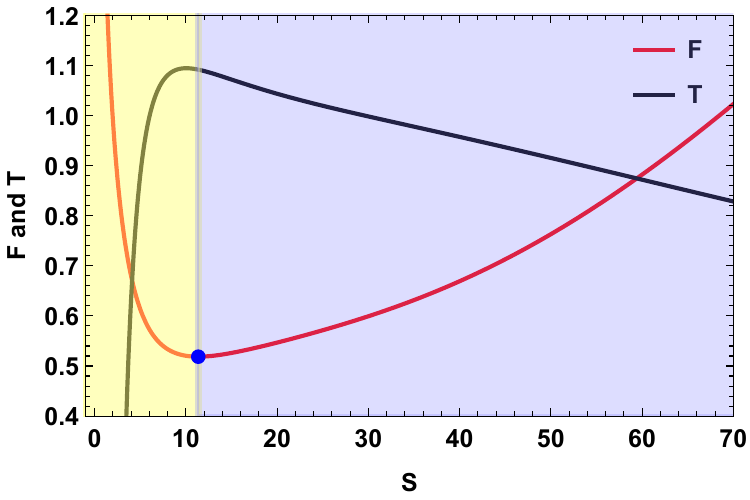}
			\caption{BH Phases in F/T vs S plot($J=0.55,l=5$)}
			\label{5d}
		\end{subfigure}
		\caption{Kerr AdS black hole: F vs S plots. 
		}
		\label{5}
	\end{figure} 
	Gibbs Free energy($F$) provides valuable insight into the characteristics of phase transitions in black holes. The free energy is calculated using the following relation:
	\begin{equation}
		F=M-TS
	\end{equation}
	here T is the horizon temperature calculated in the equation \ref{tempkerkd}. Using equation \ref{tempkerkd} and \ref{kerrmasskd}, F is calculated as :
	\begin{equation}
		F=\frac{\mathcal{C}}{\mathcal{D}}
	\end{equation}
	Where,
	\begin{multline}
		\mathcal{C}=\frac{\kappa  S \left(4 \pi ^4 \kappa ^4 J^2 l^4-\pi ^2 \kappa ^2 l^4 \sinh ^{-1}(\kappa  S)^2-4 \pi  \kappa  l^2 \sinh ^{-1}(\kappa  S)^3-3 \sinh ^{-1}(\kappa  S)^4\right)}{\sqrt{\kappa ^2 S^2+1}}\\
		+2 \sinh ^{-1}(\kappa  S) \left(\pi  \kappa  l^2+\sinh ^{-1}(\kappa  S)\right) \left(4 \pi ^3 \kappa ^3 J^2 l^2+\pi  \kappa  l^2 \sinh ^{-1}(\kappa  S)^2+\sinh ^{-1}(\kappa  S)^3\right)
	\end{multline}
	And 
	\begin{multline}
		\mathcal{D}=4 \pi ^{3/2} \kappa ^5 l^2 \left(\frac{\sinh ^{-1}(\kappa  S)}{\kappa }\right)^{3/2} \sqrt{\pi  l^2+\frac{\sinh ^{-1}(\kappa  S)}{\kappa }} \sqrt{\frac{4 \pi ^3 \kappa ^3 J^2 l^2+\pi  \kappa  l^2 \sinh ^{-1}(\kappa  S)^2+\sinh ^{-1}(\kappa  S)^3}{\kappa ^3}}
	\end{multline}
	FIG.\ref{5} represents two scenarios where from the free energy plot, the number of black hole branches can be identified. In FIG\ref{5a} and FIG.\ref{5b}, it is seen that for non-zero $\kappa$ value there are at least two black hole branches. On the other hand, the black dashed line in FIG\ref{5a} and FIG.\ref{5b} shows that there are either three or one black hole branches in GB statistics. In FIG.\ref{5c}, when $J=0.2, l=3.5, \kappa=0.015$ are taken, the four phases are visible(for better comparability $T$ vs $S$ plot is used as reference).  In FIG.\ref{5c}, black holes in the yellow region are the small black holes, black holes within the pink region are intermediate black holes, black holes within the green region are large black holes, and finally black holes in the blue region are an ultra large black hole. The phase transitioning points are shown in multicolored dots in the $F$ vs $S$ plot represented by the red solid line in FIG.\ref{5b}. Similarly in FIG.\ref{5d} we have used $J=0.2, l=3.5, \kappa=0.015$, where the black holes in the yellow region are small black holes and the black holes lie in the blue region are the large black holes. The phase transition phenomenon is also clearly visible in FIG.\ref{6}, where parametric plot between $F$ vs $T$ is shown for $J=0.2,l=3.5$ and $\kappa=0.015$
	\begin{figure}[h!]	
		\centering
		\begin{subfigure}{0.40\textwidth}
			\includegraphics[width=\linewidth]{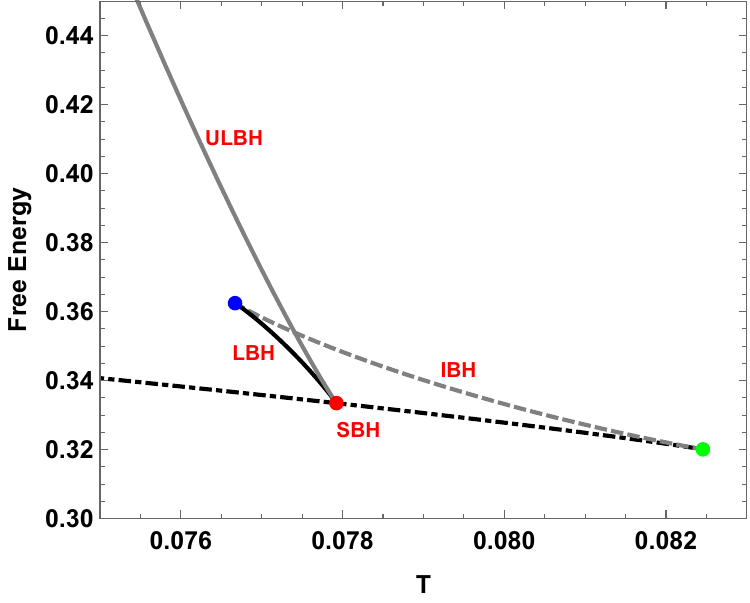}
			\caption{F vs T plot at $J=0.2$ and $l=3.5$}
			\label{6a}
		\end{subfigure}
		\hspace{0.5cm}
		\begin{subfigure}{0.40\textwidth}
			\includegraphics[width=\linewidth]{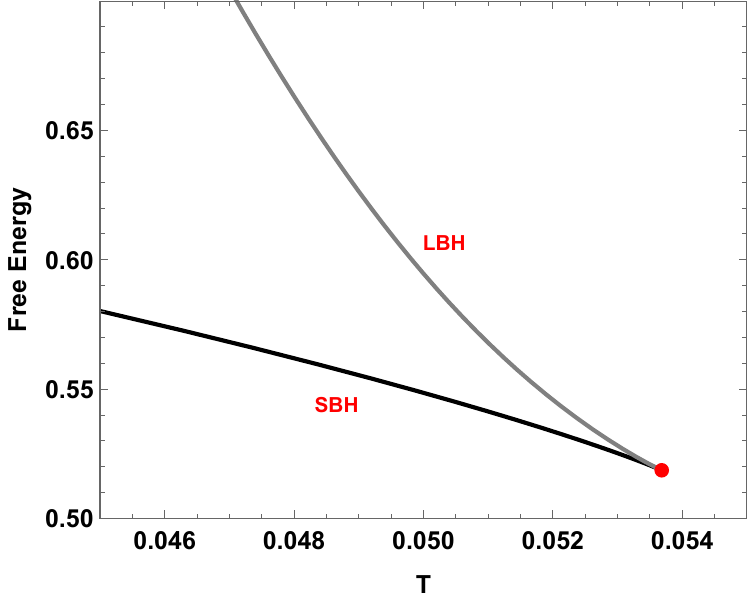}
			\caption{ F vs T plot at $J=0.55$ and $l=5$}
			\label{6b}
		\end{subfigure}
		\caption{Kerr AdS black hole: F vs T plots. 
		}
		\label{6}
	\end{figure} 
	Next, we do the free energy landscape analysis to estimate the stability of black holes from the free energy plot. To construct the free energy landscape, we write the off-shell free energy as\citep{york} :
	\begin{equation}
		F_e=F-T_e S=-\frac{2 \pi ^{3/2} l^2 S T_e \sqrt{\frac{\sinh ^{-1}(\kappa  S)}{\kappa }}-\sqrt{\frac{\pi  \kappa  l^2+\sinh ^{-1}(\kappa  S)}{\kappa }} \sqrt{\frac{4 \pi ^3 \kappa ^3 J^2 l^2+\pi  \kappa  l^2 \sinh ^{-1}(\kappa  S)^2+\sinh ^{-1}(\kappa  S)^3}{\kappa ^3}}}{2 \pi ^{3/2} l^2 \sqrt{\frac{\sinh ^{-1}(\kappa  S)}{\kappa }}}
	\end{equation}  
	Here $T_e$ is the ensemble temperature. Taking the first derivative of the off-shell or generalized free energy 
	$$\frac{\partial F_e}{\partial S}=(T-T_e)$$
	Hence, a physical black hole is only possible at the extremal point of the off-shell free energy, where $\frac{\partial F_e}{\partial S}$ is zero. Again, it can be proved that the second order derivative of $F_e$ is related to sign of specific heat through the relation :
	$$\left(\frac{\partial^2 F_e}{\partial^2 S}\right)_{T_e=T}=\frac{\partial T}{\partial M} \frac{\partial M}{\partial S}=\frac{T}{C}$$ 
	Hence at the maximum value of off-shell free energy, the black hole is thermally unstable and vice versa. The free energy landscape for Kerr ads black hole in GB statistics is analyzed in \cite{Yang_2022}. 
	We plot the free energy landscape for Kerr ads in KD statistics in the FIG.\ref{7}. For this particular analysis, we have considered five temperatures $T_1=0.076, T_2=0.077, T_3=0077924, T_4=0.07742, T_5=0.080$ and kept $J=0.2,l=3.5$ and $\kappa=0.015$ fixed. Although all the figures are almost self-deprecating, we will explain one of the scenarios in detail. For instance, if we consider the FIG.\ref{7c} where we have plot $F_e$ for $T_e=T_2=0.077$, then we see four black holes in the $F_e$ vs $S$ plot which are located at the extremum points. As it is illustrated in FIG.\ref{7a}, the $T=T_2$ line cuts the $F$ vs $T$ graph at four points: B at SBH, K at IBH, K' at LBH, and I at ULBH. These four points represent four physical black holes in FIG.\ref{7a}. The B point is located at the global minima of the graph, hence this black hole is thermally and globally stable. The K point lies in the local maxima point, this point is thermally and locally unstable. The K' point is situated in the local minima point, hence this point is thermally stable but globally unstable, and finally, the I point is at global maxima hence this point is both globally and thermally unstable. Therefore we can infer from FIG.\ref{7} that points $A, B, C, D, E$ are both thermally and globally stable, $K', H, D_3$ are thermally stable but only locally stable, $K, G, H'$ are thermally and locally unstable and finally $F, J, I, D_2, L$ are both globally and thermally unstable.
	\clearpage
	\begin{figure}[h!]	
		\centering
		\hspace{-3cm}
		\begin{subfigure}{0.4\textwidth}
			\includegraphics[height=7cm,width=10cm]{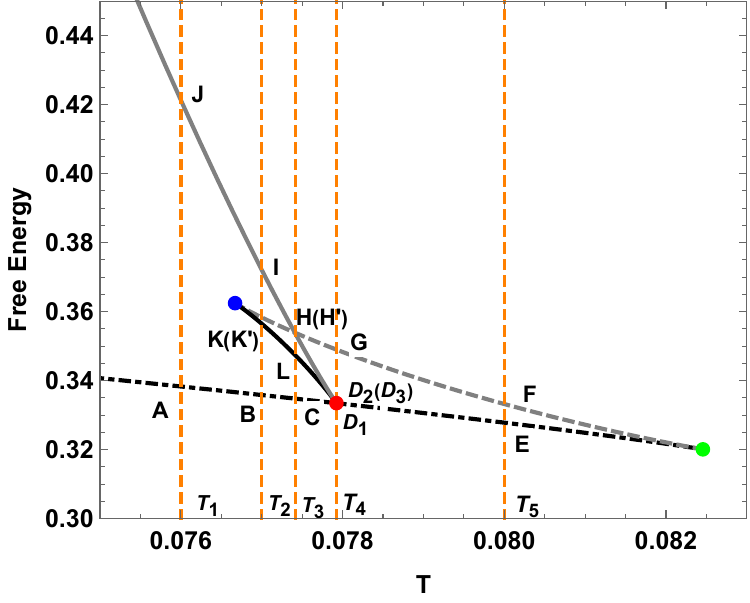}
			\caption{}
			\label{7a}
		\end{subfigure}
		\hspace{6.5cm}
		\begin{subfigure}{0.4\textwidth}
			\includegraphics[width=\linewidth]{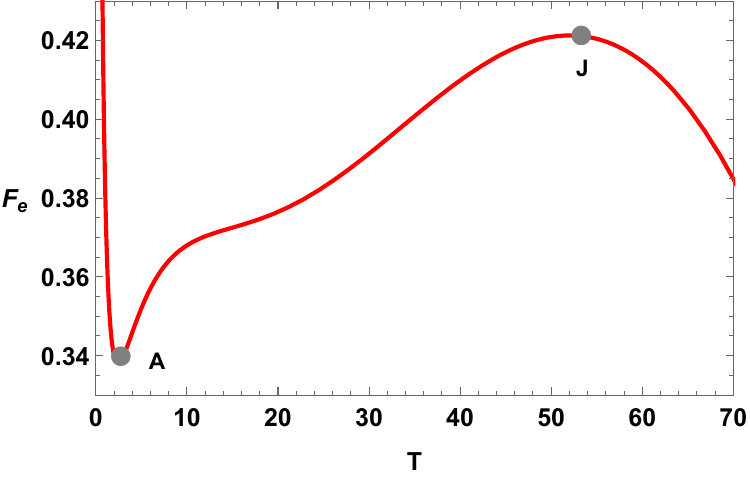}
			\caption{$T_e=0.076$}
			\label{7b}
		\end{subfigure}
		\hspace{0.5cm}
		\begin{subfigure}{0.4\textwidth}
			\includegraphics[width=\linewidth]{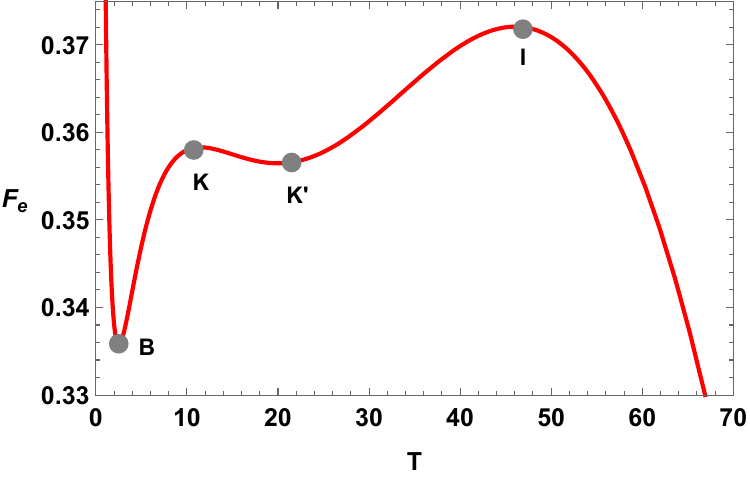}
			\caption{$T_e=0.077$ }
			\label{7c}
		\end{subfigure}
		\begin{subfigure}{0.4\textwidth}
			\includegraphics[width=\linewidth]{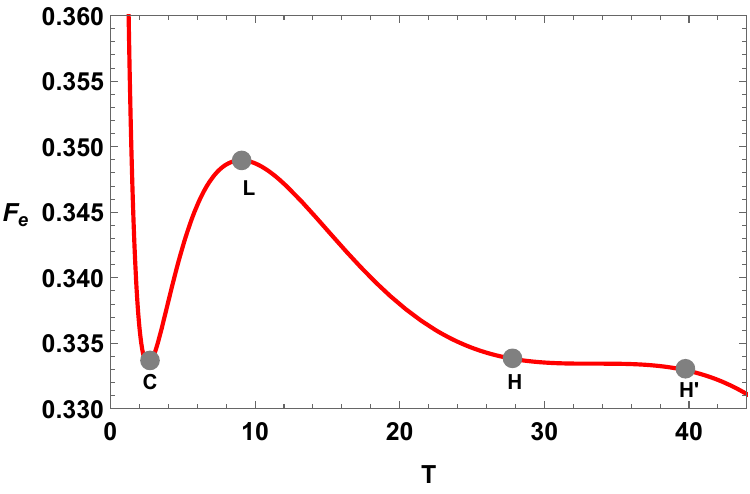}
			\caption{$T_e=0.077924$ }
			\label{7d}
		\end{subfigure}
		\hspace{0.5cm}
		\begin{subfigure}{0.4\textwidth}
			\includegraphics[width=\linewidth]{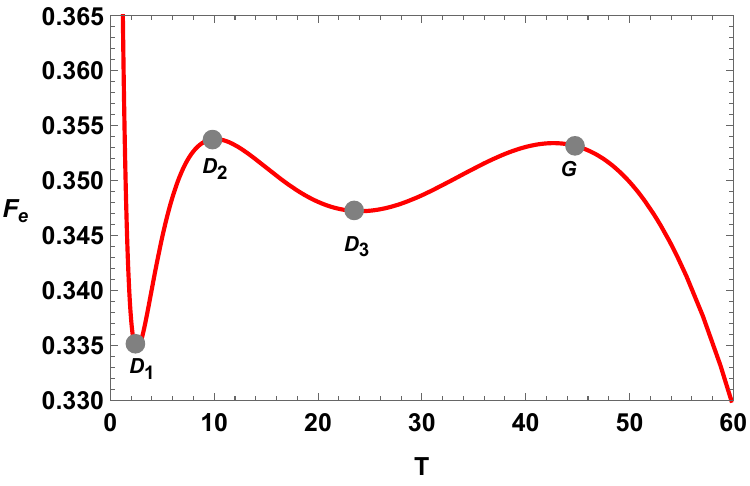}
			\caption{ $T_e=0.07742$}
			\label{7e}
		\end{subfigure}
		\begin{subfigure}{0.4\textwidth}
			\includegraphics[width=\linewidth]{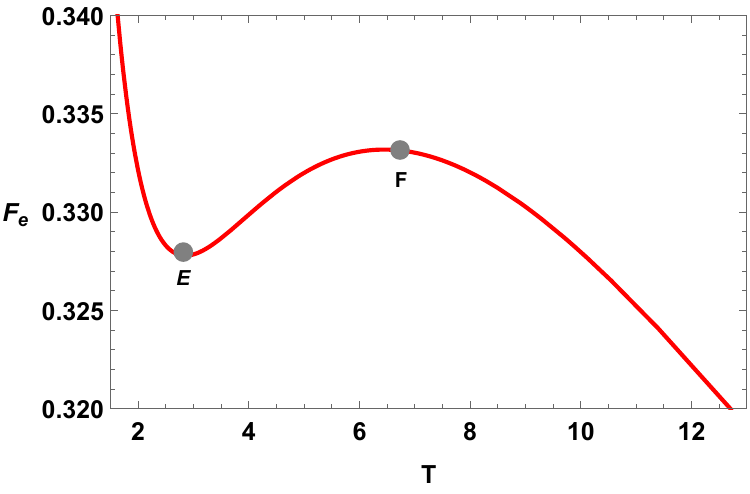}
			\caption{$T_e=0.080$ }
			\label{7f}
		\end{subfigure}
		\caption{Kerr AdS black hole: Free energy landscape. For these plots, we have kept $J=0.2, l=3.5$ and $\kappa=0.015$ fixed.
		}
		\label{7}
	\end{figure}
	\clearpage
	%%%%%%%%%%%%%%%%%%%%%%%%%%%%%%%%%%%%%%%%%%%%%%%%%%%%%%%%%%%%%%%%%%%%%%%%%%%%%%%%%%%%%%%%%%%%%%%%%%%%%%%%%%%%%%%%%%%%%%%%%%%%%%%%%%%%%%%%%%%%%%%%%%%%%%%%%%%%%%%%%%%%%%%%%%%%%%%%%%
	\section{Thermodynamic topology of Kerr-Ads black holes}
	One of the recent development in the context of understanding the criticality of black hole complex phase structure, is the introduction of thermodynamic topology.Topology is often utilized to investigate the light rings \cite{PRL119-251102,PRL124-181101,PRD102-064039,PRD103-104031,PRD105-024049,PRD108-104041,2401.05495} and the time-like circular orbits \cite{PRD107-064006,JCAP0723049,2406.13270}.The concept of topological analysis in black hole thermodynamics was introduced in Ref. \cite{28} which was inspired mainly by the works of Duan \cite{d1,d2} in the context of a relativistic particle system. This approach focusses on vectors with zero points that correspond to critical points. These zero points represent defects where the vector's direction becomes undefined. From a topological perspective, each defect can be assigned a winding number, an integer that helps classify the system into different topological classes. Systems within the same class share similar thermodynamic characteristics. Specifically, critical points are categorized into two classes based on the winding number: the conventional class and the novel class, which differ in how they generate and annihilate black hole branches. The classification of black hole systems into different topological classes also depends on the structure of these critical points. Using this method the thermodynamic topology of black holes has been extensively explored in the literature \cite{PRD105-104053,PLB835-137591,PRD107-046013,PRD107-106009,JHEP0623115,2305.05595,2305.05916,2305.15674,2305.15910,2306.16117,PRD106-064059,PRD107-044026,PRD107-064015,2212.04341,2302.06201,2304.14988,64,2309.00224,2312.12784,2402.18791,2403.14730,2404.02526},  .In this paper, we adopt the topological approach outlined in Ref. \cite{29} which is believed to be the most prominent tool to investigate thermodynamic topology. We use the off-shell free energy method, which treats black hole solutions as topological defects within their thermodynamic framework. This method involves analyzing both local and global topological aspects by calculating the winding numbers associated with these defects. These numbers help categorize black holes into some topological classes based on their overall topological charge. Importantly, a black hole's thermal stability is related to the sign of its winding number. The core idea of thermodynamic topology is to understand these topological defects and their corresponding charges. Applying this method a number of black hole system in different gravitational system has been done\cite{PRD107-064023,PRD107-024024,PRD107-084002,PRD107-084053,2303.06814,2303.13105,2304.02889,2306.13286,2304.05695,2306.05692,2306.11212,
EPJC83-365,2306.02324,PRD108-084041,2307.12873,2309.14069,AP458-169486,2310.09602,2310.09907,2310.15182,2311.04050,2311.11606,2312.04325,2312.06324,2312.13577,2312.12814,PS99-025003,2401.16756,AP463-169617,PDU44-101437,686,2404.08243,2405.02328,685,2405.20022,2406.08793,AC48-100853,682,683}.We will now outline the key mathematical procedures involved in this analysis.\\
	
To overcome negative heat capacity and imaginary energy fluctuations in the partition function of a canonical ensemble for black holes York\cite{york} defined a generalized free energy with the mass and temperature being two independent variables	.Inspired by this, a new generalized off-shell free energy is proposed in \cite{29} expressed as :
\begin{equation}
\mathcal{F} = E - \frac{S}{\tau}
\label{offshell}
\end{equation}
where $E$ represents the black hole's energy (or mass $M$), and $S$ denotes its entropy. The time scale $\tau$ is allowed to vary. which can be interpret  as the inverse of equilibrium temperature at the shell of the black hole. Based on this generalized free energy, a vector field $\phi$ is defined as \cite{29}:
\begin{equation}
\phi = \left(\phi^r, \phi^\Theta \right) = \left(\frac{\partial \mathcal{F}}{\partial S}, -\cot \Theta \, \csc \Theta \right).
\label{phi}
\end{equation}

The zero points of this vector field are significant as they mark the critical points of the black hole solutions. Specifically, these zero points occur at $(\tau, \Theta) = \left(\frac{1}{T}, \frac{\pi}{2}\right)$, where $T$ is the black hole's equilibrium temperature of  its surrounding cavity.At the locations where the vector field either becomes  undefined, these points are interpreted as defects or singularities. Essentially, black holes can be viewed as topological defects within this vector field. Each defect, or zero point, carries a topological charge that can be determined using Duan's $\phi$-mapping technique. To calculate this topological charge, we first determine the unit vector $n$ from the field $\phi$The unit vector $n^a$ must meet the following criteria:
\begin{equation}
n^a n^a = 1 \quad \text{and} \quad n^a \partial_\nu n^a = 0.
\end{equation}
We then define a topological current $j^\mu$ in the coordinate space $x^\nu = \{t, S, \Theta\}$ as:
\begin{equation}
j^\mu = \frac{1}{2\pi} \epsilon^{\mu \nu \rho} \epsilon_{ab} \partial_\nu n^a \partial_\rho n^b,
\end{equation}
where $\epsilon^{\mu \nu \rho}$ is the Levi-Civita symbol and $\partial_\nu = \frac{\partial}{\partial x^\nu}$. This current is conserved, meaning:
\begin{equation}
\partial_\mu j^\mu = 0.
\end{equation}

We can express the current $j^\mu$ in terms of the topological density as:
\begin{equation}
j^\mu = \delta^2(\phi) J^\mu \left( \frac{\phi}{x} \right),
\end{equation}
where $J^\mu$ is related to the Jacobi tensor $\epsilon^{ab} J^\mu \left( \frac{\phi}{x} \right) = \epsilon^{\mu \nu \rho} \partial_\nu \phi^a \partial_\rho \phi^b$. Using the Laplacian Green function $\triangle_{\phi^a} \ln ||\phi|| = 2 \pi \delta^2(\phi)$, we derive the above expression.

The topological charge $W$ is then obtained from the zeroth component of the current density:
\begin{equation}
W = \int_\Sigma j^0 \, d^2x = \sum_{i=1}^N w_i,
\end{equation}
where $w_i$ represents the winding number around each zero point of the vector field $\phi$, and $\Sigma$ is the region used for calculating these winding numbers. To define this region, contours are constructed as:
\begin{equation}
\begin{cases}
S = r_1 \cos \nu + r_0, \\
\Theta = r_2 \sin \nu + \frac{\pi}{2},
\end{cases}
\label{contour}
\end{equation}
where $\nu \in (0, 2\pi)$. Here, $r_1$ and $r_2$ determine the contour dimensions, and $r_0$ is the central point. The winding numbers $w_i$ describe how the vector field $n$ twists around each zero point. The relationship between the deflection angle $\Omega$ and the winding number is:
\begin{equation}
w = \frac{\Omega (2\pi)}{2\pi}
\label{winding}
\end{equation}
where $\Omega$ is computed via:
\begin{equation}
\Omega (\nu) = \int_{0}^{\nu} \epsilon_{12} n^1 \partial_\nu n^2 \, d\nu.
\label{deflection}
\end{equation}

Summing the winding numbers yields the total topological charge:
\begin{equation}
W = \sum_i w_i.
\end{equation}

This topological charge $W$ reveals the structural properties of black hole solutions within thermodynamic topology. Note that $j^\mu$ is nonzero only at the zero points of the vector field $\phi$; if no zero points are found within the parameter region, the total topological charge is zero.\\
	
\begin{figure}[h]
		\centering
		\includegraphics[width=11cm,height=8cm]{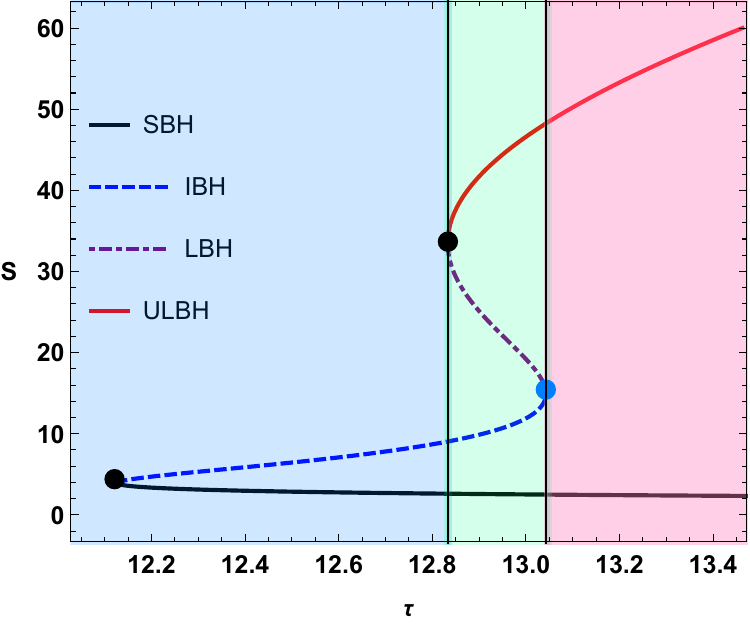}
		
		\caption{Kerr-Ads black hole : $\tau$ vs $S$ plot. Here we have taken $J=Q=0.2$ and $l=3.5$}
		\label{t0a}
	\end{figure}

\begin{figure}[h]
		\centering
		\begin{subfigure}{0.4\textwidth}
			\includegraphics[width=\linewidth]{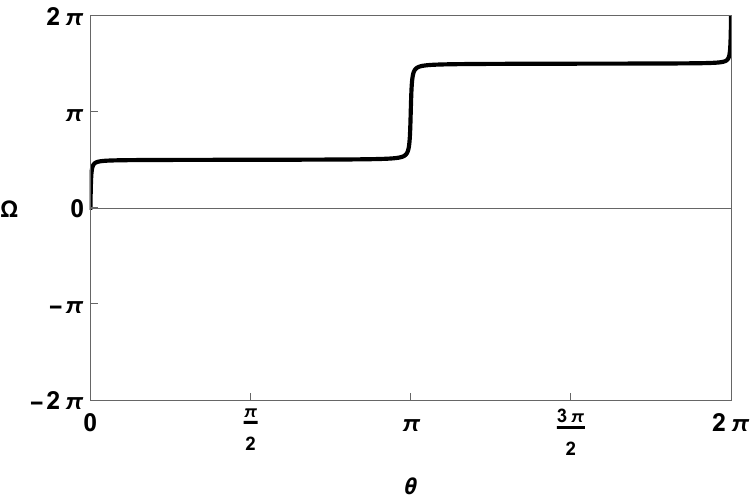}
			\caption{}
			\label{t03a}
		\end{subfigure}
		\begin{subfigure}{0.4\textwidth}
			\includegraphics[width=\linewidth]{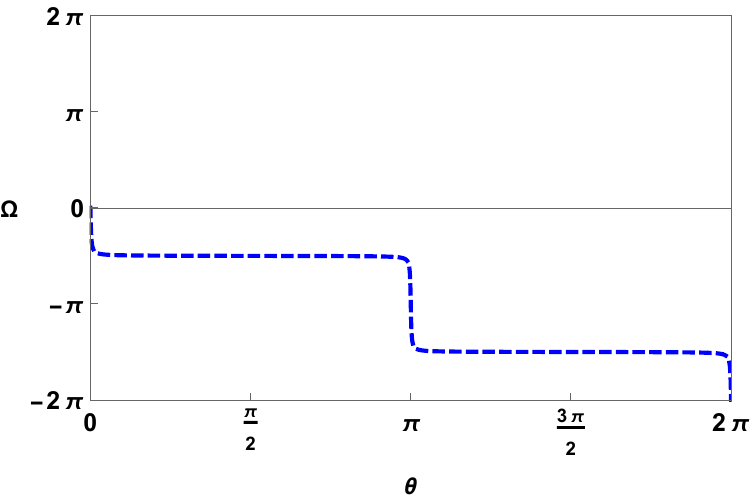}
			\caption{}
			\label{t03b}
		\end{subfigure} \hspace{0.6cm} 
		\begin{subfigure}{0.4\textwidth}
			\includegraphics[width=\linewidth]{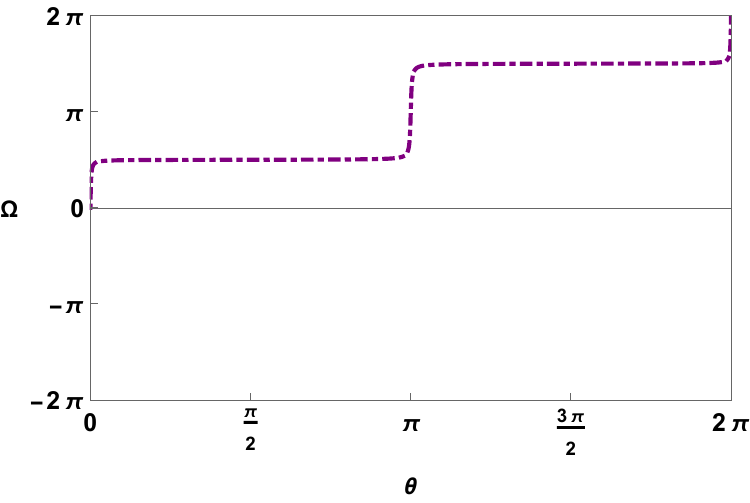}
			\caption{}
			\label{t03c}
		\end{subfigure}
		\begin{subfigure}{0.4\textwidth}
			\includegraphics[width=\linewidth]{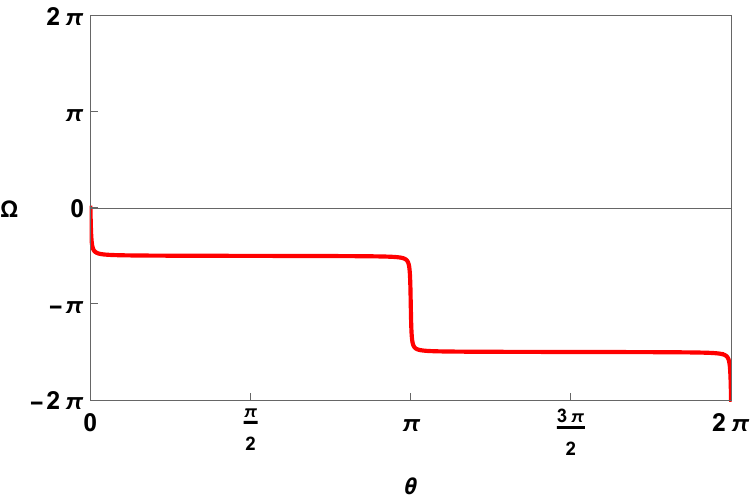}
			\caption{}
			\label{t03d}
		\end{subfigure}
		\caption{Kerr-Ads black holes: Winding number calculations for Kerr-Ads black holes. The black solid line in Fig.\ref{t03a} and purple dot dashed line in Fig.\ref{t03c} represent the winding number for SBH and LBH respectively which equals $+1$ and the blue solid line in Fig.\ref{t03b}. The red solid line in Fig.\ref{t03d} represents the winding number for IBH and ULBH respectively which equals $-1.$ }
		\label{t03}
	\end{figure}

	Now, using eq.\ref{offshell} and eq.\ref{kerrmasskd}, we write the off shell free energy as :
	\begin{multline}
		\mathcal{F}=M-\frac{S}{\tau }\\=-\frac{2 \pi ^{3/2} l^2 S \sqrt{\frac{\sinh ^{-1}(\kappa  S)}{\kappa }}-\tau  \sqrt{\frac{\pi  \kappa  l^2+\sinh ^{-1}(\kappa  S)}{\kappa }} \sqrt{\frac{4 \pi ^3 \kappa ^3 J^2 l^2+\pi  \kappa  l^2 \sinh ^{-1}(\kappa  S)^2+\sinh ^{-1}(\kappa  S)^3}{\kappa ^3}}}{2 \pi ^{3/2} l^2 \tau  \sqrt{\frac{\sinh ^{-1}(\kappa  S)}{\kappa }}}
	\end{multline}
	Using eq.\ref{phi}, components of the vector $\phi $ are found to be :
	\begin{eqnarray}
		\phi ^{r} &=&\frac{\alpha_1}{\beta_1} \\
		&&  \notag \\
		\phi ^{\Theta } &=&-\cot \Theta ~\csc \Theta .
	\end{eqnarray}
	where 
	\begin{multline}
		\alpha_1=-4 \pi ^4 \kappa ^4 J^2 l^4 \tau -4 \pi ^{3/2} \kappa ^4 l^2 \sqrt{\kappa ^2 S^2+1} \left(\frac{\sinh ^{-1}(\kappa  S)}{\kappa }\right)^{3/2} \sqrt{\pi  l^2+\frac{\sinh ^{-1}(\kappa  S)}{\kappa }}\\ \sqrt{\frac{4 \pi ^3 \kappa ^3 J^2 l^2+\pi  \kappa  l^2 \sinh ^{-1}(\kappa  S)^2+\sinh ^{-1}(\kappa  S)^3}{\kappa ^3}}+\pi ^2 \kappa ^2 l^4 \tau  \sinh ^{-1}(\kappa  S)^2+4 \pi  \kappa  l^2 \tau  \sinh ^{-1}(\kappa  S)^3+3 \tau  \sinh ^{-1}(\kappa  S)^4
	\end{multline}
	and 
	\begin{multline}
		\beta_1=4 \pi ^{3/2} \kappa ^4 l^2 \tau  \sqrt{\kappa ^2 S^2+1} \left(\frac{\sinh ^{-1}(\kappa  S)}{\kappa }\right)^{3/2} \sqrt{\pi  l^2+\frac{\sinh ^{-1}(\kappa  S)}{\kappa }} \sqrt{\frac{4 \pi ^3 \kappa ^3 J^2 l^2+\pi  \kappa  l^2 \sinh ^{-1}(\kappa  S)^2+\sinh ^{-1}(\kappa  S)^3}{\kappa ^3}}
	\end{multline}
	Next, we find the zero points or the defects of the vector field. One zero point of the vector field is always $\Theta=\frac{\pi}{2}$ due to the smart choice of the $\Theta$- component of the field. To find the other zero points, we calculate an expression for $\tau$ by solving $\phi ^{S}=0$, which is obtained as:
	\begin{equation}
		\tau =\frac{4 \pi ^{3/2} \kappa ^4 l^2 \sqrt{\kappa ^2 S^2+1} \left(\frac{\sinh ^{-1}(\kappa  S)}{\kappa }\right)^{3/2} \sqrt{\frac{\pi  \kappa  l^2+\sinh ^{-1}(\kappa  S)}{\kappa }} \sqrt{\frac{4 \pi ^3 \kappa ^3 J^2 l^2+\pi  \kappa  l^2 \sinh ^{-1}(\kappa  S)^2+\sinh ^{-1}(\kappa  S)^3}{\kappa ^3}}}{-4 \pi ^4 \kappa ^4 J^2 l^4+\pi ^2 \kappa ^2 l^4 \sinh ^{-1}(\kappa  S)^2+4 \pi  \kappa  l^2 \sinh ^{-1}(\kappa  S)^3+3 \sinh ^{-1}(\kappa  S)^4}
	\end{equation}
	One of the key aspects of the study of black hole thermodynamics from a topological perspective is the calculation of topological charge. To calculate the topological charge, we need to identify the branches of the black hole in the $\tau$ vs $S$ graph.From the thermodynamic study, we understand that there are two possible scenarios. In the first scenario, four distinct black hole phases can occur: small black hole (SBH), intermediate black hole (IBH), large black hole (LBH), and ultra-large black hole (ULBH). In the second scenario, only two black hole phases are observed: a small black hole (SBH) and a large black hole (LBH). We present the two scenarios and the corresponding topological charge in the following paragraph.\\
	In the first case, we present the plot of the KD entropy $S$ as a function of $\tau$ in Fig. \ref{t0a}, using the parameters $J=0.2$, $l=3.5$, and $\kappa=0.0015$. This plot reveals four distinct branches of black holes. 
	\begin{figure}[h]
		\centering
		\begin{subfigure}{0.4\textwidth}
			\includegraphics[width=\linewidth]{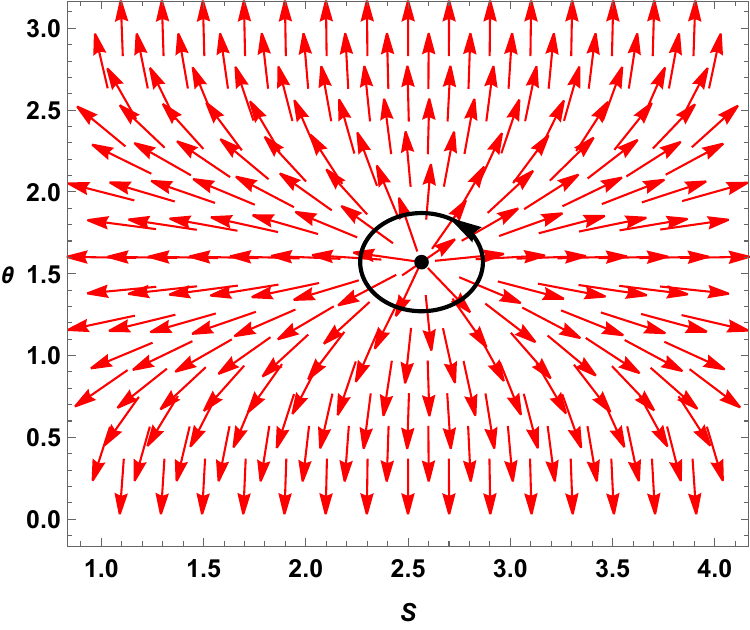}
			\caption{}
			\label{t02a}
		\end{subfigure}
		\begin{subfigure}{0.4\textwidth}
			\includegraphics[width=\linewidth]{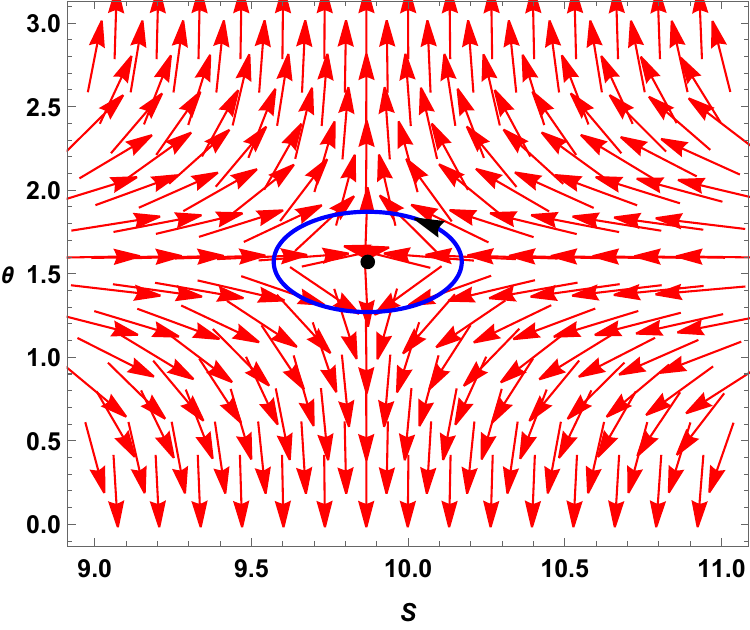}
			\caption{}
			\label{t02b}
		\end{subfigure} \hspace{0.6cm} 
		\begin{subfigure}{0.4\textwidth}
			\includegraphics[width=\linewidth]{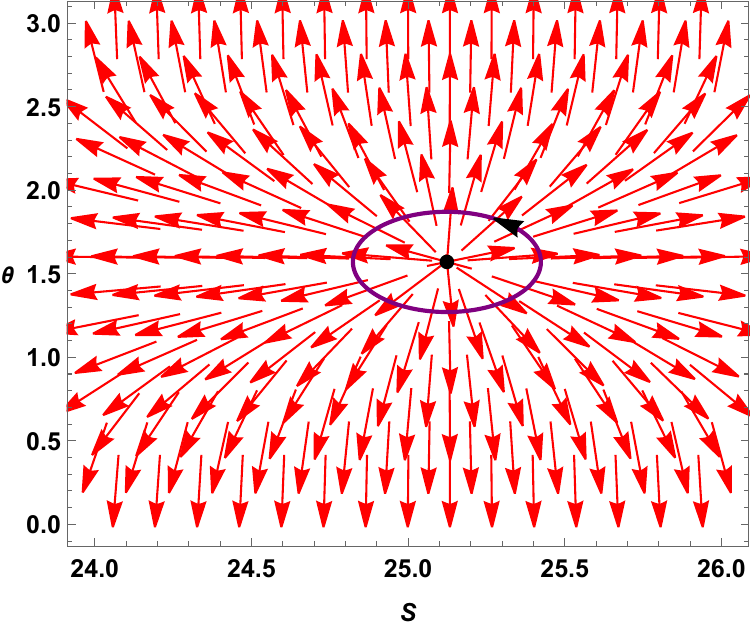}
			\caption{}
			\label{t02c}
		\end{subfigure}
		\begin{subfigure}{0.4\textwidth}
			\includegraphics[width=\linewidth]{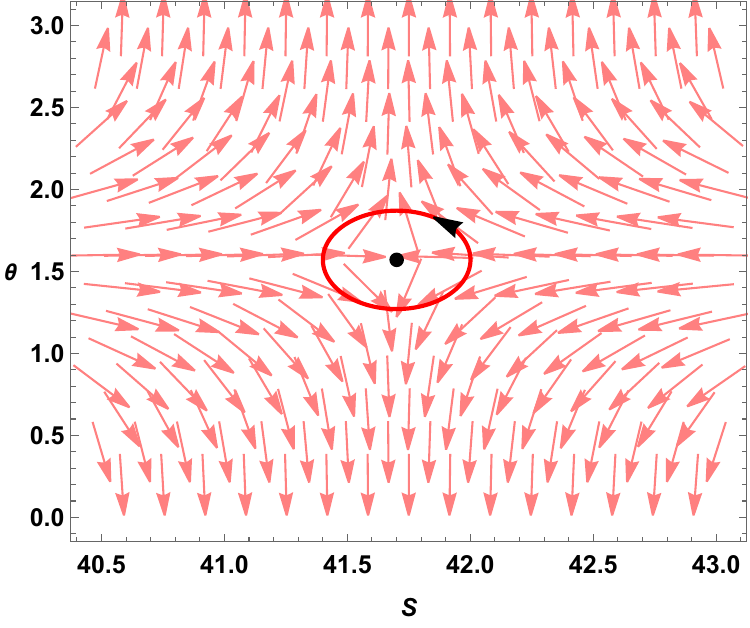}
			\caption{}
			\label{t02d}
		\end{subfigure}
		\caption{Kerr-Ads black holes: Vector plots of normalized ($\phi^S$,$\phi^\Theta$) field on the $S-\Theta$ plane }
		\label{c02}
	\end{figure}
	For the calculation of topological charge, one needs to find the winding number around each zero point of the normalized vector field $(n^1, n^2)$ for a particular value of $\tau$. For example for $\tau=12.9$, we get four zero point at  $S = 2.5657$, $S = 9.8711$, $S = 25.1238$, and $S = 41.7004$. 
	In Figs. \ref{t02a}, \ref{t02b}, \ref{t02c}, and \ref{t02d}, we provide vector plots illustrating the normalized components $\phi^S$ and $\phi^\Theta$ for $\tau = 12.9$. Next, contours are formed around each zero point using the eq.\ref{contour} as :
	\begin{equation}
		\begin{cases}
			S=0.3 cos\nu +r_{0}, \\ 
			\\ 
			\theta =0.2 sin\nu +\frac{\pi }{2},%
		\end{cases}%
		\label{contour1}
	\end{equation}%
	where we have set $r_1=0.3$ and $r_2=0.2.$ and $r_0$ is the zero point around which contours are formed. Substituting the value of $S$ and $\Theta$ in the expression of the normalized field $(n^1, n^2)$ we find the deflection using  eq.\ref{deflection}. This is called the parameterization of the contours. After solving the contour integration, we substitute the value of deflection in the eq.\ref{winding} and find the winding number. The winding number around a zero point represents the winding number for the entire branch in which the zero point lies.  For example, the zero point $S=2.5657$ is on the SBH, hence the winding number around this zero point represents the winding number for the whole branch.\\
	Fig. \ref{t03} displays four contour plots, each illustrating the computation of the winding number around the identified zero points. The solid black and red colored contour represents the winding number for SBH and ULBH respectively. On the other hand, the dashed blue colored contour and the dot-dashed purple colored contour represent the winding number for IBH and LBH respectively. As depicted by the black contour in Fig. \ref{t03a}, the winding number for $S = 2.5657$ (SBH branch) is $+1$. In contrast, Fig. \ref{t03b} shows that the winding number for $S = 9.8711$ (IBH branch) is $-1$. Similarly, the dot-dashed purple contour in Fig. \ref{t03c} indicates that the winding number for $S = 25.1238$ (LBH branch) is $+1$. Finally, the red contour in Fig. \ref{t03d} demonstrates that the winding number for $S=41.7004$ (ULBH branch) is $-1$.
	A positive winding number implies stability for the small and large black hole branches. Conversely, the intermediate and ultra-large black hole branches, having negative winding numbers, are unstable. To find the topological charge, all the individual winding number of each branch is added. Hence the total topological charge of the Kerr-AdS black hole in KD statistics is calculated as $\Sigma_i w_i =1 - 1 + 1 - 1 = 0$. A creation or generation point is the exact location where an unstable black hole branch vanishes and a stable black hole branch begins. Conversely, an annihilation point is where a stable black hole branch vanishes and an unstable black hole branch starts. Here two annihilation point is identified at $(\tau_c, S) = ( 12.1276, 3.9115), ( 12.833, 33.4983)$, marked by black dots in Fig. \ref{t0a} and a generation point is observed at $(\tau_c, S) = (13.0428, 14.9326)$ marked by the blue dot.
	
	%%%%%%%%%%%%%%%%%%%%%%%%%%%%%%%%%%%%%%%5
	
\begin{figure}[h]
		\centering
		\begin{subfigure}{0.4\textwidth}
			\includegraphics[width=\linewidth]{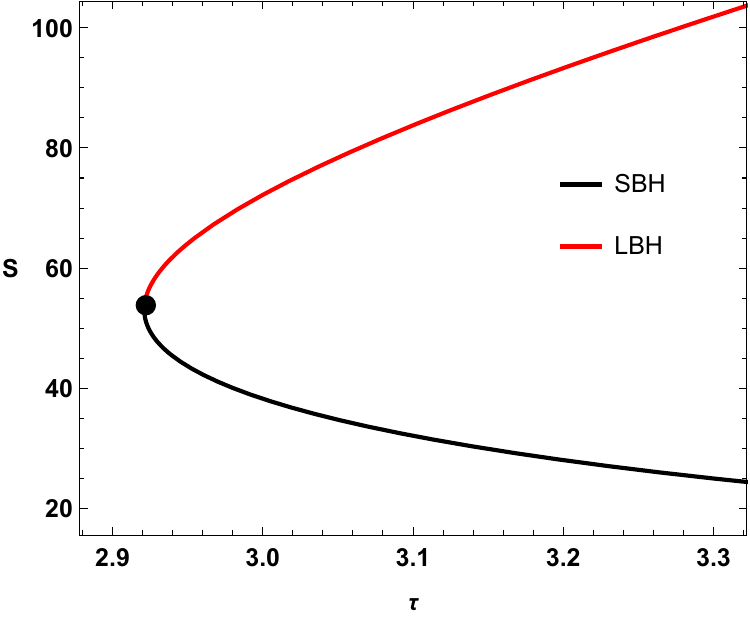}
			\caption{}
			\label{c012a}
		\end{subfigure}
		\begin{subfigure}{0.4\textwidth}
			\includegraphics[width=\linewidth]{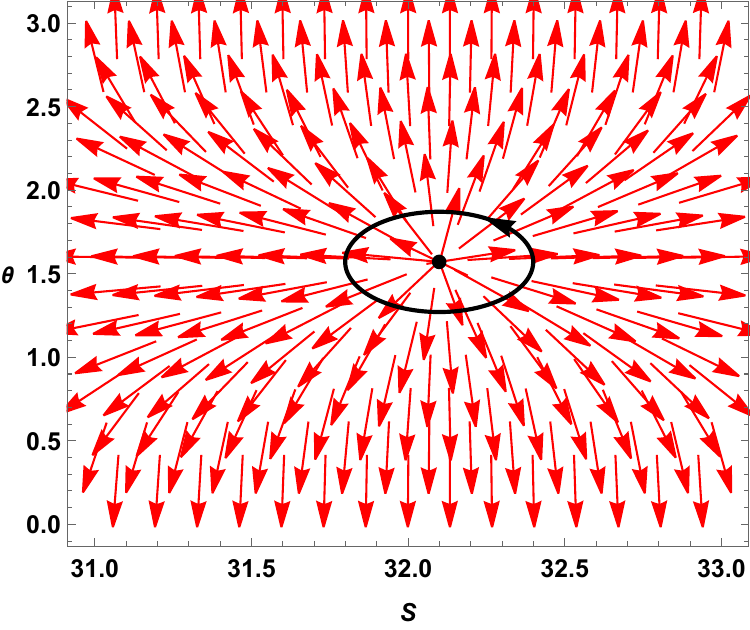}
			\caption{}
			\label{c012b}
		\end{subfigure} \hspace{0.6cm} 
		\begin{subfigure}{0.4\textwidth}
			\includegraphics[width=\linewidth]{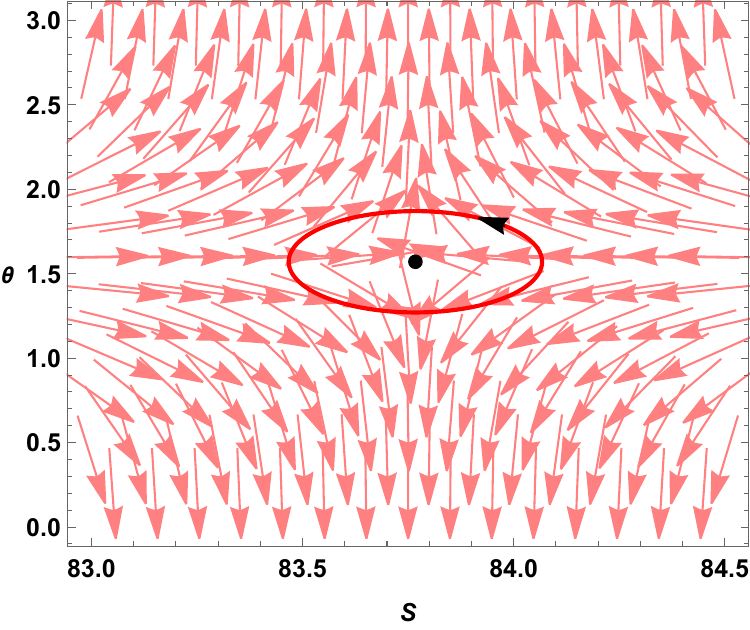}
			\caption{}
			\label{c012c}
		\end{subfigure}
		\begin{subfigure}{0.4\textwidth}
			\includegraphics[width=\linewidth]{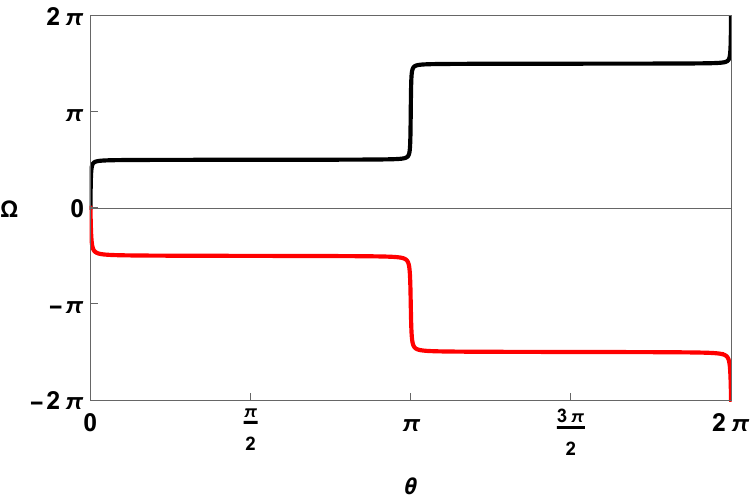}
			\caption{}
			\label{c012d}
		\end{subfigure}
		\caption{Kerr-Ads black holes: In Fig.\ref{c012a}, $\tau$ vs $S$ is plotted for Kerr-Ads black holes when $J=Q=0.5,l=1$ and $\kappa=0.015$ is taken. In Fig.\ref{c012b} and Fig.\ref{c012c} vector plots are shown in $S-\Theta$ plane where the zero point  (S,$\Theta$)= $(35.5982,\pi/2)$ in Fig.\ref{c012b} is on the SBH and the zero point  (S,$\Theta$)= $(82.9878,\pi/2)$ in Fig.\ref{c012c} is on the LBH. The winding number calculation is shown in Fig.\ref{c12d} where the black solid line   in Fig.\ref{c012d} represents winding number for SBH which equals to $+1$ and the red solid line in Fig.\ref{c012d} represents winding number for LBH  which equals to $-1.$ }
		\label{c012}
	\end{figure}

\begin{figure}[h]
		\centering
		\begin{subfigure}{0.4\textwidth}
			\includegraphics[width=\linewidth]{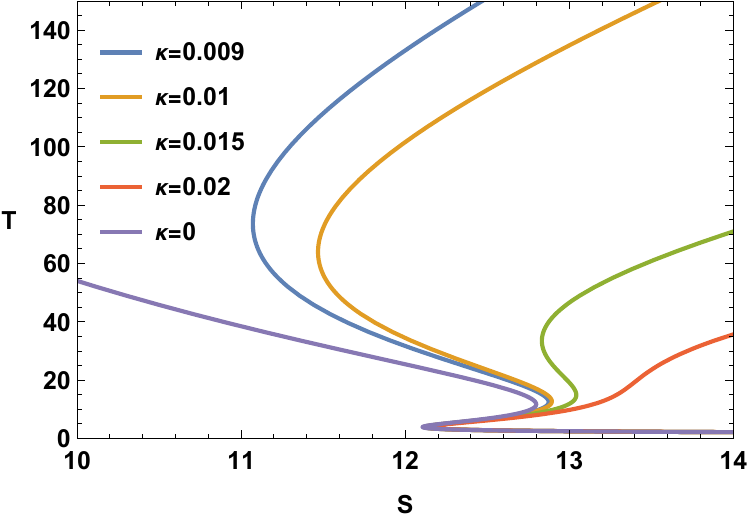}
			\caption{}
			\label{t04a}
		\end{subfigure}
		\begin{subfigure}{0.4\textwidth}
			\includegraphics[width=\linewidth]{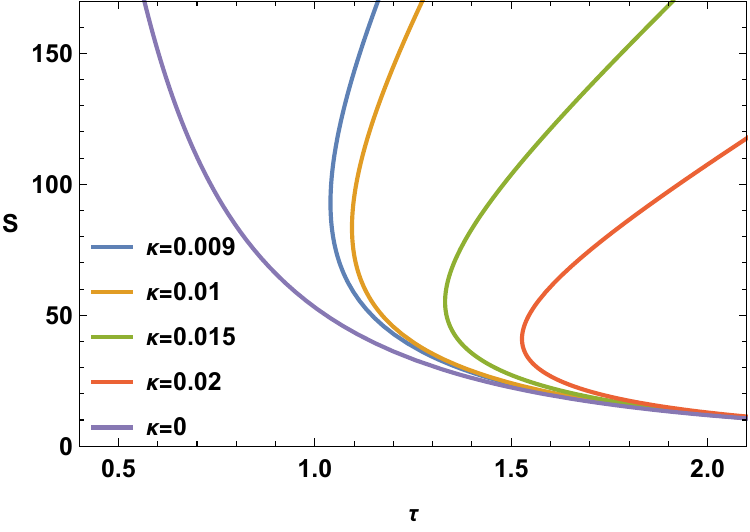}
			\caption{}
			\label{t04b}
		\end{subfigure} \hspace{0.6cm} 
		\begin{subfigure}{0.4\textwidth}
			\includegraphics[width=\linewidth]{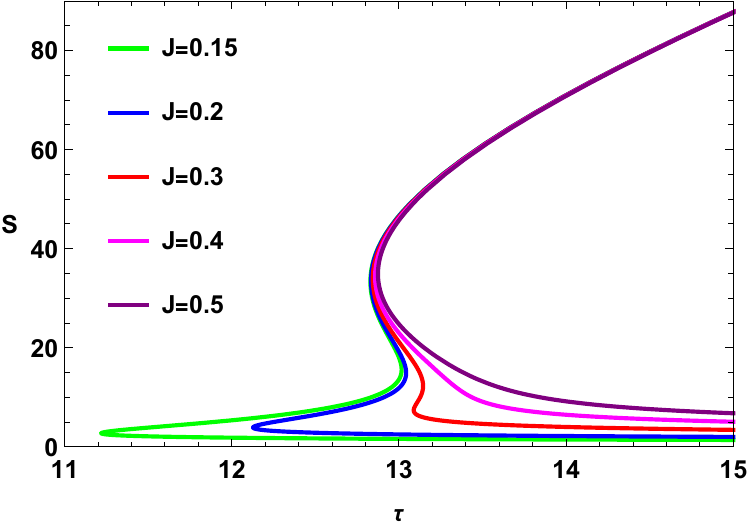}
			\caption{}
			\label{t04c}
		\end{subfigure}
		\begin{subfigure}{0.4\textwidth}
			\includegraphics[width=\linewidth]{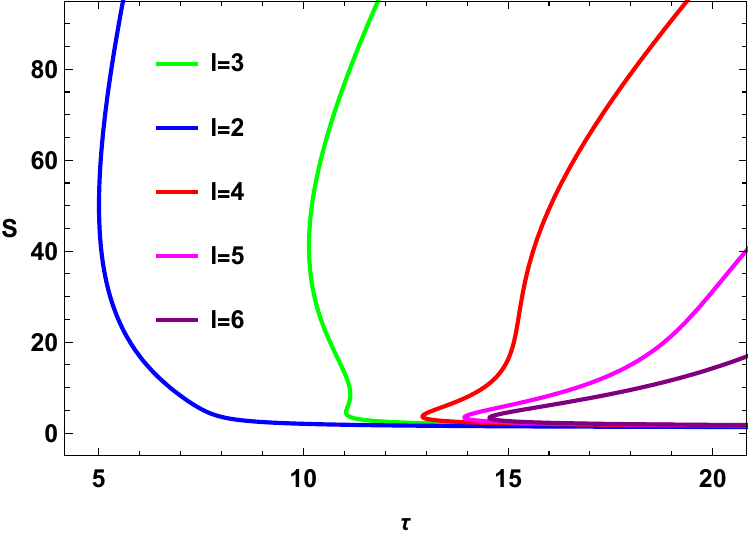}
			\caption{}
			\label{t04e}
		\end{subfigure}
		\caption{Variation of $\protect\tau $ vs $S$ plots for Kerr-Ads black holes with different thermodynamic parameter. }
		\label{c4}
	\end{figure}

	%%%%%%%%%%%%%%%%%%%%%%%%%%%%%%%%%%%%%%%%%%%%%%%%%%%%%%%%%%%%%%%%%%
	Moving on to the second case, where two black hole branches are observed as shown in Fig.\ref{c012a}: a small black hole branch(SBH) and a large black hole branch (LBH). The winding number for SBH is $+1$ which means it is stable and that for LBH is found to be $-1$ which suggests it is an unstable branch. Moreover, we found an annihilation point in  $(\tau=2.9211, S=53.1069)$ represented by the black dot in the Fig.\ref{c012a}. 
	In Fig. \ref{t04a} and Fig. \ref{t04b}, we show the variation of $\kappa$ with $J=0.2$, $l=3.5$, and with $J=0.5$, $l=1$, respectively. Fig. \ref{t04c} illustrates the variation of $J$ with $l=3.5$, and $\kappa=0.015$ held constant. Fig. \ref{t04e},  depict the variation of $l$ with $J=0.2$ and $\kappa=0.015$ fixed.\\
	The number of branches in the $\tau$ vs $S$ graph changes with variations in the values of $J$ and $l$. We observe either four or two branches depending on the specific values of these thermodynamic parameters. However, the topological charge remains invariant when the values of $J$, $l$, and $\kappa$ are varied. Most importantly, a change in the topological charge of the Kerr-Sen AdS black hole is observed when transitioning from the GB statistics framework to the KD statistics framework. In GB statistics, the topological charge of the black hole is $+1$, whereas in the KD statistics formalism, it is found to be $0$.\\
	%%%%%%%%%%%%%%%%%%%%%%%%%%%%%%%%%%%%%%%%%%%%%%%
	%%%%%%%%%%%%%%%%%%%%%%%%%%%%%%%%%%%%%%%%%%%%%%%%%%%%
	In conclusion, for the Kerr-ads black hole in KD statistics, the topological charge is $0$ and it is observed that, the topological charge
	is independent of other thermodynamic parameters.Our topological study of thermodynamics of  Kerr-AdS black holes within the Kaniadakis entropy framework reveals key differences from the Gibbs-Boltzmann case. A crucial finding is the emergence of an entropy bound beyond which the black hole becomes unstable. In contrast to the GB framework, where black holes remain stable across an infinite entropy range, the KD framework introduces additional annihilation points that set an upper limit on stability. This entropy bound is controlled by the Kaniadakis parameter \(\kappa\), with smaller values of \(\kappa\) extending the stability range. Moreover, the introduction of KD entropy alters the thermodynamic topology by modifying the number of black hole branches, generation and annihilation points, and ultimately, the local phase structure. While the total topological charge remains unchanged, the presence of additional unstable branches signifies a fundamental shift in the black hole's phase space, distinguishing Kaniadakis entropy from traditional thermodynamic descriptions.In the upcoming sections on thermodynamic topology, we will demonstrate that a similar topological shift is observed in the other two black hole systems studied. This further establishes that the entropy bound introduced by Kaniadakis entropy is a universal feature of all rotating black holes considered in this work.
	%%%%%%%%%%%%%%%%%%%%%%%%%%%%%%%%%%%%%%%%%%%%%%%%%%%%%%%%%%%%%%%%%%%%%%%%%%%%%%%%%%%%%%%%%%%%%%%%%%%%%%%%%%%%%%%%%%%%%%%%%%%%%%%%%%%%%%%%%%%%%%%%%%%%%%%%%%%%%%%%%%%%%%%%%%%%%%%%%%
		\section{Thermodynamic geometry of the Kerr-AdS black holes}
Another interesting aspect in the arena of black hole thermodynamic is the study of thermodynamic geometry\cite{Weinhold,Ruppeiner1,Ruppeiner2,Mrugala,Aman,Que1,Que2}.The Riemannian curvature is a measure of  the  complexity  in the underlying microscopic interactions prevalent in a particular thermodynamic system. The thermodynamic geometry establishes a link between statistical mechanics and thermodynamics in which the most important thing is an adequately robust and useful metric in the state space of the thermodynamic system.
Weinhold\cite{Weinhold} defined a Riemannian metric in the equilibrium state space of a thermodynamic system. He defined the metric to be the second derivative of internal energy with respect to the other extensive variables. The metric was defined as:
$$g_{ij}^{W} = \partial_{i} \partial_{j} U(S,X)$$  
where the internal energy `$U(S,X)$' is a function of entropy S and other extensive variables X where for a charged rotating  black hole system $X= (J,Q)$. This metric could not provide a proper understanding of the thermodynamic geometry in pure equilibrium thermodynamics. So in the late 1970s Ruppeiner\cite{Ruppeiner1,Ruppeiner2} introduced a Riemannian metric which was defined as the negative hessian of the entropy with respect to the other extensive variables. The metric was defined as:   
$$g_{ij}^{R} =- \partial_{i} \partial_{j} S(U,X)$$
where the entropy `$S(U,X)$' is a function of internal energy, U and other extensive variables, X. This metric is useful in interpreting the distance between the equilibrium states in order to study the thermodynamics of those systems. The Riemannian scalar which is defined here as `$R_{rupp}$' is a scalar invariant function in thermodynamic geometry. This scalar is useful in predicting the nature of microscopic interactions. For the negative and positive scalar curvatures, the interaction is predominantly attractive or repulsive accordingly and the null curvature would mean no interaction and thus signify a flat thermodynamic geometry.\\
The Ruppeiner geometry is found to be conformally related to the Weinhold geometry by the following relation \cite{Mrugala}:
$$g_{ij}^{R} dU^i dU^j = \frac{1}{T}  g_{ij}^{W} dS^i dS^j$$
where $T=\frac{\partial{U}}{\partial{S}}$ is the Hawking temperature of the black hole with $U^i =(U,X)$ and $S^i =(S,X)$.In the rotating black hole hole systems here we have replaced the internal energy, U with the system's mass, M. We will use this conformal relation here to find the Ruppeiner metric and then investigate the thermodynamic geometry of the black hole systems. \\
	
	Although Weinhold and Ruppeiner geometry have provided us with useful insights on the phase structure of various thermodynamic systems. It was found for the case of Kerr AdS black hole \cite{Aman} that the Weinhold metric predicts no phase transitions which is in direct conflict with the results obtained from standard black hole thermodynamics whereas the Ruppeiner metric, would predict phase transitions, but only for a specific thermodynamic potential. Such inconsistencies in the Weinhold and Ruppeiner geometry are removed in a new geometric formalism proposed by Quevedo known as the geometrothermodynamics (GTD)\cite{Que1,
 Que2} where properties of both the phase space and the space of equilibrium states could be unified. The GTD metric unlike the Weinhold and Ruppeiner metric is a legendre invariant metric and is therefore not dependent on any particular choice of thermodynamic potential. The phase transitions that are obtained from the heat capacity of the black hole are properly incorporated in the scalar curvature of the GTD metric, such that a curvature singularity in the GTD scalar `$R_{GTD}$' would be in accordance with the phase transitions  obtained from the heat capacity curves. The general form of the GTD metric is given as: 
$$g = \left(E^{c} \frac{\partial{\Phi}}{\partial{E^{c}}} \right)
\left(\eta_{ab} \delta^{bc} \frac{\partial^2 \Phi}{\partial E^{c} \partial E^{d}} dE^{a} dE^{d}\right)$$     
where `$\Phi$' is the thermodynamic potential and `$E^a$' is an extensive thermodynamic variable with $a=1,2,3....$. Where $\eta_{ab}$=diag$(-1,1,1....)$ and $\delta^{bc}$=diag$(1,1,1....)$. \\

\begin{figure}[h]	
	\centering
	\begin{subfigure}{0.4\textwidth}
		\includegraphics[width=\linewidth]{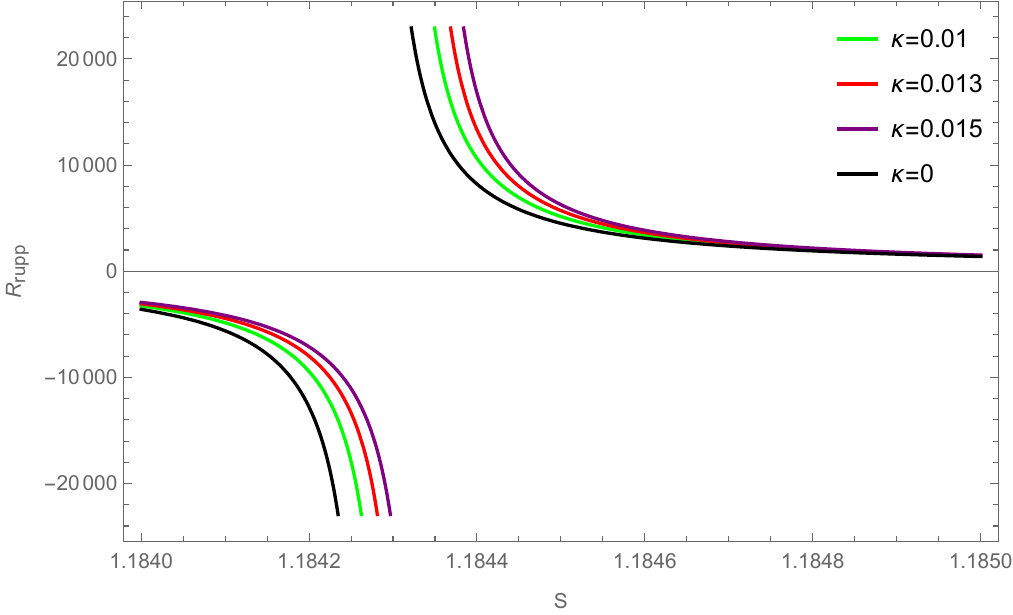}
		\caption{$3.8<S<4.1$ for all $\kappa$}
		\label{a1a}
		\end{subfigure}
		\begin{subfigure}{0.4\textwidth}
		\includegraphics[width=\linewidth]{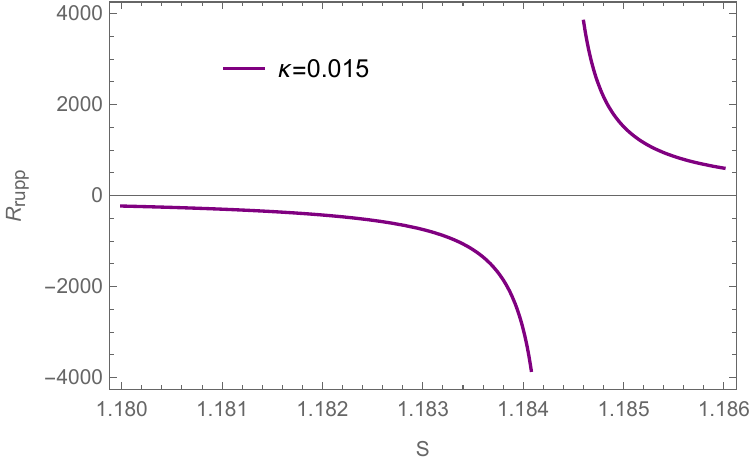}
		\caption{For $\kappa=0.015$ }
		\label{a1b}
		\end{subfigure}
		\begin{subfigure}{0.4\textwidth}
		\includegraphics[width=\linewidth]{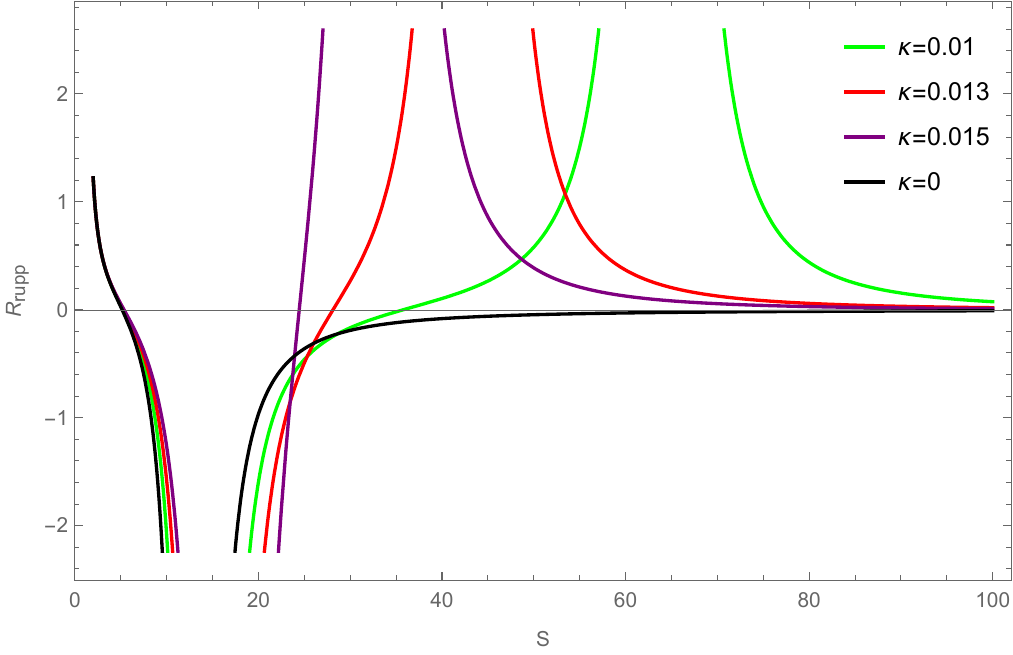}
		\caption{$S>4.1$ for all $\kappa$ }
		\label{a1c}
		\end{subfigure}
		\begin{subfigure}{0.4\textwidth}
		\includegraphics[width=\linewidth]{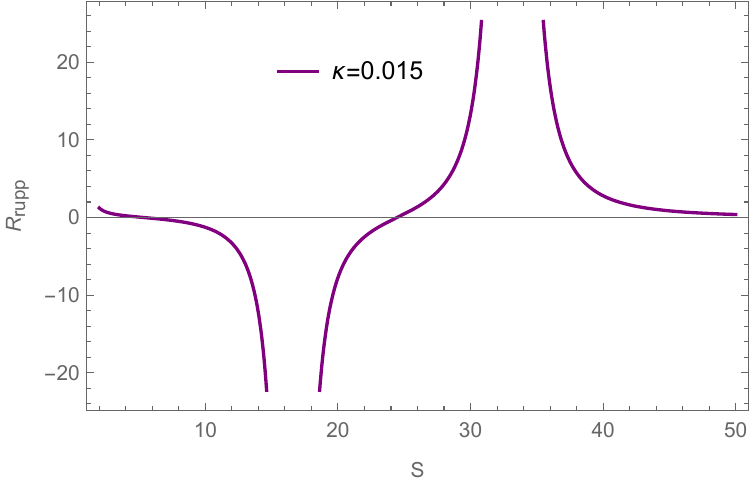}
		\caption{For $\kappa=0.015$ }
		\label{a1d}
		\end{subfigure}
		\begin{subfigure}{0.4\textwidth}
		\includegraphics[width=\linewidth]{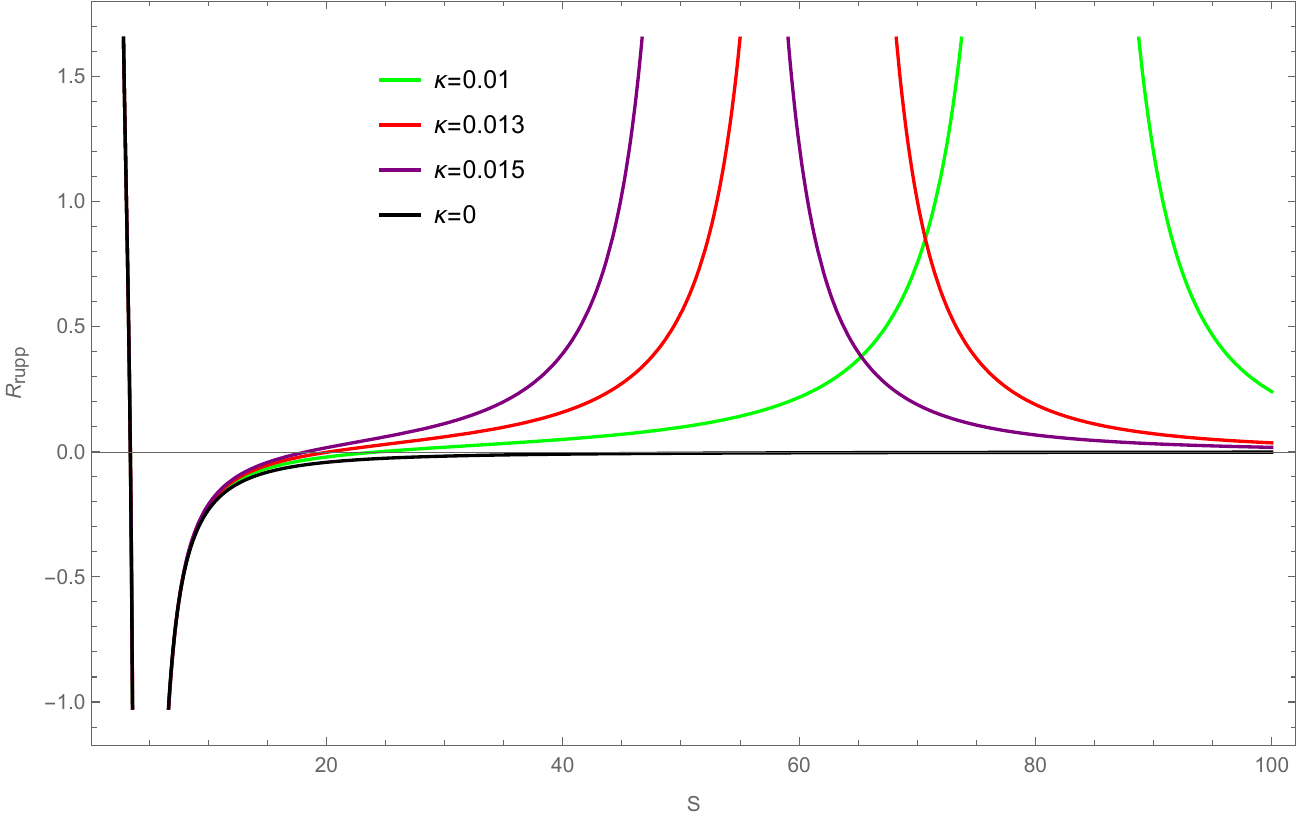}
		\caption{For all $\kappa$ }
		\label{a2a}
		\end{subfigure}
		\begin{subfigure}{0.4\textwidth}
		\includegraphics[width=\linewidth]{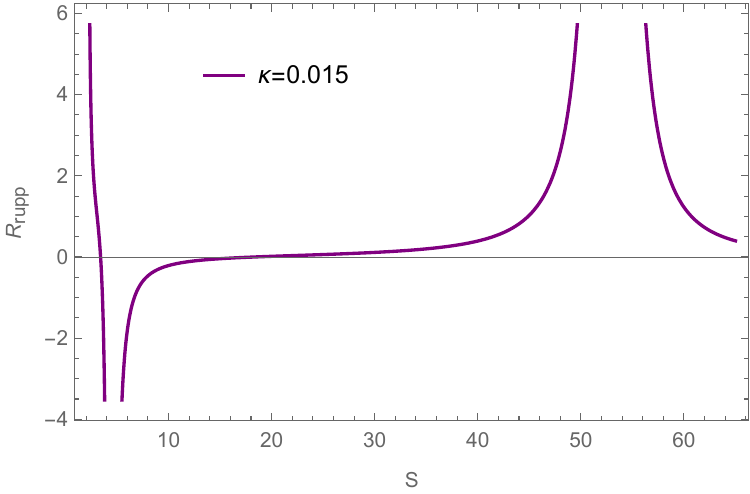}
		\caption{For $\kappa=0.015$ }
		\label{a2b}
		\end{subfigure}
	\caption{ The Ruppeiner scalar versus Entropy plot for Kerr-AdS black hole for $J=0.2$ and $L=3.5$.Fig.\ref{a2a} and Fig.\ref{a2b} shows the Ruppeiner scalar versus Entropy plot for Kerr-AdS black hole for $J=0.55$ and $L=1.5$.}
	\label{a1}
 \end{figure}

\begin{figure}[h]	
	\centering
	\begin{subfigure}{0.4\textwidth}
		\includegraphics[width=\linewidth]{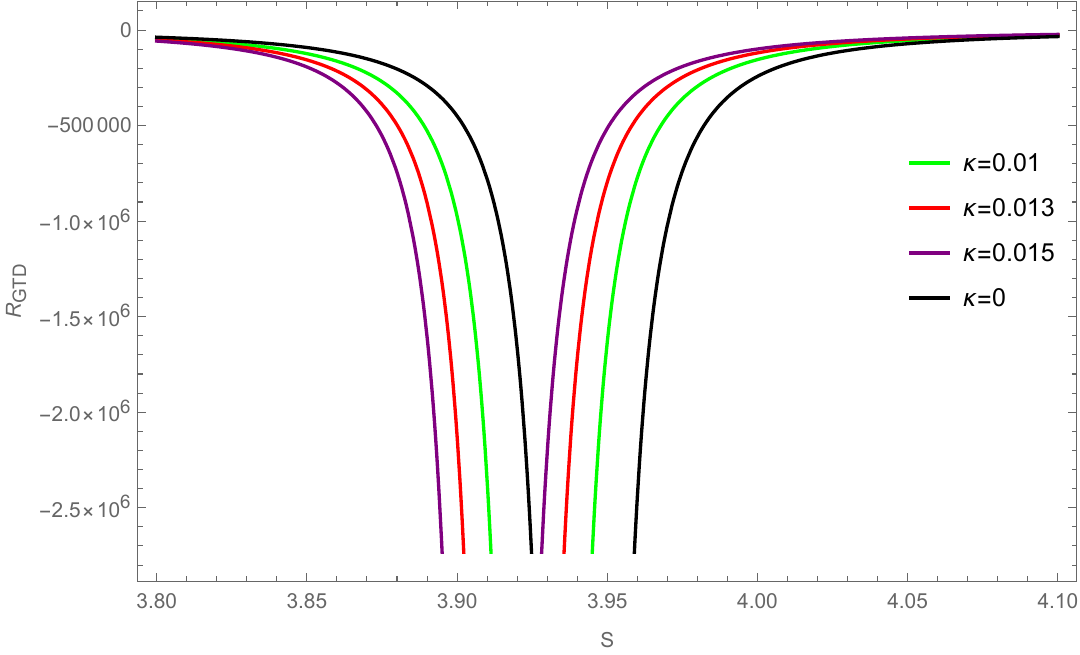}
		\caption{$3.8<S<4.1$ for all $\kappa$}
		\label{a3a}
		\end{subfigure}
		\begin{subfigure}{0.4\textwidth}
		\includegraphics[width=\linewidth]{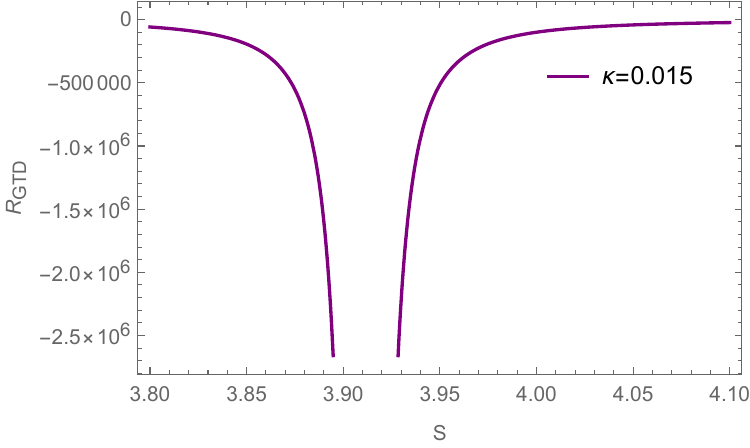}
		\caption{For $\kappa=0.015$ }
		\label{a3b}
		\end{subfigure}
		\begin{subfigure}{0.4\textwidth}
		\includegraphics[width=\linewidth]{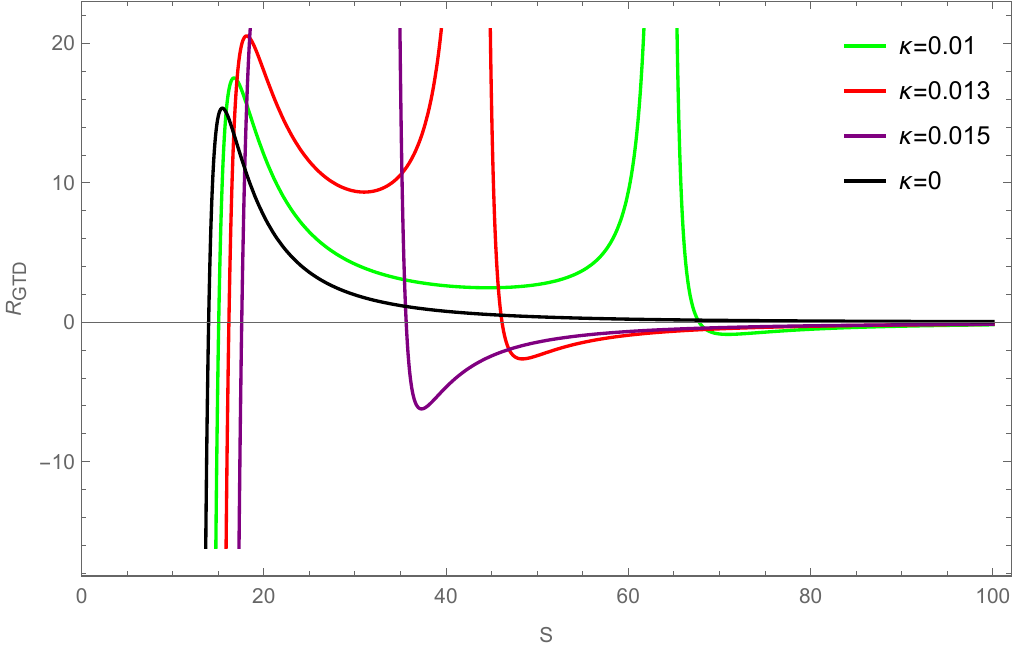}
		\caption{$S>4.1$ for all $\kappa$ }
		\label{a3c}
		\end{subfigure}
		\begin{subfigure}{0.4\textwidth}
		\includegraphics[width=\linewidth]{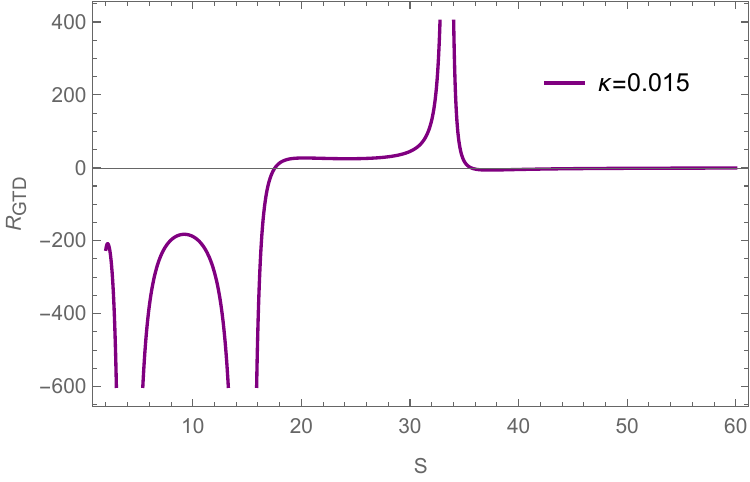}
		\caption{For $\kappa=0.015$ }
		\label{a3d}
		\end{subfigure}
		\begin{subfigure}{0.4\textwidth}
		\includegraphics[width=\linewidth]{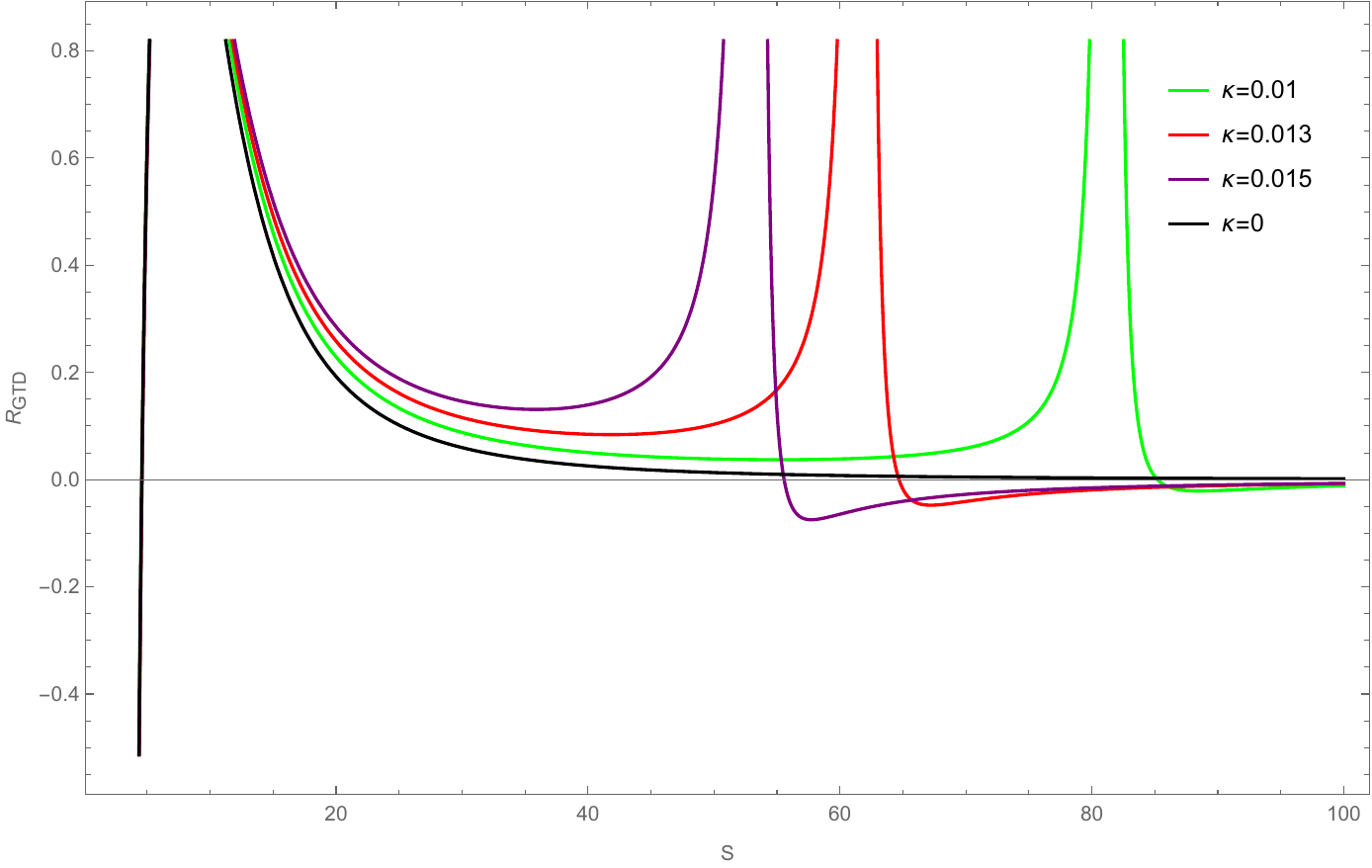}
		\caption{For all $\kappa$ }
		\label{a4a}
		\end{subfigure}
		\begin{subfigure}{0.4\textwidth}
		\includegraphics[width=\linewidth]{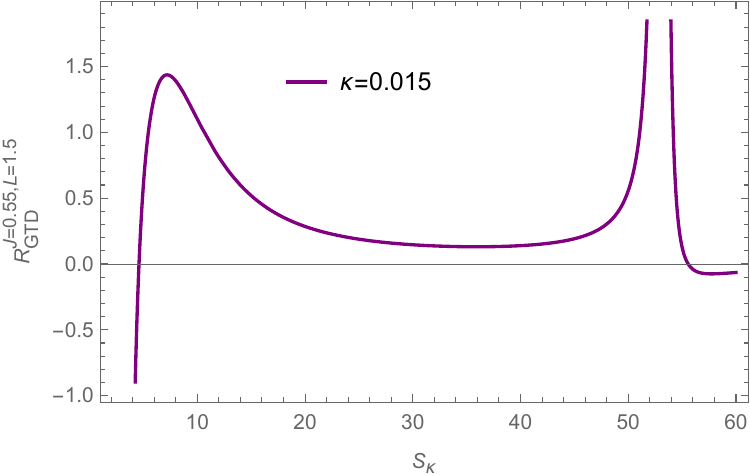}
		\caption{For $\kappa=0.015$ }
		\label{a4b}
		\end{subfigure}
	\caption{ The GTD scalar versus Entropy plot for Kerr-AdS black hole for $J=0.2$ and $L=3.5$.Fig.\ref{a4a} and Fig.\ref{a4b} represents the GTD scalar versus Entropy plot for Kerr-AdS black hole for $J=0.55$ and $L=1.5$.}
	\label{a3}
 \end{figure}

 The line element for the Ruppeiner metric in the Kerr-AdS black hole could be written as :
$$dS_{R}^{2}= \frac{1}{T} \left(\frac{\partial^2 M}{\partial S^2}dS^2 + \frac{\partial^2 M}{\partial J^2} dJ^2 + 2\frac{\partial^2 M}{\partial S \partial J} dS dJ \right)$$
and the line element for the GTD metric in the Kerr-AdS black hole is given by:
$$dS_{G}^{2} =S \left(\frac{\partial M}{\partial S}\right)\left(- \frac{\partial^2 M}{\partial S^2} dS^2 + \frac{\partial^2 M}{\partial J^2} dJ^2\right) $$
 We see here from Fig.\ref{a1} that the Ruppeiner scalar is plotted against the Kaniadakis entropy for $J=0.2$ and $L=3.5$. There are curvature singularities in the Ruppeiner scalar curves for every value of the Kaniadakis parameter and for $\kappa=0.015$ the Ruppeiner curve is separately shown in Fig.\ref{a1b} which shows the presence of  a  singularity at $S=1.184$. From Fig.\ref{a1d} we see that there are two more curvature singularities at $S=16.492$ and $33.147$ for the same J and L values for $\kappa=0.015$. Similarly from Fig.\ref{a2b} for $J=0.55$ and $L=1.5$ we see only one singularity at $S=53.013$. These singularities found in the Ruppeiner scalar curves do not match with the points of Davies type phase transitions found in the specific heat capacity curves for $\kappa=0.015$ for the same J and L values. The GTD curves shown in Fig.\ref{a3b} and Fig.\ref{a3d} show a curvature singularity at $S=3.911$ and two more singularities at   $S=14.933$ and $33.498$ respectively for $J=0.2$ and $L=3.5$. For $J=0.55$ and $L=1.5$ we see from Fig.\ref{a4b} that there happens to be only one singularity at $S=53.107$. Unlike Ruppeiner scalar curves the singularities in the GTD scalar curves match exactly with the points of Davies type phase transitions obtained in the heat capacity curves. 

	\section{Thermodynamic Topology  of Kerr-Sen-Ads Black holes}
	The ADM mass of the Kerr-sen-Ads black hole in terms of entropy is obtained to be 
	\begin{equation}
		M=\frac{\sqrt{S_{\text{BH}}+\pi  l^2} \sqrt{\pi  l^2 S_{\text{BH}} \left(S_{\text{BH}}+2 \pi  Q^2\right)+S_{\text{BH}}^3+4 \pi ^3 J^2 l^2}}{2 \pi ^{3/2} l^2 \sqrt{S_{\text{BH}}}}
	\end{equation}
	The mass is converted to the KD mass as :
	\begin{equation}
		M_{KD}=\frac{\sqrt{\pi  \kappa  l^2+\sinh ^{-1}(\kappa  S)} \sqrt{4 \pi ^3 \kappa ^3 J^2 l^2+\pi  \kappa  l^2 \sinh ^{-1}(\kappa  S) \left(2 \pi  \kappa  Q^2+\sinh ^{-1}(\kappa  S)\right)+\sinh ^{-1}(\kappa  S)^3}}{2 \pi ^{3/2} \kappa ^{3/2} l^2 \sqrt{\sinh ^{-1}(\kappa  S)}}
		\label{kerrsenmasskd}
	\end{equation}
	The off-shell free energy is constructed as :
	\begin{multline}
		\mathcal{F}=M-\frac{S}{\tau }\\=-\frac{2 \pi ^{3/2} l^2 S \sqrt{\frac{\sinh ^{-1}(\kappa  S)}{\kappa }}-\tau  \sqrt{\frac{4 \pi ^3 \kappa ^3 J^2 l^2 \left(\pi  \kappa  l^2+\sinh ^{-1}(\kappa  S)\right)+\left(\pi  \kappa  l^2 \left(\pi  \kappa  Q^2+\sinh ^{-1}(\kappa  S)\right)+\sinh ^{-1}(\kappa  S)^2\right)^2}{\kappa ^4}}}{2 \pi ^{3/2} l^2 \tau  \sqrt{\frac{\sinh ^{-1}(\kappa  S)}{\kappa }}}
	\end{multline}
	The components of the vector $\phi $ are 
	\begin{eqnarray}
		\phi ^{S} &=&\frac{\alpha_2}{\beta_2} \\
		&&  \notag \\
		\phi ^{\Theta } &=&-\cot \Theta ~\csc \Theta .
	\end{eqnarray}
	
	where 
	\begin{figure}[h!]
		\centering
		\begin{subfigure}{0.4\textwidth}
			\includegraphics[width=\linewidth]{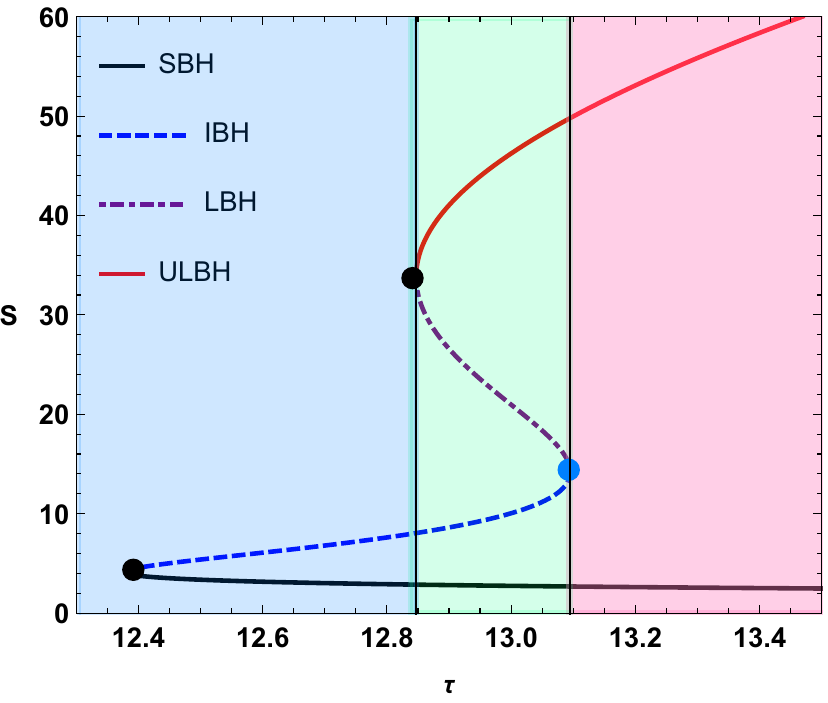}
			\caption{}
			\label{t1a}
		\end{subfigure}
		\begin{subfigure}{0.4\textwidth}
			\includegraphics[width=\linewidth]{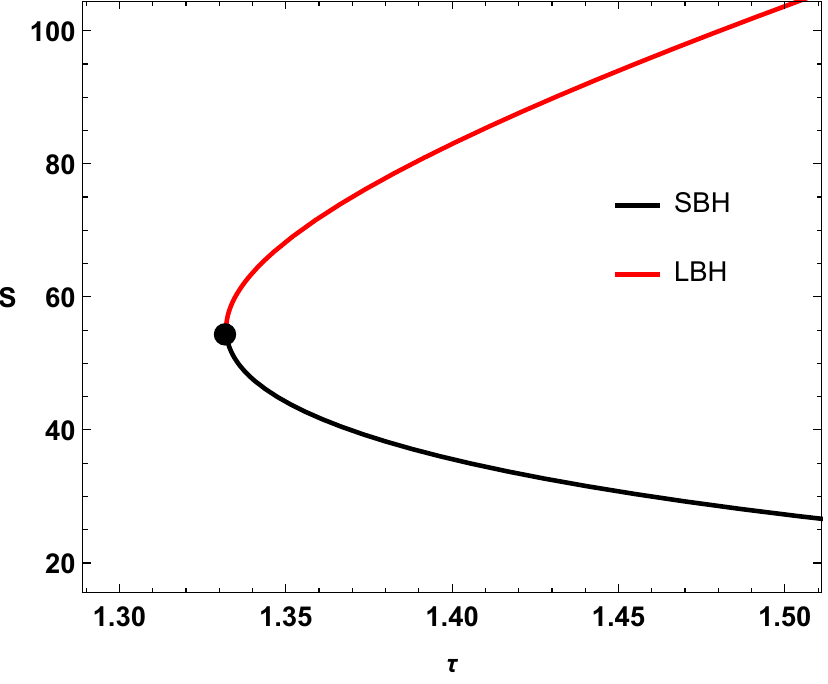}
			\caption{}
			\label{c12a}
		\end{subfigure}
		\caption{Kerr Sen Ads black hole : $\tau$ vs $S$ plot.  In  \ref{t1a}, we have taken $J=Q=0.2$ and $l=3.5$ and  In Fig.\ref{c12a},  $J=Q=0.5,l=1$ and $\kappa=0.015$ is taken. }
	\end{figure}
	\begin{multline}
		\alpha_2=-4 \pi ^4 \kappa ^4 J^2 l^4 \tau -4 \pi ^{3/2} \kappa ^4 l^2 \sqrt{\kappa ^2 S^2+1} \left(\frac{\sinh ^{-1}(\kappa  S)}{\kappa }\right)^{3/2} \sqrt{\pi  l^2+\frac{\sinh ^{-1}(\kappa  S)}{\kappa }}\\ \sqrt{\frac{4 \pi ^3 \kappa ^3 J^2 l^2+2 \pi ^2 \kappa ^2 l^2 Q^2 \sinh ^{-1}(\kappa  S)+\pi  \kappa  l^2 \sinh ^{-1}(\kappa  S)^2+\sinh ^{-1}(\kappa  S)^3}{\kappa ^3}}\\+\pi ^2 \kappa ^2 l^2 \tau  \left(l^2+2 Q^2\right) \sinh ^{-1}(\kappa  S)^2+4 \pi  \kappa  l^2 \tau  \sinh ^{-1}(\kappa  S)^3+3 \tau  \sinh ^{-1}(\kappa  S)^4
	\end{multline}
	and 
	\begin{multline}
		\beta_2=4 \pi ^{3/2} \kappa ^4 l^2 \tau  \sqrt{\kappa ^2 S^2+1} \left(\frac{\sinh ^{-1}(\kappa  S)}{\kappa }\right)^{3/2} \sqrt{\pi  l^2+\frac{\sinh ^{-1}(\kappa  S)}{\kappa }}\\ \sqrt{\frac{4 \pi ^3 \kappa ^3 J^2 l^2+2 \pi ^2 \kappa ^2 l^2 Q^2 \sinh ^{-1}(\kappa  S)+\pi  \kappa  l^2 \sinh ^{-1}(\kappa  S)^2+\sinh ^{-1}(\kappa  S)^3}{\kappa ^3}}
	\end{multline}
	The expression for $\tau$ for which $\phi ^{S}=0$, is calculated as:
	\begin{multline}
		\tau =\\\frac{4 \pi ^{3/2} \kappa ^4 l^2 \sqrt{\kappa ^2 S^2+1} \left(\frac{\sinh ^{-1}(\kappa  S)}{\kappa }\right)^{3/2} \sqrt{\frac{\pi  \kappa  l^2+\sinh ^{-1}(\kappa  S)}{\kappa }} \sqrt{\frac{4 \pi ^3 \kappa ^3 J^2 l^2+2 \pi ^2 \kappa ^2 l^2 Q^2 \sinh ^{-1}(\kappa  S)+\pi  \kappa  l^2 \sinh ^{-1}(\kappa  S)^2+\sinh ^{-1}(\kappa  S)^3}{\kappa ^3}}}{-4 \pi ^4 \kappa ^4 J^2 l^4+2 \pi ^2 \kappa ^2 l^2 Q^2 \sinh ^{-1}(\kappa  S)^2+\pi ^2 \kappa ^2 l^4 \sinh ^{-1}(\kappa  S)^2+4 \pi  \kappa  l^2 \sinh ^{-1}(\kappa  S)^3+3 \sinh ^{-1}(\kappa  S)^4}
	\end{multline}
Next, we present the plot of the KD entropy $S$ as a function of $\tau$ in Fig. \ref{t1a} for the parameters $J=Q=0.2$, $l=3.5$, and $\kappa=0.0015$. This plot reveals four distinct branches of black holes. For $\tau = 13$, the  zero points are identified  at $S = 2.7377$, $S = 10.0419$, $S = 20.9662$, and $S = 46.2368$. Now, we can calculate the winding number around these zero points by applying Duan's $\phi$ mapping technique as shown in the previous section. The winding number associated with $S = 2.7377$ and $S= 20.9662$ is $+1$, while that for $S = 10.0419$ and $S = 46.2368$ is $-1$. As the positive winding number suggests, the small and large black hole branches are stable. Conversely, the intermediate and ultra-large black hole branches are unstable as they have negative winding numbers. The total topological charge of the Kerr-Sen-AdS black hole in KD statistics is found to be $1 - 1 + 1 - 1 = 0$. Additionally, two annihilation points are observed at $(\tau_c, S) = (12.3799, 4.1193), (12.8470, 33.8221)$, which is represented by black dots and one generation point is observed at  $(\tau_c, S) = (13.0952, 14.1983)$  represented by a blue dot in Fig. \ref{t1a}.Another case is presented in Fig.\ref{c12a}, where we see two black hole branches : a small black hole branch(SBH) and a large black hole branch (LBH). The winding number for SBH is $+1$ which means it is stable and that for LBH is found to be $-1$ which suggests it is an unstable branch. Moreover we found a annihilation point in  $(\tau=1.33207,S=55.1569)$ represented by the black dot in the Fig.\ref{c12a} .The number of branches in the $\tau$ vs $S$ graph changes with a change in the value of $J, Q$ and $l.$ We observe either four or two branches depending on the values of thermodynamic parameters. However, the topological charge remains invariant when the values of $J, Q, l$ and $\kappa$ are varied. In conclusion, for the Kerr-Sen-ads black hole in KD statistics, the topological charge is $0$ and it is observed that, the topological charge is independent of other thermodynamic parameters.Here again, we see a change in the topological charge of Kerr-Sen Ads black hole when we change the framework from GB  to KD statistics. In GB statistics, the topological charge of the black hole is $+1$  whereas  the same for the black hole is found to be $0$ in KD statistics formalism.Here again we observe  that Kaniadakis (KD) statistics introduce an entropy bound beyond which black hole stability changes. In KD statistics, we find a stable black hole branch up to a certain entropy value, marked by the presence of an additional annihilation point. Beyond this point, an extra unstable ultra-large black hole branch appears. In contrast, in Gibbs-Boltzmann (GB) statistics, no such entropy bound exists, and the black hole remains stable across the entire entropy range. This difference suggests that the thermodynamic behavior of black holes depends on the chosen entropy framework, affecting the stability and possible phase transitions.\\

	%%%%%%%%%%%%%%%%%%%%%%%%%%%%%%%%%%%%%%%%%%%%%%%%%%%%%%%%%%%%%%%%%%%%%%%%%%%%%%%%%%%%%%%%%%%%%%%%%%%%%%%%%%%%%%%%%%%%%%%%
	
\section{Thermodynamic geometry of Kerr-Sen-AdS black holes} 
\begin{figure}[h]	
	\centering
	\begin{subfigure}{0.4\textwidth}
		\includegraphics[width=\linewidth]{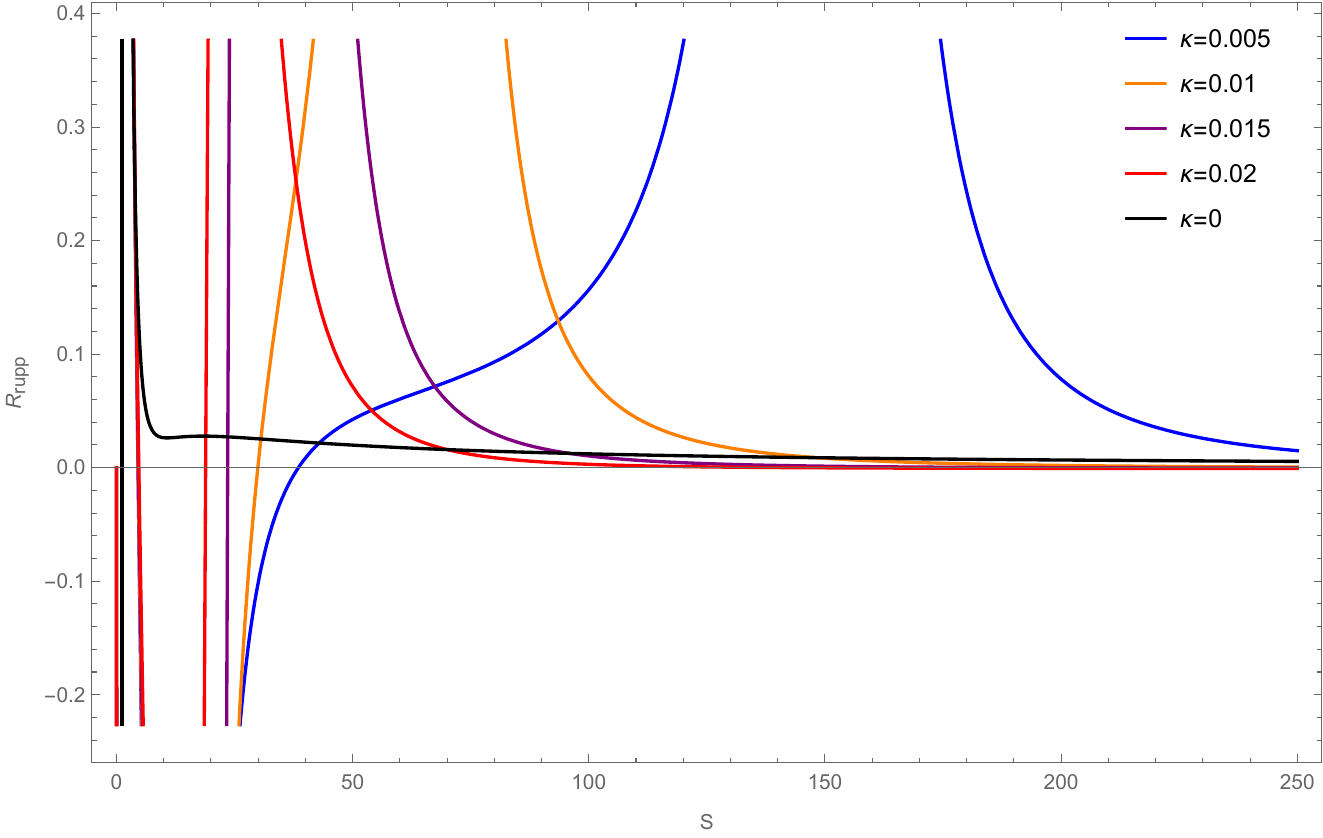}
		\caption{For all $\kappa$}
		\label{a5a}
		\end{subfigure}
		\begin{subfigure}{0.4\textwidth}
		\includegraphics[width=\linewidth]{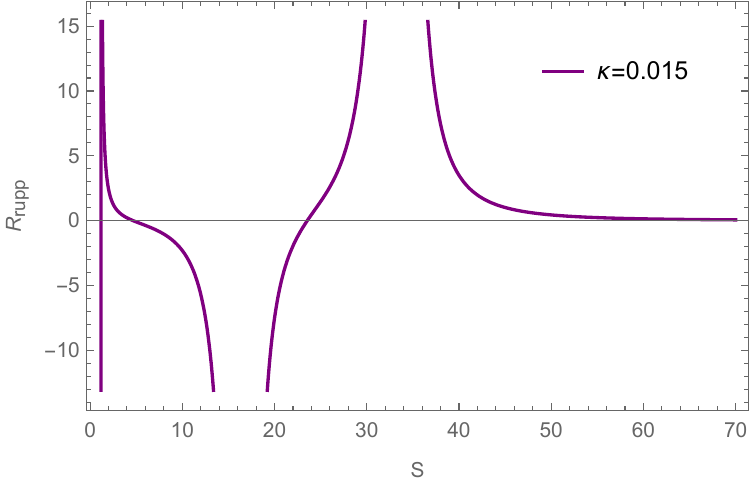}
		\caption{For $\kappa=0.015$ }
		\label{a5b}
		\end{subfigure}
		\begin{subfigure}{0.4\textwidth}
		\includegraphics[width=\linewidth]{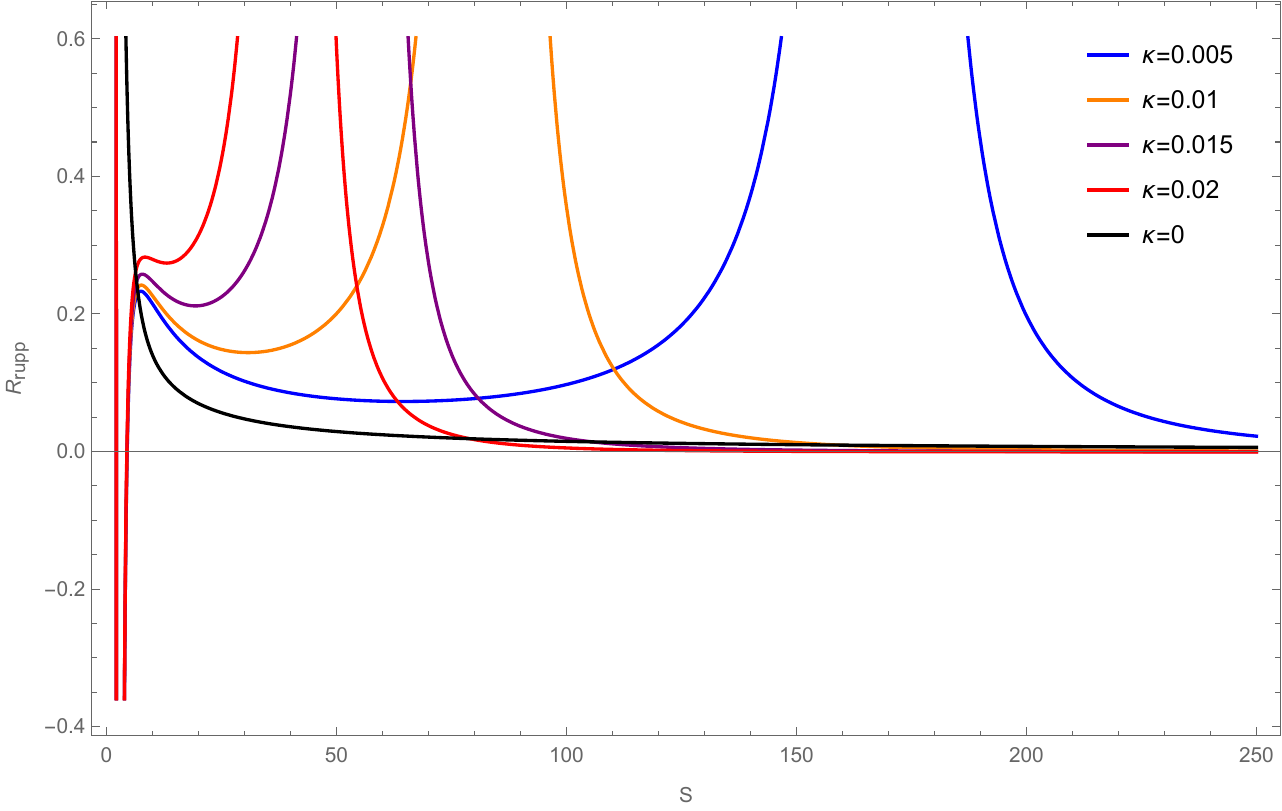}
		\caption{For all $\kappa$ }
		\label{a6a}
		\end{subfigure}
		\begin{subfigure}{0.4\textwidth}
		\includegraphics[width=\linewidth]{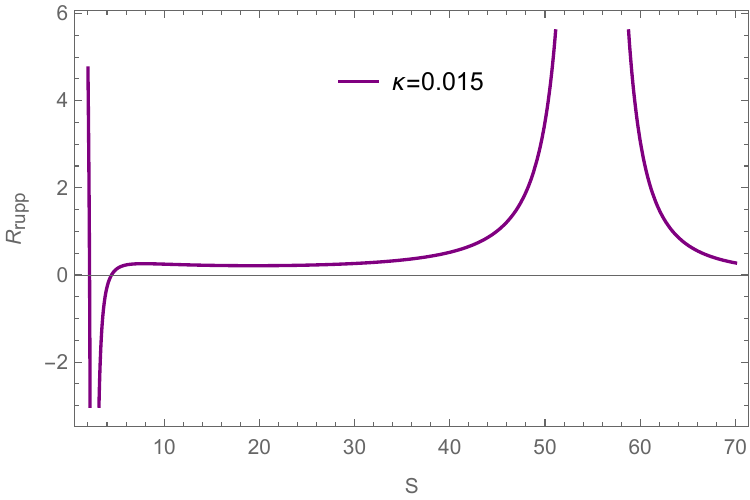}
		\caption{For $\kappa=0.015$ }
		\label{a6b}
		\end{subfigure}
	\caption{ The first panel represents the Ruppeiner scalar versus Entropy plot for Kerr-Sen-AdS black hole for $J=Q=0.2$ and $L=3.5$ and the second panel illustrates the Ruppeiner scalar versus Entropy plot for Kerr-Sen-AdS black hole for $J=Q=0.5$ and $L=1$.}
	\label{a5}
 \end{figure}

\begin{figure}[h]	
	\centering
	\begin{subfigure}{0.4\textwidth}
		\includegraphics[width=\linewidth]{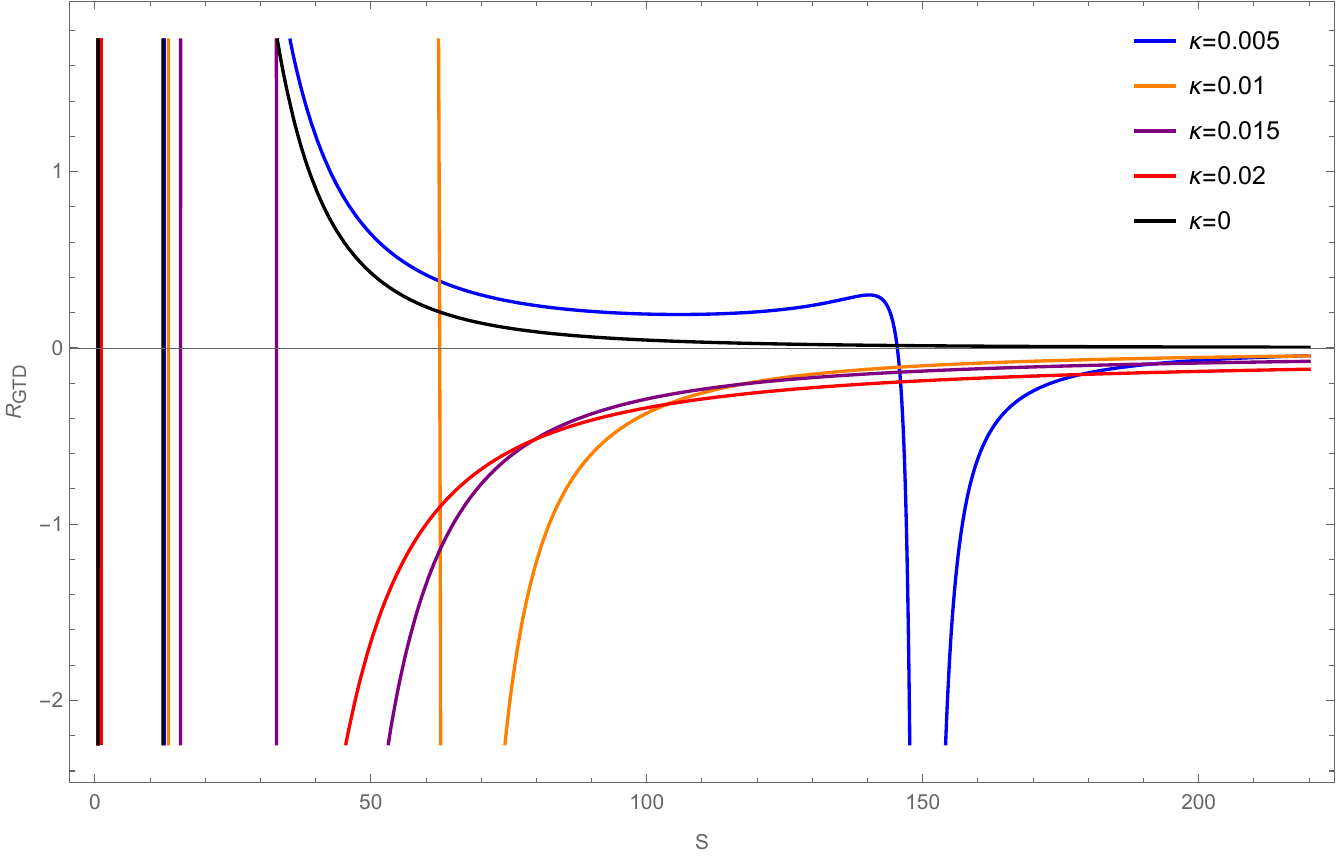}
		\caption{For all $\kappa$}
		\label{a7a}
		\end{subfigure}
		\begin{subfigure}{0.4\textwidth}
		\includegraphics[width=\linewidth]{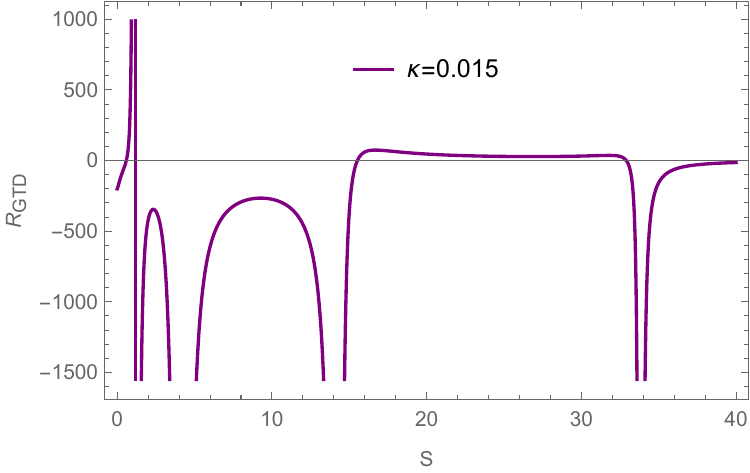}
		\caption{For $\kappa=0.015$ }
		\label{a7b}
		\end{subfigure}
		\begin{subfigure}{0.4\textwidth}
		\includegraphics[width=\linewidth]{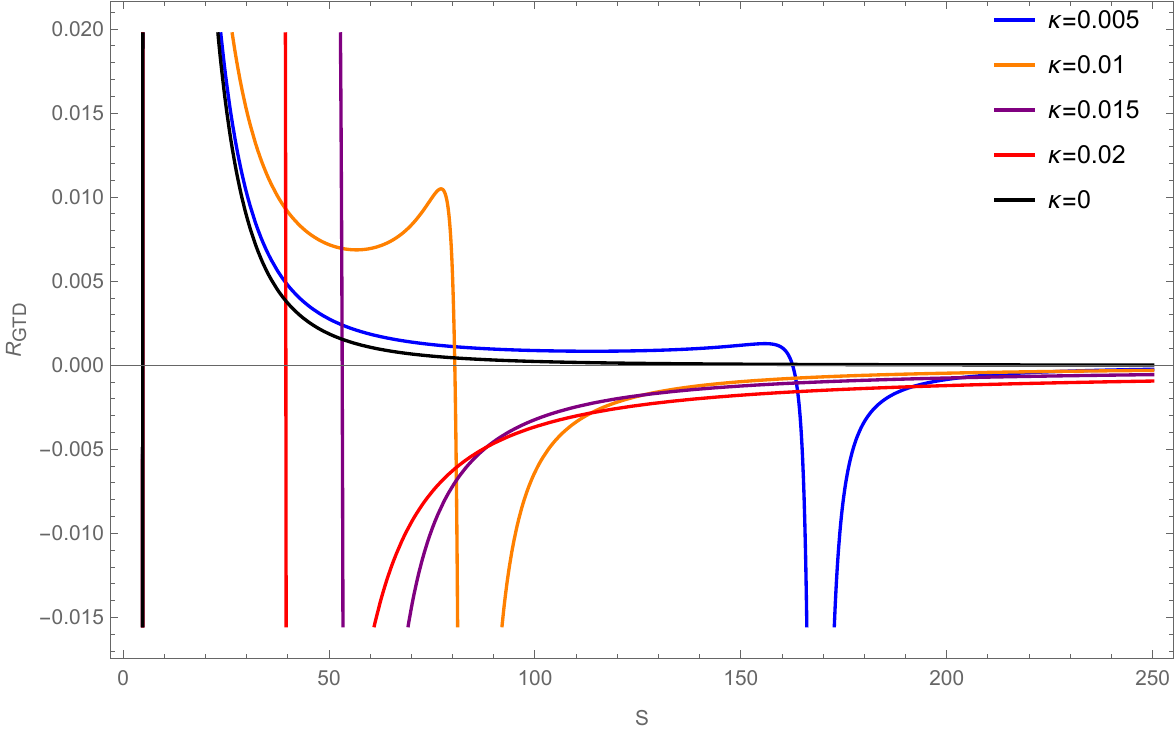}
		\caption{For all $\kappa$ }
		\label{a8a}
		\end{subfigure}
		\begin{subfigure}{0.4\textwidth}
		\includegraphics[width=\linewidth]{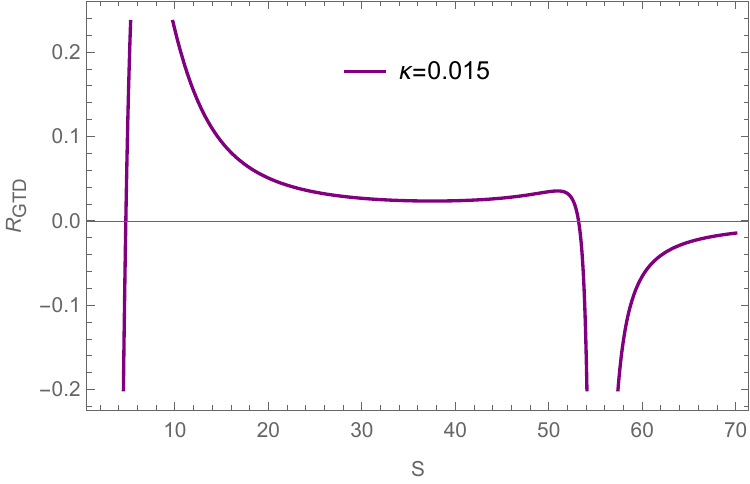}
		\caption{For $\kappa=0.015$ }
		\label{a8b}
		\end{subfigure}
	\caption{ The first panel shows GTD scalar versus Entropy plot for Kerr-Sen-AdS black hole for $J=Q=0.2$ and $L=3.5$ and the second panel illustrates the GTD scalar versus Entropy plot for Kerr-Sen-AdS black hole for $J=Q=0.5$ and $L=1$.}
	\label{a7}
 \end{figure}

 The line element for the Ruppeiner metric in the Kerr-Sen-AdS black hole could be written as :
$$dS_{R}^{2}= \frac{1}{T} \left(\frac{\partial^2 M}{\partial S^2}dS^2 + \frac{\partial^2 M}{\partial J^2} dJ^2 +\frac{\partial^2 M}{\partial Q^2}  dQ^2 + 2\frac{\partial^2 M}{\partial S \partial J} dS dJ + 2 \frac{\partial^2 M}{\partial J \partial Q}  dJ dQ + 2\frac{\partial^2 M}{\partial S \partial Q} dS dQ\right)$$

and the line element for the GTD metric in the Kerr-Sen-AdS black hole is given by:
$$dS_{G}^{2}=S \left(\frac{\partial M}{\partial S}\right)\left(- \frac{\partial^2 M}{\partial S^2} dS^2 + \frac{\partial^2 M}{\partial J^2} dJ^2 +  \frac{\partial^2 M}{\partial Q^2}dQ^2\right) $$
 We see from Fig.\ref{a5b} that the Ruppeiner scalar is plotted against the Kaniadakis entropy for $J=Q=0.2$ and $L=3.5$ which has three singularities at $S=1.193, 16.184$ and $33.297$ for $\kappa=0.015$ whereas in Fig.\ref{a6b} we see that for $J=Q=0.5$ and $L=1$ there is only one singularity at $S=55.100$ for $\kappa=0.015$. The GTD scalar on the other hand has singularities at  $S=4.119, 14.200$ and $33.822$ for $J=Q=0.2$ and $L=3.5$ as can be seen in Fig.\ref{a7b}. We see from Fig.\ref{a8b} that for $J=Q=0.5$ and $L=1$ there is only one singularity at $S=55.157$. It is worth mentioning that unlike Ruppeiner scalar it is only the GTD scalar that produce singularities which agree with the points of Davies type phase transitions found in the heat capacity curves. 
 
		\section{Thermodynamic Topology of Kerr-Newman-AdS black hole}
	The ADM mass of the Kerr-Newman-Ads black hole in terms of entropy is obtained to be 
	\begin{equation}
		M=\frac{\sqrt{4 \pi ^3 J^2 l^2 \left(\pi  l^2+S\right)+\left(\pi  l^2 \left(\pi  Q^2+S\right)+S^2\right)^2}}{2 \pi ^{3/2} l^2 \sqrt{S}}
		\label{knmass}
	\end{equation}
	In KD statistics the mass is converted to the KD mass as :
	\begin{equation}
		M_{KD}=\frac{\sqrt{\frac{4 \pi ^3 \kappa ^3 J^2 l^2 \left(\pi  \kappa  l^2+\sinh ^{-1}(\kappa  S)\right)+\left(\pi  \kappa  l^2 \left(\pi  \kappa  Q^2+\sinh ^{-1}(\kappa  S)\right)+\sinh ^{-1}(\kappa  S)^2\right)^2}{\kappa ^4}}}{2 \pi ^{3/2} l^2 \sqrt{\frac{\sinh ^{-1}(\kappa  S)}{\kappa }}}
		\label{kerrnmmasskd}
	\end{equation}
	
	The off-shell free energy for Kerr-Newman-AdS black hole is  :
	\begin{multline}
		\mathcal{F}=M-\frac{S}{\tau }\\=-\frac{2 \pi ^{3/2} l^2 S \sqrt{\frac{\sinh ^{-1}(\kappa  S)}{\kappa }}-\tau  \sqrt{\frac{\pi  \kappa  l^2+\sinh ^{-1}(\kappa  S)}{\kappa }} \sqrt{\frac{4 \pi ^3 \kappa ^3 J^2 l^2+2 \pi ^2 \kappa ^2 l^2 Q^2 \sinh ^{-1}(\kappa  S)+\pi  \kappa  l^2 \sinh ^{-1}(\kappa  S)^2+\sinh ^{-1}(\kappa  S)^3}{\kappa ^3}}}{2 \pi ^{3/2} l^2 \tau  \sqrt{\frac{\sinh ^{-1}(\kappa  S)}{\kappa }}}
	\end{multline}
	The components of the vector $\phi $ are 
	\begin{eqnarray}
		\phi ^{S} &=&\frac{\alpha_3}{\beta_3} \\
		&&  \notag \\
		\phi ^{\Theta } &=&-\cot \Theta ~\csc \Theta .
	\end{eqnarray}
	where 
	\begin{multline}
		\alpha_3=-4 \pi ^4 \kappa ^4 J^2 l^4 \tau -4 \pi ^{3/2} \kappa ^4 l^2 \sqrt{\kappa ^2 S^2+1} \left(\frac{\sinh ^{-1}(\kappa  S)}{\kappa }\right)^{3/2} \\\sqrt{\frac{4 \pi ^3 \kappa ^3 J^2 l^2 \left(\pi  \kappa  l^2+\sinh ^{-1}(\kappa  S)\right)+\left(\pi  \kappa  l^2 \left(\pi  \kappa  Q^2+\sinh ^{-1}(\kappa  S)\right)+\sinh ^{-1}(\kappa  S)^2\right)^2}{\kappa ^4}}\\-\pi ^4 \kappa ^4 l^4 Q^4 \tau +\pi ^2 \kappa ^2 l^4 \tau  \sinh ^{-1}(\kappa  S)^2+2 \pi ^2 \kappa ^2 l^2 Q^2 \tau  \sinh ^{-1}(\kappa  S)^2+4 \pi  \kappa  l^2 \tau  \sinh ^{-1}(\kappa  S)^3+3 \tau  \sinh ^{-1}(\kappa  S)^4
	\end{multline}
	and 
	\begin{multline}
		\beta_3=4 \pi ^{3/2} \kappa ^4 l^2 \tau  \sqrt{\kappa ^2 S^2+1} \left(\frac{\sinh ^{-1}(\kappa  S)}{\kappa }\right)^{3/2} \\\sqrt{\frac{4 \pi ^3 \kappa ^3 J^2 l^2 \left(\pi  \kappa  l^2+\sinh ^{-1}(\kappa  S)\right)+\left(\pi  \kappa  l^2 \left(\pi  \kappa  Q^2+\sinh ^{-1}(\kappa  S)\right)+\sinh ^{-1}(\kappa  S)^2\right)^2}{\kappa ^4}}
	\end{multline}
	\begin{figure}[h!]
		\centering
		\includegraphics[width=11cm,height=8cm]{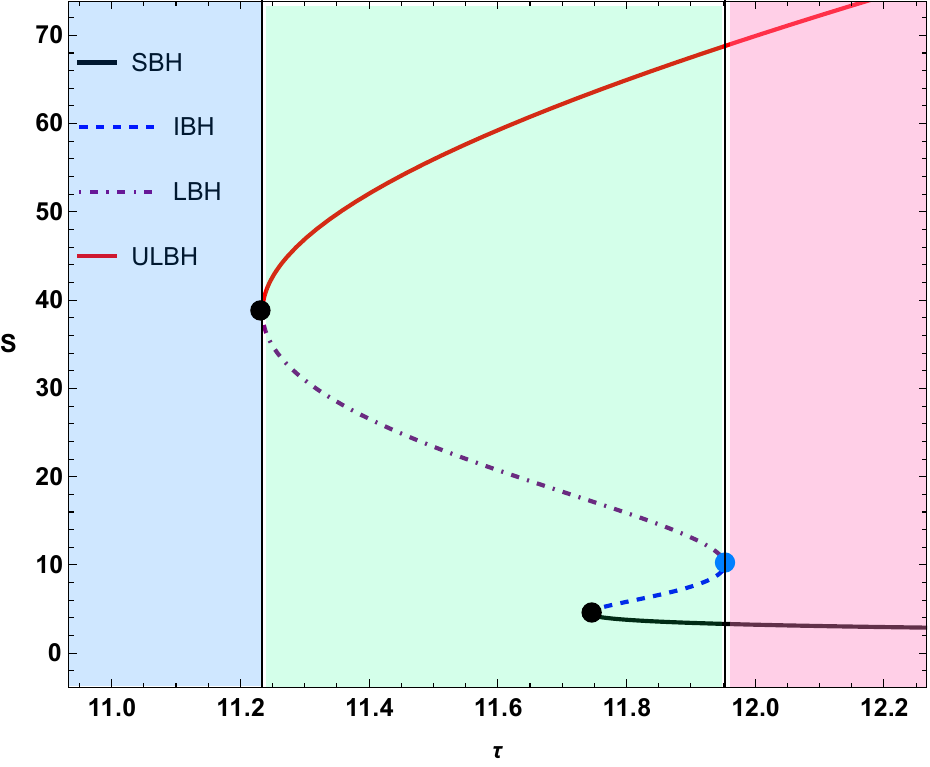}
		\caption{Kerr Newman black hole : $\tau$ vs $S$ plot. Here we have taken $J=Q=0.2$ and $l=3.2$}
		\label{30}
	\end{figure}
	The expression for $\tau$ for which $\phi ^{S}=0$, is calculated as:
	\begin{multline}
		\tau =\\\frac{4 \pi ^{3/2} \kappa ^4 l^2 \sqrt{\kappa ^2 S^2+1} \left(\frac{\sinh ^{-1}(\kappa  S)}{\kappa }\right)^{3/2} \sqrt{\frac{4 \pi ^3 \kappa ^3 J^2 l^2 \left(\pi  \kappa  l^2+\sinh ^{-1}(\kappa  S)\right)+\left(\pi  \kappa  l^2 \left(\pi  \kappa  Q^2+\sinh ^{-1}(\kappa  S)\right)+\sinh ^{-1}(\kappa  S)^2\right)^2}{\kappa ^4}}}{-4 \pi ^4 \kappa ^4 J^2 l^4-\pi ^4 \kappa ^4 l^4 Q^4+\pi ^2 \kappa ^2 l^4 \sinh ^{-1}(\kappa  S)^2+2 \pi ^2 \kappa ^2 l^2 Q^2 \sinh ^{-1}(\kappa  S)^2+4 \pi  \kappa  l^2 \sinh ^{-1}(\kappa  S)^3+3 \sinh ^{-1}(\kappa  S)^4}
	\end{multline}
	
	The KD entropy $S$ as a function of $\tau$ is presented in Fig. \ref{30} for the parameters $J=Q=0.2$, $l=3.2$, and $\kappa=0.0015$. As expected there are four distinct branches of black holes. in the $\tau$ vs $S$ plot.  For $\tau = 11.8$, we identify zero points of the vector field at $S = 3.7899$, $S = 5.8937$, $S = 15.8937$, and $S = 64.9428$. The winding number associated with $S = 3.7899$ and $S= 15.8937$ is $+1$, while that for $S = 5.8186$ and $S = 15.9428$ is $-1$. As the positive winding number suggests, the small and large black hole branches are stable. Conversely, the intermediate and ultra-large black hole branches are unstable as they have negative winding numbers. The total topological charge of the Kerr-Newman-AdS black hole in KD statistics is found to be $1 - 1 + 1 - 1 = 0$. Additionally, two annihilation points are observed at $(\tau_c, S) = (11.7419, 4.5450), (11.2342, 38.6597)$, which is represented by black dots and one generation point is observed at  $(\tau_c, S) = (11.9527, 10.0505)$  represented by a blue dot in Fig. \ref{30}.

For $J = Q = 0.5$, $l = 1.5$, and $\kappa = 0.015$, we observe the presence of two black hole branches: a small black hole (SBH) branch and a large black hole (LBH) branch. The winding number for the SBH branch is calculated to be $+1$, indicating its stability, whereas for the LBH branch, it is found to be $-1$, signifying its instability. Additionally, we identify the existence of an annihilation point.  
Notably, as we vary the thermodynamic parameters, the number of black hole branches reduces from four (as seen in previous cases also) to two. However, the overall topological charge remains unchanged at $W = 0$. Despite this, the number of annihilation and generation points changes, indicating a shift in the local topology of the black hole system.  It is important to emphasize that when we set $\kappa = 0$, the number of branches in Gibbs-Boltzmann (GB) statistics is consistently one less than in Kaniadakis (KD) statistics. Specifically, when four branches exist in KD statistics, only three are observed in GB statistics, and when two branches exist in KD statistics, only one is present in GB statistics. This also suggests that the thermodynamic topology of the black hole undergoes a transformation, as the topological charge in GB statistics remains $+1$, while in KD statistics, it is $0$.  From a local topological perspective, the introduction of KD statistics consistently introduces an additional annihilation point. In GB statistics, when there are four branches, there are two annihilation points, and when there are two branches, there is one annihilation point. Conversely, in KD statistics, for the same parameter values, three branches exist with one annihilation point, and when only one branch remains, there are no annihilation points. This distinction highlights the significant differences between the two frameworks from a topological perspective. \\
	\section{Thermodynamic geometry of Kerr-Newman-AdS black holes} 
\begin{figure}[h]	
	\centering
	\begin{subfigure}{0.4\textwidth}
		\includegraphics[width=\linewidth]{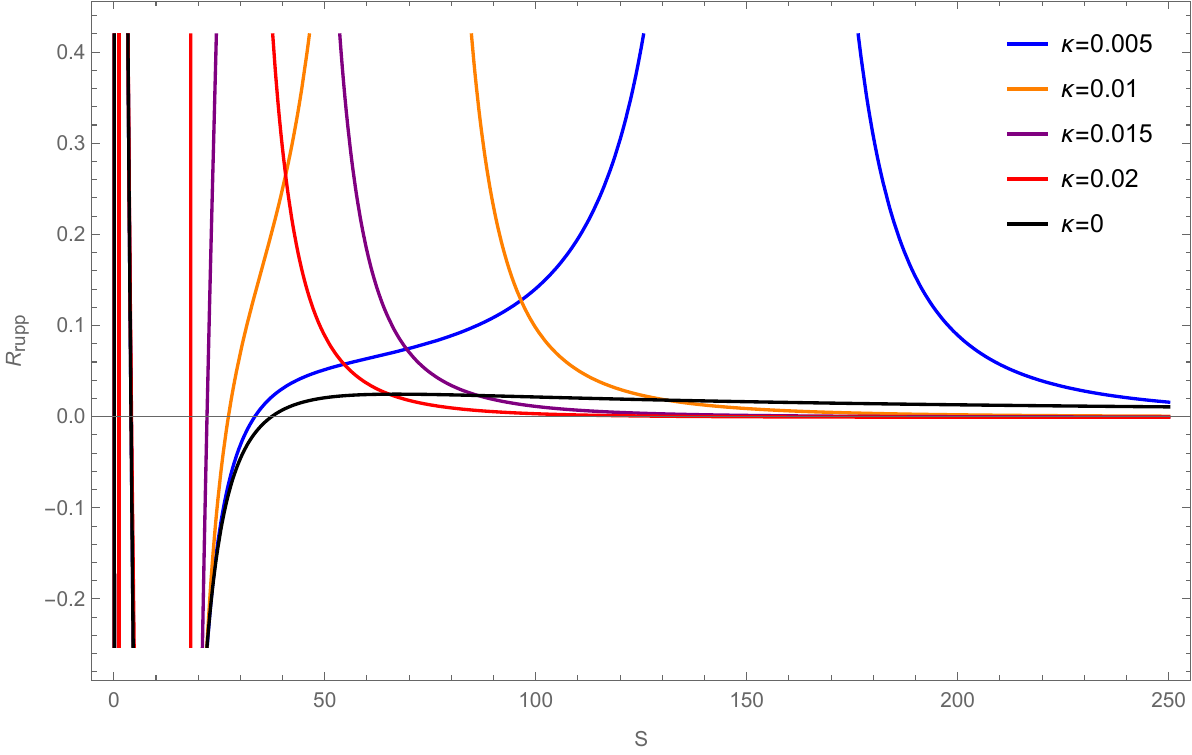}
		\caption{For all $\kappa$}
		\label{a9a}
		\end{subfigure}
		\begin{subfigure}{0.4\textwidth}
		\includegraphics[width=\linewidth]{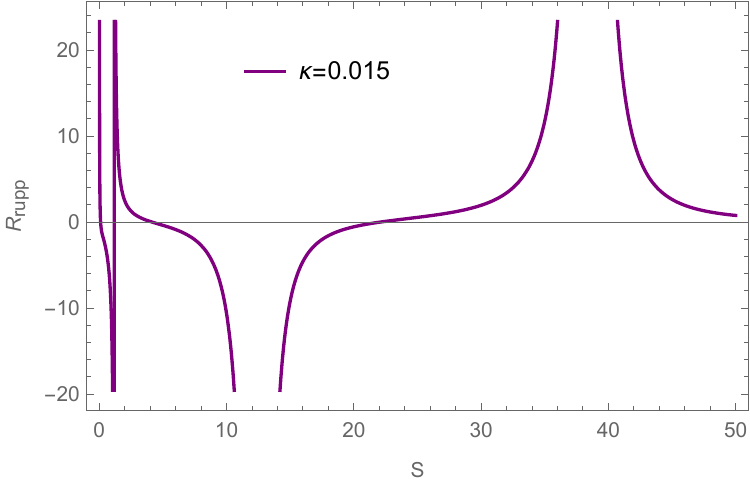}
		\caption{For $\kappa=0.015$ }
		\label{a9b}
		\end{subfigure}
 \begin{subfigure}{0.4\textwidth}
		\includegraphics[width=\linewidth]{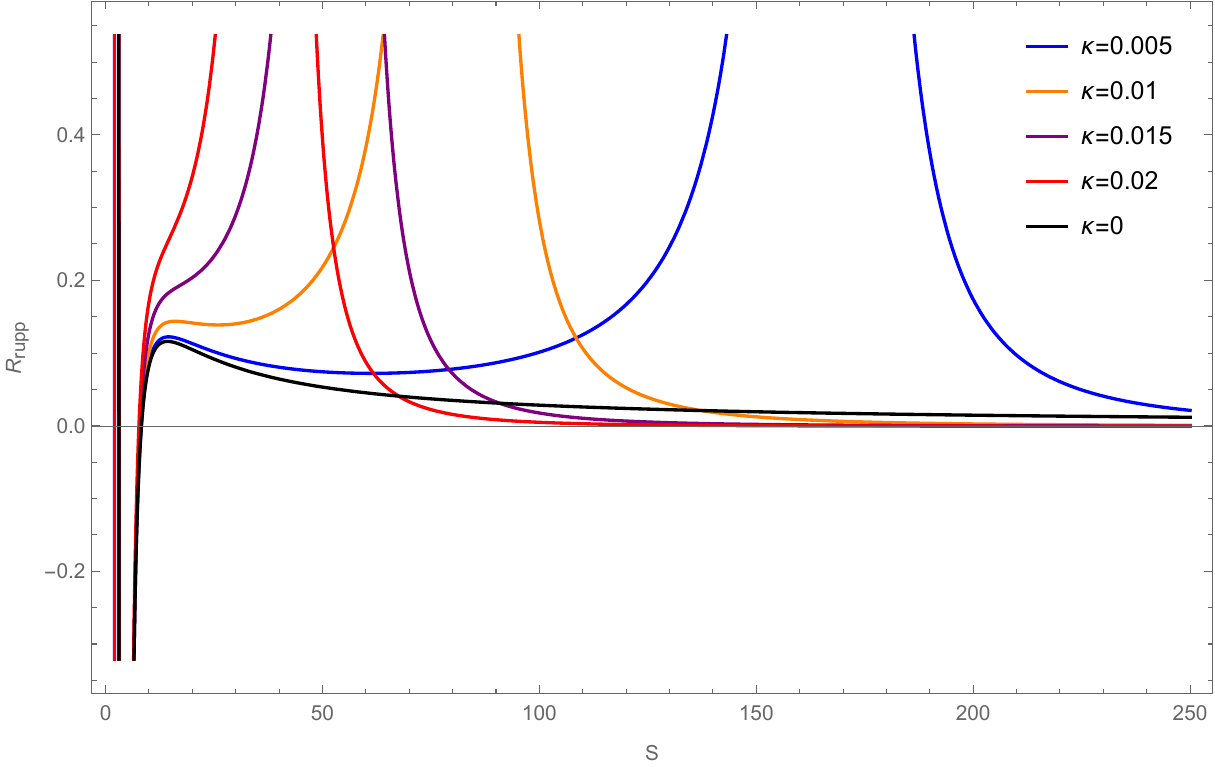}
		\caption{For all $\kappa$ }
		\label{a10a}
		\end{subfigure}
		\begin{subfigure}{0.4\textwidth}
		\includegraphics[width=\linewidth]{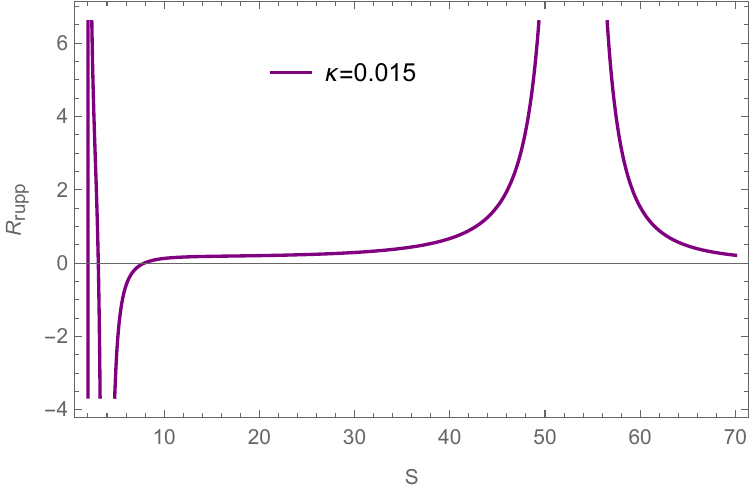}
		\caption{For $\kappa=0.015$ }
		\label{a10b}
		\end{subfigure}
	\caption{ The first panel corresponds to the Ruppeiner scalar versus Entropy plot for Kerr-Newman-AdS black hole for $J=Q=0.2$ and $L=3.2$ and the second panel represents the Ruppeiner scalar versus Entropy plot for Kerr-Newman-AdS black hole for $J=Q=0.5$ and $L=1.5$.}
	\label{a10}
 \end{figure}

\begin{figure}[h]	
	\centering
	\begin{subfigure}{0.4\textwidth}
		\includegraphics[width=\linewidth]{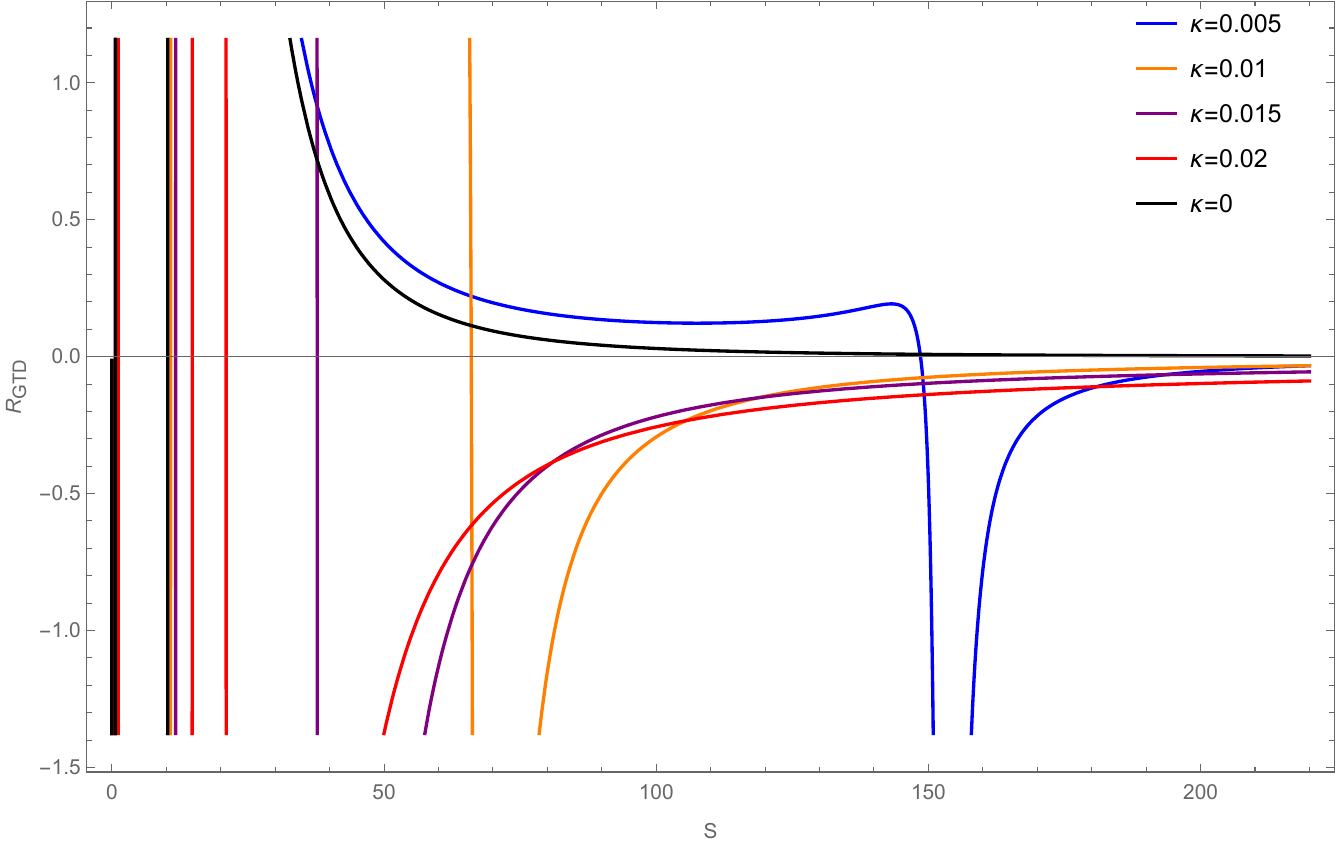}
		\caption{For all $\kappa$}
		\label{a11a}
		\end{subfigure}
		\begin{subfigure}{0.4\textwidth}
		\includegraphics[width=\linewidth]{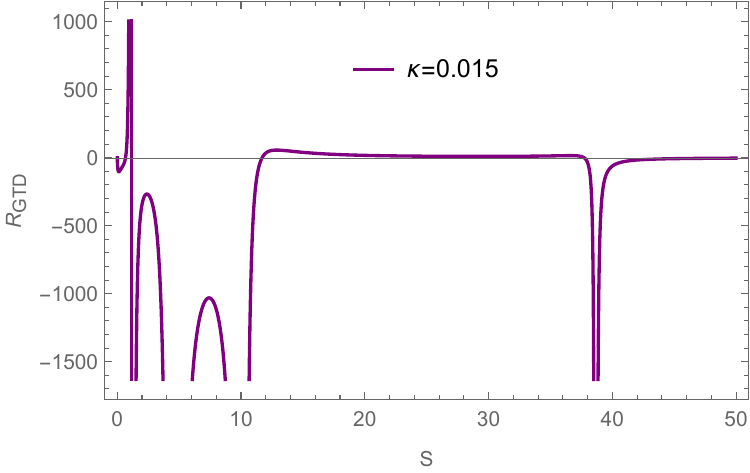}
		\caption{For $\kappa=0.015$ }
		\label{a11b}
		\end{subfigure}
 \begin{subfigure}{0.4\textwidth}
		\includegraphics[width=\linewidth]{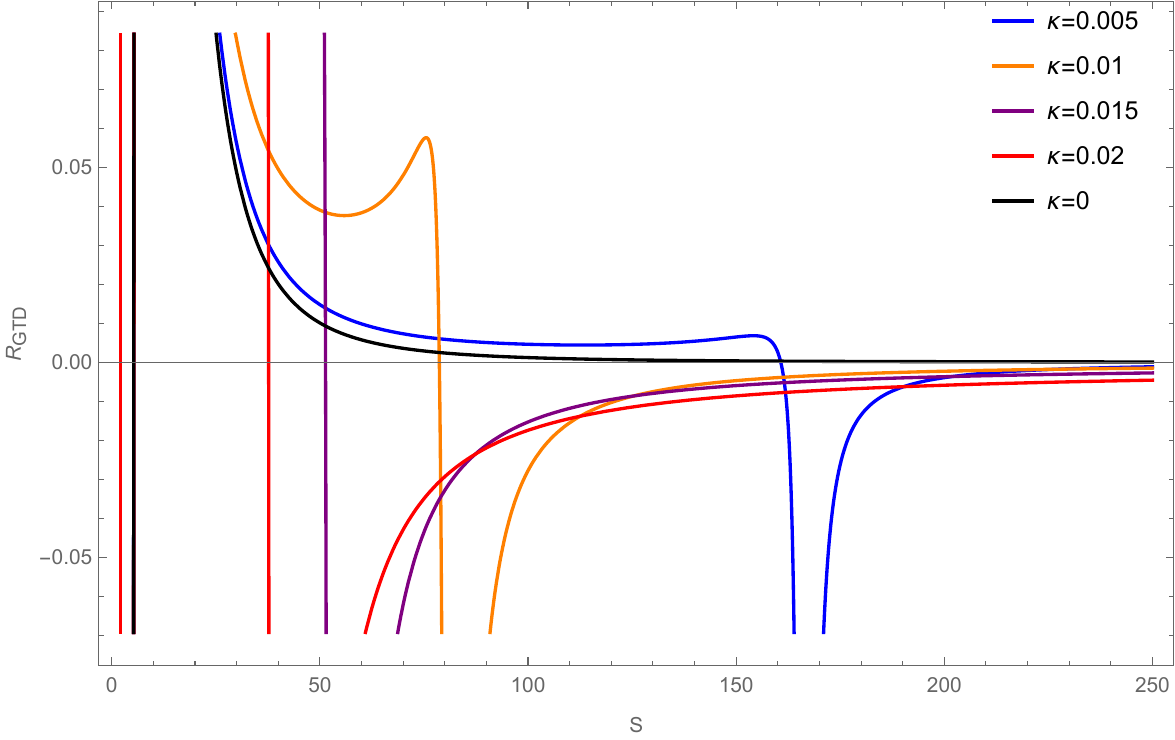}
		\caption{For all $\kappa$ }
		\label{a12a}
		\end{subfigure}
		\begin{subfigure}{0.4\textwidth}
		\includegraphics[width=\linewidth]{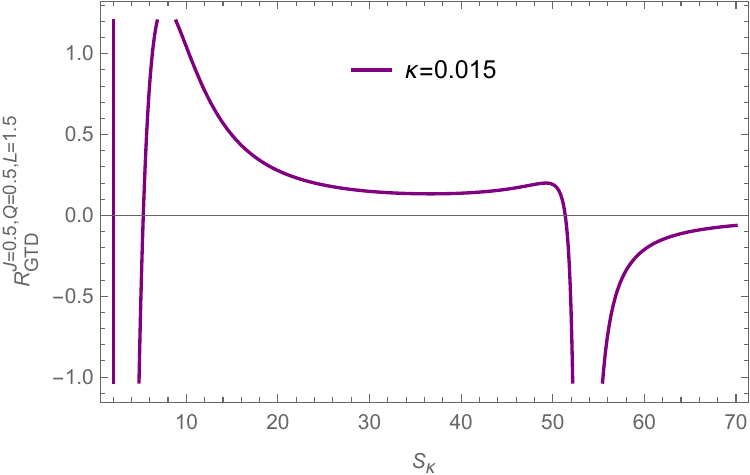}
		\caption{For $\kappa=0.015$ }
		\label{a12b}
		\end{subfigure}
	\caption{ The first panel corresponds to the GTD scalar versus Entropy plot for Kerr-Newman-AdS black hole for $J=Q=0.2$ and $L=3.2$ and the second panel shows the GTD scalar versus Entropy plot for Kerr-Newman-AdS black hole for $J=Q=0.5$ and $L=1.5$.}
	\label{a12}
 \end{figure}

 The line element for the Ruppeiner metric in the Kerr-Newman-AdS black hole could be written as :
$$dS_{R}^{2}= \frac{1}{T} \left(\frac{\partial^2 M}{\partial S^2}dS^2 + \frac{\partial^2 M}{\partial J^2} dJ^2 +\frac{\partial^2 M}{\partial Q^2}  dQ^2 + 2\frac{\partial^2 M}{\partial S \partial J} dS dJ + 2 \frac{\partial^2 M}{\partial J \partial Q}  dJ dQ + 2\frac{\partial^2 M}{\partial S \partial Q} dS dQ\right)$$

and the line element for the GTD metric in the Kerr-Newman-AdS black hole is given by:
$$dS_{G}^{2}=S \left(\frac{\partial M}{\partial S}\right)\left(- \frac{\partial^2 M}{\partial S^2} dS^2 + \frac{\partial^2 M}{\partial J^2} dJ^2 +  \frac{\partial^2 M}{\partial Q^2}dQ^2\right) $$
 We see from Fig.\ref{a9b} that the Ruppeiner scalar is plotted against the Kaniadakis entropy for $J=Q=0.2$ and $L=3.2$. We see the Ruppeiner scalar curve has three singularities at $S=1.174, 12.392$ and $38.407$ for $\kappa=0.015$ whereas in Fig.\ref{a10b} we see that for $J=Q=0.5$ and $L=1.5$ there is only one singularity at $S=53.090$ for $\kappa=0.015$. The GTD scalar on the other hand has singularities at  $S=4.545, 10.051$ and $38.659$ for $J=Q=0.2$ and $L=3.2$ as can be seen in Fig.\ref{a11b}. We see from Fig.\ref{a12b} that for $J=Q=0.5$ and $L=1.5$ there is only one singularity at $S=53.214$. The singularities in the GTD scalar curves agree with the points of Davies type phase transitions obtained from the heat capacity curves. 

 \section{Conclusions}
 In this paper,  we studied the thermodynamic properties and phase transitions in rotating anti-de Sitter (AdS) black holes by applying the Kaniadakis entropy framework. We analyzed three rotating AdS black hole systems: the Kerr-AdS-black hole, the Kerr-Sen-AdS black hole, and the Kerr-Newman-AdS black hole. We assessed their thermodynamic quantities, phase transitions,  thermodynamic topology and  thermodynamic geometry within the Kaniadakis statistical framework. \\
 
For all three black holes, the impact of the Kaniadakis parameter $\kappa$ on the black hole mass and temperature becomes more pronounced at higher values of Kaniadakis entropy $S$. The range of $S$ within which the black hole mass in Kaniadakis statistics approximates the mass in Gibbs-Boltzmann statistics varies with $\kappa$. In the $T$ vs. $S$ diagram, Gibbs-Boltzmann statistics typically show either one or three branches: one branch indicates no phase transition, while three branches correspond to a Van der Waals-like small-intermediate-large black hole phase transition. In contrast, when using Kaniadakis entropy, we observe either two or four branches. Besides the small-intermediate-large black hole branches, an additional branch, the ultra-large black hole branch (ULBH), emerges. The number of branches in the $T$ vs. $S$ diagram depended on the thermodynamic parameters of the black hole system.\\

The study of heat capacity plots of all three black hole systems. revealed that, in Gibbs-Boltzmann statistics, the number of discontinuities in the  heat capacity versus entropy is either zero or two. Conversely, in Kaniadakis statistics, the number of discontinuities is either one or three. The sign of the  heat capacity indicates the stability of the system. In Kaniadakis statistics, we find either two stable and two unstable branches or one stable and one unstable branch. In Gibbs-Boltzmann statistics, we found either two stable and one unstable branch or just one stable branch.\\

We also studied the free energy $F$ as a function of $S$ and $T$ for each black hole system. Moreover, we used the free energy landscape to estimate the stability of the black hole solutions. By constructing an off-shell free energy with an ensemble temperature $T_e$, we determined that a physical black hole can only exist at the extremal points of the off-shell free energy. Thus, at the maximum point of the off-shell free energy curve, a black hole is considered unstable, whereas at the minimum point, the black hole is considered stable. This approach allows us to analyze both the global and local stability of each black hole.\\

Next,we investigated the thermodynamic topology of each black hole system. In our analysis, using the off shell free energy method, all the black holes are treated as topological defects.The winding numbers at those defects is calculated in order to understand the topology of the  thermodynamic spaces of these black holes.Our primary motivation was to explore the  impact of KD entropy on the  topological charge of the black hole.Our analysis revealed that, under KD statistics, the topological charge for all three black hole systems is $0$, whereas in Gibbs-Boltzmann (GB) statistics, the charge is $+1$. Furthermore, in KD statistics, we observed one or zero generation point and two or one annihilation point, while in GB statistics, we found one or zero annihilation point and generation point. That means we will always encounter a annihilation point in KD entropy framework regardless of the values of other thermodynamic quantities.This further indicates that transitioning from GB to KD statistics results in change of  the topological charge and the topological class of these rotating black holes.The topological charge is found to be independent of all the thermodynamic parameter in both the statistics.\\

This paper also explored  the thermodynamic geometry of these rotating AdS black holes by applying the Ruppeiner and geometrothermodynamic (GTD) formalisms. Our analysis revealed that singularities in the scalar curvature are present in both statistical frameworks. Notably, in the Ruppeiner formalism, these singularities do not align with the Davies points observed in the heat capacity curves previously analyzed. Conversely, in the GTD formalism, the singularities in the scalar curvature matches exactly with the Davies points found in the heat capacity curves.\\

Our analysis reveals that in all three cases, there are clear deviations from the Gibbs-Boltzmann (GB) statistical framework. A significant insight from this study is the identification of an entropy bound introduced by Kaniadakis entropy, beyond which the black hole solution becomes unstable, as evidenced by our study of thermodynamic topology. Unlike the GB framework, where black holes remain globally stable across an infinite range of entropy, the stability in the Kaniadakis framework is restricted by this new entropy bound, governed by the Kaniadakis parameter \(\kappa\). Notably, smaller values of \(\kappa\) correspond to higher entropy bounds, implying that the stability range expands as \(\kappa\) decreases. This inherent instability of rotating black holes under Kaniadakis entropy is closely tied to the unique formulation of the entropy, whose mathematical structure is consistent with both Einstein's special theory of relativity and the second law of thermodynamics. The profound impact of Kaniadakis entropy on the thermodynamic phase space of rotating AdS black holes is further validated through our study of thermodynamic geometry and thermodynamic topology. To put it briefly, the introduction of the Kaniadakis entropy results in additional unstable black hole branches as can be seen from the heat capacity curves. These additional branches were not to be seen for the traditional black hole entropy and are entirely responsible for the changes that take place in both the thermodynamic topology and the thermodynamic geometry of these rotating AdS black holes. We see a change in topological charge along with the number of generation and anihilation points. Also additional curvature singularities are observed in the curvature scalar of these black holes which are a result of the discontinuity in phase structure of the black holes brought about by the KD framework.A comparison of the thermodynamic geometry and topology in the Bekenstein-Hawking and Kaniadakis entropy frameworks for rotating AdS black holes is presented in Table (\ref{table1}), which also summarizes the new insights gained from this study.

\begin{table}[h!]
\centering
\renewcommand{\arraystretch}{1.2}
\setlength{\tabcolsep}{8pt}
\scriptsize
\begin{tabular}{|p{4cm}|p{5cm}|p{5cm}|}
\hline
\textbf{Aspect} & \textbf{Bekenstein-Hawking \hspace{-0.04cm}entropy Framework} & \textbf{Kaniadakis entropy Framework} \\ \hline
\textbf{Topological Charge} & $+1$ & $0$ \\ \hline
\textbf{Thermodynamic Topology (Generation/Annihilation Points)} & One or zero generation and annihilation point & Two or one annihilation point and one or zero generation point, indicating a shift in topological behavior due to entropy bound \\ \hline
\textbf{Entropy Bound} & Infinite stability range & Finite stability range governed by \(\kappa\) \\ \hline
\textbf{Number of Scalar Curvature Singularities} & Two or zero & Three or one \\ \hline
\textbf{Singularities in Ruppeiner Formalism} & Do not match Davies points & Do not match Davies points \\ \hline
\textbf{Singularities in GTD Formalism} & Match Davies points & Match Davies points \\ \hline
\textbf{Stability Behavior} & Globally stable across all entropy values & Stability limited by critical entropy bound \\ \hline
\textbf{Effect of \(\kappa\)} & Not applicable & Lower \(\kappa\) leads to higher entropy bounds and expanded stability range \\ \hline
\end{tabular}
\caption{Comparison of thermodynamic geometry and topology in the GB and KD frameworks for rotating AdS black holes.}
\label{table1}
\end{table}
It is important to emphasize that our results remain continuous in the limit$ \kappa \to 0$ thus smoothly recovering the standard thermodynamic behavior governed by Bekenstein-Hawking entropy. The nontrivial deviations introduced by Kaniadakis entropy appears only for $\kappa \neq 0$, which implies the modifications that the deformed statistical framework inherently contains. Therefore, the qualitative differences that has been discussed in this work should be stringently understood as emerging purely as a result of the nonzero deformation parameter $(\kappa \neq 0)$ and not by any discontinuity in the formalism.\\

We also found that Kaniadakis statistics introduces similar modifications in modified gravity frameworks as well. Specifically, the entropy bound induced by Kaniadakis entropy leads to shifts in the stability pattern and phase transitions of black holes, analogous to what is observed in certain modified gravity scenarios such as \cite{GGL1,GGL2}. This further highlights the universal role of entropy modifications in black hole thermodynamics.\\

Apart from the field of black holes thermodynamics,  the Kaniadakis framework, offering a generalized form of entropy, has demonstrated effectiveness in diverse high-energy contexts, including astrophysics, cosmology, gravitation\cite{ka1,ka11,ka2,ka3,ka4,ka5,ka6,ka7,ka8}, and particle physics \cite{particle1, particle2}.  Its applicability extends beyond these realms, encompassing fields like epidemiology \cite{epid}, seismology \cite{seis}, economics \cite{eco}, and natural sciences \cite{neu}. This broad success suggests its potential for extending conventional statistical mechanics to systems exhibiting intricate correlations, such as those encountered in  the AdS/CFT correspondence\cite{ka1,ka11} and other stellar phenomenon\cite{GGL3,app} and  therefore, the Kaniadakis entropy offers a promising avenue for investigating complex physical systems.\\

It is to be noted that  Kaniadakis statistics was originally proposed in the context of relativistic statistical mechanics and its application to black hole thermodynamics is still very recent\cite{ka0,Sekhmani:2024kfj,Sadeghi:2024pme}. The validity of  Kaniadakis entropy for black hole thermodynamics  remains an open question, and our work here represents an attempt to explore its consequences in the context of rotating 
AdS black hole.  Moreover, as has already been alluded to previously  in introduction,  generalized entropies such as Kaniadakis, Tsallis, and Rényi often lead to inconsistencies in black hole energy or temperature.\cite{gen2,gen4}. Contrary to that  for more complicated BHs non-extreme entropies may give correct thermodynamic quantities as shown in Ref. \cite{r2}.  It will  be interesting to explore these open questions in the context of non-extensive entropy framework of black hole thermodynamics.%%%%%%%%%%%

\section{Acknowledgments}
BH would like to thank DST-INSPIRE, Ministry of Science and Technology fellowship program, Govt. of India for awarding the DST/INSPIRE Fellowship[IF220255] for financial support.

\end{document}